\documentclass[12pt]{article}
\usepackage{latexsym}
\usepackage{amsmath,amsfonts}
\usepackage{times}
\allowdisplaybreaks[4]

\catcode`\@=11

\newcount\hour
\newcount\minute
\newtoks\amorpm \hour=\time\divide\hour by 60\minute
=\time{\multiply\hour by 60 \global\advance\minute by-\hour}
\edef\standardtime{{\ifnum\hour<12 \global\amorpm={am}%
        \else\global\amorpm={pm}\advance\hour by-12 \fi
        \ifnum\hour=0 \hour=12 \fi
        \number\hour:\ifnum\minute<10
        0\fi\number\minute\the\amorpm}}
\edef\militarytime{\number\hour:\ifnum\minute<10
0\fi\number\minute}

\def\draftlabel#1{{\@bsphack\if@filesw {\let\thepage\relax
   \xdef\@gtempa{\write\@auxout{\string
      \newlabel{#1}{{\@currentlabel}{\thepage}}}}}\@gtempa
   \if@nobreak \ifvmode\nobreak\fi\fi\fi\@esphack}
        \gdef\@eqnlabel{#1}}
\def\@eqnlabel{}
\def\@vacuum{}
\def\marginnote#1{}
\def\draftmarginnote#1{\marginpar{\raggedright\scriptsize\tt#1}}
\def\draft{
        \pagestyle{plain}
        \overfullrule=2pt
        \oddsidemargin -.5truein
        \def\@oddhead{\sl \phantom{\today\quad\militarytime} \hfil
        \smash{\Large\sl DRAFT} \hfil \today\quad\militarytime}
        \let\@evenhead\@oddhead
        \let\label=\draftlabel
        \let\marginnote=\draftmarginnote
        \def\ps@empty{\let\@mkboth\@gobbletwo
        \def\@oddfoot{\hfil \smash{\Large\sl DRAFT} \hfil}
        \let\@evenfoot\@oddhead}
        \def\@eqnnum{(\theequation)\rlap{\kern\marginparsep\tt\@eqnlabel}%
        \global\let\@eqnlabel\@vacuum}  }

\newcommand{\rf}[1]{(\ref{#1})}
\renewcommand{\theequation}{\thesection.\arabic{equation}}
\renewcommand{\thefootnote}{\fnsymbol{footnote}}
\newcommand{\newsection}{    
\setcounter{equation}{0}\section}

\def\appendix#1{\addtocounter{section}{1}\setcounter{equation}{0}
\renewcommand{\thesection}{\Alph{section}}
\section*{Appendix \thesection\protect\indent \parbox[t]{11.15cm}{#1}}
\addcontentsline{toc}{section}{Appendix \thesection\ \ \ #1}}

\def\nline{\,\nabla\kern -0.7em\raise0.2ex\hbox{/}\,\,}
\def\yline{\,y\kern -0.47em /}
\def\aline{\,a\kern -0.49em /}
\def\parline{\,\partial\kern -0.55em /\,\,}

\newcommand{\Mo}{\mathbb{M}}
\newcommand{\No}{\mathbb{N}}
\newcommand{\Po}{\mathbb{P}}

\def\be{\begin{equation}}
\def\ee{\end{equation}}
\def\beq{\begin{eqnarray}}
\def\eeq{\end{eqnarray}}

\def\smCSF{{\scriptscriptstyle CSF}}

\def\smpt{{\scriptscriptstyle [2]}}
\def\smp3{{\scriptscriptstyle [3]}}

\def\smpn{{\scriptscriptstyle [n]}}

\def\Jbf{{\bf J}}

\def\Mbf{{\bf M}}
\def\Pbf{{\bf P}}

\def\alphabf{{\boldsymbol{\alpha}}}
\def\betabf{{\boldsymbol{\beta}}}
\def\gammabf{{\boldsymbol{\gamma}}}

\def\ibf{{\bf i}}
\def\iibf{{\bf ii}}
\def\iiibf{{\bf iii}}
\def\ivbf{{\bf iv}}

\def\Asf{{\sf A}}
\def\Bsf{{\sf B}}
\def\Csf{{\sf C}}

\def\vph{{\vphantom{5pt}}}

\def\half{\frac{1}{2}}

\def\alphab{\bar{\alpha}}
\def\upsilonb{\bar{\upsilon}}
\def\zetab{\bar{\zeta}}

\def\gb{{\bar{g}}}

\def\phik{|\phi\rangle}

\def\irm{{\rm i}}

\def\dyn{{\rm dyn}}

\def\betach{\check{\beta}}

\begin{document}


\begin{flushright}
FIAN-TD-2018-19 \ \  \ \ \ \\
arXiv: 1809-09075V3
\end{flushright}

\vspace{1cm}

\begin{center}

{\Large \bf Cubic interaction vertices for massive/massless

\medskip
continuous-spin fields and arbitrary spin fields}

\vspace{2.5cm}

R.R. Metsaev\footnote{ E-mail: metsaev@lpi.ru }

\vspace{1cm}

{\it Department of Theoretical Physics, P.N. Lebedev Physical
Institute, \\ Leninsky prospect 53,  Moscow 119991, Russia }

\vspace{3cm}

{\bf Abstract}

\end{center}

We use light-cone gauge formalism to study interacting massive and massless continuous-spin fields and finite component arbitrary spin fields propagating in the flat space. Cubic interaction vertices for such fields are considered. We obtain parity invariant cubic vertices for coupling of one continuous-spin field to two arbitrary spin fields and cubic vertices for coupling of two continuous-spin fields to one arbitrary spin field. Parity invariant cubic vertices for self-interacting massive/massless continuous-spin fields are also obtained. We find the complete list of parity invariant cubic vertices for continuous-spin fields and arbitrary spin fields.

\vspace{3cm}


\newpage
\renewcommand{\thefootnote}{\arabic{footnote}}
\setcounter{footnote}{0}

\section{ \large Introduction}

Continuous-spin field propagating in flat space is associated with continuous-spin representation of the Poincar\'e algebra (for review, see Refs.\cite{Bekaert:2006py}-\cite{Brink:2002zx}). From the point of view of field theory, the continuous-spin field provides interesting example of relativistic dynamical system which involves infinite number of coupled finite component fields. In this respect the continuous-spin field theory has some features in common with string theory and higher-spin theory.
For example, we note that the continuous-spin field is decomposed into an infinite chain of coupled scalar, vector, and totally symmetric tensor fields which consists of every spin just once. We recall then a similar infinite chain of scalar, vector and totally symmetric fields appears in the theory of higher-spin gauge field in AdS space \cite{Vasiliev:1990en}.
We note also the intriguing discussions about possible interrelations between continuous-spin field theory and the string theory in Refs.\cite{Savvidy:2003fx}.%
\footnote{Note that, in Ref.\cite{Font:2013hia}, it was observed that the continuous-spin massless fields do not appear in perturbative string theory spectrum. We believe that continuous-spin fields could be appeared in the full-fledged string theory.}
In view of just mentioned and other interesting features, the continuous-spin field theory has attracted some interest  recently (see, e.g., Refs.\cite{Bekaert:2005in}-\cite{Najafizadeh:2017tin}).

In Ref.\cite{Metsaev:2017cuz}, we developed the light-cone gauge formulation of  massless and massive continuous-spin fields propagating in the flat space $R^{d-1,1}$ with arbitrary $d\geq 4$.%
\footnote{ Light-cone gauge free continuous-spin massless fields in $R^{3,1}$ and $R^{4,1}$ were discussed in Ref.\cite{Brink:2002zx}. For $d\geq 4$, the discussion of UIR of the Poincar\'e algebra $iso(d-1,1)$ may be found in Ref.\cite{Bekaert:2006py}. We think that our  continuous-spin massive field is associated with the tachyonic UIR of the Poincar\'e algebra which is discussed in Sec.3 in Ref.\cite{Bekaert:2006py}.}
Also, in Ref.\cite{Metsaev:2017cuz}, we applied our formulation to study parity invariant cubic vertices for coupling of massless continuous-spin fields to massive arbitrary spin fields and obtained complete list of such cubic vertices.%
\footnote{ We note that continuous-spin massless/massive fields are infinite component fields. In this paper, finite component massless/massive fields with arbitrary but fixed values of spin will be referred to as arbitrary spin fields.}
Cubic vertices involving continuous-spin fields have one interesting and intriguing feature in common with interaction vertices of string theory and higher-spin theory. It turns out that the interaction vertices for coupling of continuous-spin fields to arbitrary spin fields involve infinite number of derivatives. It seems therefore highly likely that the continuous-spin field theory is the interesting and promising direction to go.

This paper is a continuation of the investigation of cubic interaction vertices for continuous-spin fields and arbitrary spin field begun in Ref.\cite{Metsaev:2017cuz}. In Ref.\cite{Metsaev:2017cuz}, we studied cubic vertices which involve massless continuous-spin fields and massive/massless arbitrary spin fields, while, in this paper, we study cubic vertices which involve massive/massless continuous-spin fields and massive/massless arbitrary spin fields. We recall that, in general, a continuous-spin field is labelled by mass parameter, which we denote by $m$, $m^2 \leq 0$,  and continuous-spin parameter, which we denote by $\kappa$, $\kappa>0$. To indicate such continuous-spin field we use the shortcut $(m,\kappa)_\smCSF^\vph$. A finite-component arbitrary spin field is labeled by mass parameter, denoted by $m$, $m^2\geq 0$, and spin value denoted by integer $s$, $s\geq 0$. Such finite component field will be denoted as $(m,s)$. We now note that cubic vertices involving massless continuous-spin fields and massive/massless arbitrary spin fields can be classified as

\vspace{-0.8cm}
\beq
&& \hspace{-1.5cm} \hbox{\bf Cubic vertices with one massless continuous-spin field}.
\nonumber\\
\label{14092018-man03-01-int-a1} && (0,s_1)\hbox{-}(0,s_2)\hbox{-}(0,\kappa_3)_\smCSF \hspace{1.5cm}  -
\\
\label{14092018-man03-01-int-q} && (0,s_1)\hbox{-}(m_2,s_2)\hbox{-}(0,\kappa_3)_\smCSF \hspace{1.3cm}  m_2^2 > 0\,,
\\
\label{14092018-man03-01-int} && (m_1,s_1)\hbox{-}(m_2,s_2)\hbox{-}(0,\kappa_3)_\smCSF \hspace{1cm}   m_1 = m\,, \hspace{0.5cm} m_2 = m\,, \hspace{0.4cm} m^2 > 0\,,
\\
\label{15092018-man03-01-int} && (m_1,s_1)\hbox{-}(m_2,s_2)\hbox{-}(0,\kappa_3)_\smCSF \hspace{1cm} m_1 \ne m_2\,, \hspace{0.4cm} m_1^2 > 0\,, \hspace{0.6cm} m_2^2 > 0\,;
\\
&& \hspace{-1.5cm} \hbox{\bf Cubic vertices with two massless continuous-spin field}.
\nonumber\\
\label{14092018-man03-01-int-a2} && (0,\kappa_1)_\smCSF\hbox{-}(0,\kappa_2)_\smCSF\hbox{-}(0,s_3)  \hspace{1.1cm}  -
\\
\label{16092018-man03-01-int} && (0,\kappa_1)_\smCSF\hbox{-}(0,\kappa_2)_\smCSF\hbox{-}(m_3,s_3), \hspace{0.5cm} m_3^2 > 0\,,
\\
&& \hspace{-1.5cm} \hbox{\bf Cubic vertices with three massless continuous-spin field}.
\nonumber\\
\label{14092018-man03-01-int-a3} && (0,\kappa_1)_\smCSF\hbox{-}(0,\kappa_2)_\smCSF\hbox{-}(0,\kappa_3)_\smCSF  \hspace{1cm}  ?
\eeq
Cubic vertices for fields in \rf{14092018-man03-01-int-a1}-\rf{16092018-man03-01-int} were studied in Ref.\cite{Metsaev:2017cuz}. Namely, in Ref.\cite{Metsaev:2017cuz}, we built the complete list of parity invariant cubic vertices for fields in \rf{14092018-man03-01-int},\rf{15092018-man03-01-int}, \rf{16092018-man03-01-int}.%
\footnote{ Cubic vertices for fields in \rf{14092018-man03-01-int-q} were lost while preparing paper in Ref.\cite{Metsaev:2017cuz}. We present our result for the cubic vertices \rf{14092018-man03-01-int-q} in Appendix E in this paper.}
Also, in Ref.\cite{Metsaev:2017cuz}, we demonstrated that, there are no parity invariant vertices for fields in \rf{14092018-man03-01-int-a1},\rf{14092018-man03-01-int-a2}. This is to say that, in the framework of $so(d-2)$ covariant light-cone gauge formalism, there are no parity invariant cubic vertices for coupling of continuous-spin massless fields to finite component arbitrary spin massless fields.
It remains to investigate whether cubic vertices describing self interacting continuous-spin massless field \rf{14092018-man03-01-int-a3} do exist.%
\footnote{ Cubic vertex for three massless continuous-spin fields \rf{14092018-man03-01-int-a3} is a problem for the future research.}

In this paper, we study cubic vertices which involve, among other fields, at least one massive continuous-spin field. Such cubic vertices we separate in the following three groups.

\vspace{-0.8cm}
\beq
&& \hspace{-2cm}  \hbox{\bf Cubic vertices with one continuous-spin field:}
\nonumber\\
\label{03092018-man03-01-int} &&  \hspace{-1cm} (0,s_1)\hbox{-}(0,s_2)\hbox{-}(m_3,\kappa_3)_\smCSF^\vph\,, \hspace{2.4cm} m_3^2 <  0;
\\
\label{04092018-man03-01-int} && \hspace{-1cm} (0,s_1)\hbox{-}(m_2,s_2)\hbox{-}(m_3,\kappa_3)_\smCSF^\vph\,, \hspace{2.1cm} m_2^2 > 0\,,  \hspace{1cm}   m_3^2 <  0\,,
\\
\label{05092018-man03-01-int} && \hspace{-1cm} (m_1,s_1)\hbox{-}(m_2,s_2)\hbox{-}(m_3,\kappa_3)_\smCSF^\vph\,, \hspace{1.8cm}   m_1^2 > 0\,, \hspace{1cm}    m_2^2 > 0\,,  \hspace{0.7cm}  m_3^2 <  0\,;
\\
&& \hspace{-2cm}  \hbox{\bf Cubic vertices with two continuous-spin fields:}
\nonumber\\
\label{06092018-man03-01-int} && \hspace{-1cm} (m_1,\kappa_1)_\smCSF\hbox{-}(0,\kappa_2)_\smCSF\hbox{-}(0,s_3)\,, \hspace{1.8cm}    m_1^2 < 0\,,
\\
\label{07092018-man03-01-int} && \hspace{-1cm} (m_1,\kappa_1)_\smCSF\hbox{-}(m_2,\kappa_2)_\smCSF\hbox{-}(0,s_3)\,,  \hspace{1.5cm}    m_1=m\,, \hspace{0.7cm}  m_2=m\,, \hspace{0.7cm}  m^2 < 0\,,
\\
\label{08092018-man03-01-int} && \hspace{-1cm} (m_1,\kappa_1)_\smCSF\hbox{-}(m_2,\kappa_2)_\smCSF\hbox{-}(0,s_3)\,, \hspace{1.5cm}     m_1^2 < 0\,,\qquad    m_2^2 < 0\,,  \hspace{0.8cm}  m_1\ne m_2\,,
\\
\label{08092018-man03-01-int-q} && \hspace{-1cm} (0,\kappa_1)_\smCSF\hbox{-}(m_2,\kappa_2)_\smCSF\hbox{-}(m_3,s_3)\,,  \hspace{1.5cm}     m_2^2 < 0\,,\qquad m_3^2>0\,,
\\
\label{09092018-man03-01-int} && \hspace{-1cm} (m_1,\kappa_1)_\smCSF\hbox{-}(m_2,\kappa_2)_\smCSF\hbox{-}(m_3,s_3)\,, \hspace{1.2cm}     m_1^2 < 0\,,\qquad    m_2^2 < 0\,,\qquad m_3^2 > 0 \,;
\\
&& \hspace{-2cm} \hbox{\bf Cubic vertices with three continuous-spin fields:}
\nonumber\\
\label{10092018-man03-01-int} && \hspace{-1cm} (0,\kappa_1)_\smCSF\hbox{-}(0,\kappa_2)_\smCSF\hbox{-}(m_3,\kappa_3)_\smCSF\,, \hspace{1.2cm}    m_3^2 < 0 \,,
\\
\label{11092018-man03-01-int} && \hspace{-1cm} (m_1,\kappa_1)_\smCSF\hbox{-}(m_2,\kappa_2)_\smCSF\hbox{-}(0,\kappa_3)_\smCSF\,, \qquad     m_1=m\,, \hspace{0.6cm} m_2=m\,, \qquad m^2 < 0 \,,
\\
\label{12092018-man03-01-int} && \hspace{-1cm} (m_1,\kappa_1)_\smCSF\hbox{-}(m_2,\kappa_2)_\smCSF\hbox{-}(0,\kappa_3)_\smCSF\,, \qquad     m_1^2<0\,, \qquad m_2^2<0\,, \qquad m_1^2 \ne m_2^2 \,,
\\
\label{13092018-man03-01-int} && \hspace{-1cm} (m_1,\kappa_1)_\smCSF\hbox{-}(m_2,\kappa_2)_\smCSF\hbox{-}(m_3,\kappa_3)_\smCSF\,, \hspace{0.5cm}   m_1^2 < 0\,, \hspace{0.9cm}     m_2^2 < 0\,,\hspace{0.9cm} m_3^2 < 0 \,.
\eeq
In this paper, we build the complete list of parity invariant cubic vertices for fields in  \rf{03092018-man03-01-int}-\rf{13092018-man03-01-int}.%
\footnote{ Lorentz covariant cubic vertex for coupling of massless continuous-spin field to two massive scalar fields (case \rf{15092018-man03-01-int} with $s_1=s_2=0$) was studied in Ref.\cite{Bekaert:2017xin}, while Lorentz covariant cubic vertex for coupling of massive  continuous-spin field to two massive scalar fields (case \rf{05092018-man03-01-int} with $s_1=s_2=0$) was studied in Ref.\cite{Rivelles:2018tpt}.}
With exception of vertices \rf{14092018-man03-01-int-a3}, results in this paper together with the ones in Ref.\cite{Metsaev:2017cuz} provide the exhaustive solution to the problem of description of all parity invariant cubic vertices for massive/massless continuous spin fields and massive/massless arbitrary spin fields in \rf{14092018-man03-01-int-a1}-\rf{13092018-man03-01-int}.

We now note the general features of our vertices. \ibf) As in string field theory, our vertices involve infinite number of derivatives. Also, as in light-cone gauge string field theory, our vertices are nonlocal wrt the spatial light-cone coordinate $x^-$ and expandable in derivatives wrt the transverse spatial transverse coordinates $x^i$.
\iibf) Our vertices for massive fields are singular when the masses tend to zero.  \iiibf) In contrast to the massless continuous-spin fields, the massive continuous-spin fields admit coupling to arbitrary spin massless fields.

This paper is organized as follows. In Sec.\ref{sec-02}, we briefly review the manifestly $so(d-2)$
covariant light-cone gauge formulation of free continuous-spin massless and massive fields in $R^{d-1,1}$ developed in Ref.\cite{Metsaev:2017cuz}. For the reader convenience, we also recall the textbook  light-cone gauge formulation of finite-component arbitrary spin massless and massive fields. In Sec.\ref{sec-03}, we review the setup for studying cubic vertices which was developed in Ref.\cite{Metsaev:2017cuz}.
We present the complete system of equations required to determine cubic vertices uniquely.   For the case of parity invariant cubic vertices, we present the particular form of the complete system of equations which is convenient for our study.

In Sec.\ref{sec-one-cont}, we present our result
for parity invariant cubic vertices describing coupling of one continuous-spin massive field to two massive/massless arbitrary spin fields,
while Sec.\ref{sec-two-cont} is devoted to study of cubic vertices for coupling of two massive/massless continuous-spin field to one massive/massless arbitrary spin field.
Section \ref{sec-three-cont} is devoted to cubic vertices for self-interacting massive/massless continuous-spin fields. Namely, we consider vertices for coupling of one massless continuous-spin field to two massive continuous-spin fields and vertices for coupling of one massive  continuous-spin field to two massless continuous-spin fields.  Cubic vertices for self-interacting massive continuous-spin field are also studied.

In Sec.\ref{sec-07}, we summarize our conclusions and suggest directions for future research.

In Appendix A, we describe notation we use in this paper. In Appendices B,C,D we outline some details of the derivation of cubic vertices. Namely, in Appendix B, we discuss vertices for coupling of one continuous-spin field to two arbitrary spin fields, while, in Appendix C, we discuss vertices for coupling of two continuous-spin fields to one arbitrary spin fields.  In Appendix D, we outline details of the derivation of vertices for self-interacting massive/massless continuous-spin fields. In appendix E, we study cubic vertex for one  massless continuous-spin field, one arbitrary spin massless field, and one arbitrary spin massive field.

\newsection{ \large Light-cone gauge formulation of free continuous-spin fields and arbitrary spin massive and massless fields}\label{sec-02}

Detailed discussion of light-cone formulation to free continuous-spin fields and arbitrary spin massive/massless fields may be found in Sec.2 in Ref.\cite{Metsaev:2017cuz}. Here, in order to make our presentation self-contained, we briefly review the most important ingredients of the light-cone formulation.

\noindent {\bf Poincar\'e algebra in light-cone frame}. We use the method in Ref.\cite{Dirac:1949cp} which reduces the problem of finding a new dynamical system  to a problem
of finding a new solution for commutators of a basic symmetry algebra. For the case under consideration, basic symmetries are associated with the Poincar\'e algebra.%
\footnote{ Application of light-cone formalism for studying vertices in string theory and arbitrary spin field theories may be found in Refs.\cite{Green:1983hw} and
\cite{Bengtsson:1983pd}-\cite{Skvortsov:2018jea}, while discussion of vertices of 11d SUGRA is given in Refs.\cite{Metsaev:2004wv,Ananth:2005vg}.
}
Therefore we start with the presentation of a realization of the Poincar\'e algebra symmetries on a space of light-cone gauge fields. The commutation relations of the Poincar\'e algebra $iso(d-1,1)$ are given by%
\footnote{ Indices $\mu,\nu,\rho,\sigma = 0,1,\ldots,d-1$ are vector indices of the Lorentz algebra $so(d-1,1)$.}
\be \label{01092018-man03-01}
{} [P^\mu,\,J^{\nu\rho}]=\eta^{\mu\nu} P^\rho - \eta^{\mu\rho} P^\nu\,,
\qquad {} [J^{\mu\nu},\,J^{\rho\sigma}] = \eta^{\nu\rho} J^{\mu\sigma} + 3\hbox{ terms}\,,
\ee
where $P^\mu$ are the translation generators, while $J^{\mu\nu}$ are generators of
the $so(d-1,1)$ Lorentz algebra. The $P^\mu$ taken to be hermitian, while the $J^{\mu\nu}$ are considered to be anti-hermitian. Flat metric $\eta^{\mu\nu}$ is taken to be mostly positive.

For the discussion of light-cone formulation, we usee, in place of
the Lorentz basis coordinates $x^\mu$, the light-cone basis
coordinates $x^\pm$, $x^i$, where the $so(d-2)$ algebra vector indices $i,j$ take values $i,j=1,\ldots,d-2$, while the coordinates $x^\pm$ are defined as $x^\pm \equiv (x^{d-1}  \pm x^0)/\sqrt{2}$. The coordinate $x^+$ is considered as an evolution parameter. We note then that the $so(d-1,1)$ Lorentz algebra vector $X^\mu$ is decomposed as $X^+,X^-,X^i$, while non vanishing elements of the flat metric are given by $\eta_{+-} = \eta_{-+}=1$, $\eta_{ij} = \delta_{ij}$. We note also that, in
light-cone approach, generators of the Poincar\'e algebra are
separated into kinematical and dynamical generators defined as
\beq
\label{01092018-man03-04} && P^+,\quad P^i,\quad J^{+i},\quad J^{+-},\quad J^{ij}, \qquad
\hbox{ kinematical generators};
\\
\label{01092018-man03-05}  && P^-,\quad J^{-i}\,, \hspace{4.3cm} \hbox{ dynamical generators}.
\eeq
In the field theory, the kinematical generators are quadratic in fields when $x^+ =0$%
\footnote{ The generators $P^i$, $P^+$, $J^{ij}$ do not depend on $x^+$, while the generators $J^{+i}$, $J^{+-}$  take the form $G= G_1 + x^+ G_2$, where $G_1$ is quadratic in fields, while $G_2$ involves quadratic and higher order terms in fields.}. The dynamical generators involve quadratic and, in general, higher order terms in fields.
Before discussion of the field theoretical realization of the kinematical and dynamical generators we discuss light-cone gauge description of the continuous-spin field, and arbitrary spin massive and massless fields.

\noindent {\bf Continuous-spin field, arbitrary spin massive field, and arbitrary spin massless field}. In light-cone gauge approach, continuous-spin field, spin-$s$ massive field, and spin-$s$ massless field, are described by the following set of scalar, vector, and tensor fields of the $so(d-2)$ algebra:
\beq
\label{01092018-man03-06} && \sum_{n=0}^\infty \oplus \,\,\phi^{i_1\ldots i_n}\,, \hspace{2cm} \hbox{for continuous-spin  field};
\\[-2pt]
\label{01092018-man03-07} && \sum_{n=0}^s \oplus \,\,\phi^{i_1\ldots i_n}\,, \hspace{2cm} \hbox{for spin-$s$ massive field};
\\[-2pt]
\label{01092018-man03-08} && \phi^{i_1\ldots i_s}\,, \hspace{3.2cm} \hbox{for spin-$s$ massless field}.
\eeq
In \rf{01092018-man03-06},\rf{01092018-man03-08}, fields $\phi^{i_1\ldots i_n}$ ($\phi^{i_1\ldots i_s}$) with $n=0$ ($s=0$) and $n=1$ ($s=1$) are the respective scalar and vector fields of the $so(d-2)$ algebra, while fields with $n\geq 2$  ($s\geq 2$) are totally symmetric traceless tensor fields of the $so(d-2)$ algebra, $\phi^{iii_3\ldots i_n} = 0$ ($\phi^{iii_3\ldots i_s} = 0$). Traceless constraint for massive fields \rf{01092018-man03-07} will be described below.

To simplify our presentation, we use oscillators  $\alpha^i$, $\zeta$, $\upsilon$ to collect fields \rf{01092018-man03-06}-\rf{01092018-man03-08} into respective ket-vectors defined as
\beq
\label{01092018-man03-09} && \hspace{-1cm} |\phi(p,\alpha)\rangle = \sum_{n=0}^\infty \frac{\upsilon^n}{n!\sqrt{n!}} \alpha^{i_1} \ldots \alpha^{i_n} \phi^{i_1\ldots i_n}(p) |0\rangle\,, \hspace{2cm} \hbox{for continuous-spin  field};
\\
\label{01092018-man03-10} && \hspace{-1cm} |\phi_s(p,\alpha)\rangle = \sum_{n=0}^s  \frac{\zeta^{s-n}}{n!\sqrt{(s-n)!}} \alpha^{i_1} \ldots \alpha^{i_n}
\phi^{i_1\ldots i_n}(p) |0\rangle\,, \hspace{0.7cm} \hbox{for spin-$s$ massive field};
\\
\label{01092018-man03-11} && \hspace{-1cm} |\phi_s(p,\alpha)\rangle = \frac{1}{s!} \alpha^{i_1} \ldots \alpha^{i_s} \phi^{i_1\ldots i_s}(p) |0\rangle\,, \hspace{3.3cm} \hbox{for spin-$s$ massless field};
\eeq
where the argument $p$ stands for the momenta $p^i$, $\beta$. Argument $\alpha$ of ket-vectors in \rf{01092018-man03-06} and \rf{01092018-man03-07} stands for the respective set of oscillators $\alpha^i$, $\upsilon$ and $\alpha^i$, $\zeta$, while the argument $\alpha$ in \rf{01092018-man03-08} stands for $\alpha^i$.
Ket-vectors \rf{01092018-man03-09}-\rf{01092018-man03-11} satisfy the following algebraic constraints
\beq
\label{01092018-man03-12} && \hspace{-2cm} (N_\alpha - N_\upsilon)\phik = 0 \,, \hspace{2cm} \alphab^2 \phik = 0\,, \hspace{2.3cm} \hbox{for continuous-spin  field};\qquad
\\
\label{01092018-man03-13} && \hspace{-2cm} (N_\alpha + N_\zeta -s) |\phi_s\rangle  = 0\,, \hspace{1.1cm}   (\alphab^2 + \zetab^2) |\phi_s\rangle =0 \,, \hspace{1cm} \hbox{for spin-$s$ massive field};
\\
\label{01092018-man03-14} && \hspace{-2cm}  (N_\alpha - s) |\phi_s\rangle  = 0\,, \hspace{2.2cm}  \alphab^2  |\phi_s\rangle =0 \,, \hspace{2.1cm} \hbox{for spin-$s$ massless field}.\qquad
\eeq
From the first constraints in \rf{01092018-man03-13},\rf{01092018-man03-14}, we learn that ket-vector $\phik$ \rf{01092018-man03-10} is  degree-$s$ homogeneous polynomials in the $\alpha^i$, $\zeta$, while the ket-vector \rf{01092018-man03-11} is degree-$s$ homogeneous polynomial in the $\alpha^i$. The second constraints in \rf{01092018-man03-12}, \rf{01092018-man03-14} amount to tracelessnes constraints for the respective tensor fields in \rf{01092018-man03-06},\rf{01092018-man03-08}. The second constraint in \rf{01092018-man03-13} is required for the massive fields \rf{01092018-man03-07} to be associated with irreps of the $so(d-1)$ algebra. Sometimes we prefer to use an infinite chain of massive/massless fields which consists of every spin just once. Such chain of massive/massless fields is described by the ket-vector
\be  \label{01092018-man03-15}
|\phi(p,\alpha)\rangle = \sum_{s=0}^\infty |\phi_s(p,\alpha)\rangle\,,
\ee
where, in \rf{01092018-man03-15}, the ket-vector $|\phi_s(p,\alpha)\rangle$ stands for the ket-vector of spin-$s$ massive/massless field given in \rf{01092018-man03-10}/\rf{01092018-man03-11}.

\noindent {\bf Field-theoretical realization of Poincar\'e algebra}. To get a field theoretical realization of the Poincar\'e algebra we need the realization of generators \rf{01092018-man03-04},\rf{01092018-man03-05} in terms of differential operators acting on ket-vector $|\phi\rangle$ \rf{01092018-man03-09}-\rf{01092018-man03-11}. The realization in term of differential operators is given by,%
\footnote{ In this paper, without loss of generality, the generators of the Poincar\'e algebra are analysed for $x^+=0$.}

\vspace{-0.8cm}
\beq
&& \hspace{-2cm} \hbox{\it Kinematical generators}:
\nonumber\\
\label{01092018-man03-16} && P^i=p^i\,, \qquad \qquad\quad P^+=\beta\,, \qquad J^{+i}=\partial_{p^i} \beta\,, \qquad \quad \ J^{+-} = \partial_\beta \beta\,,\qquad
\\
\label{01092018-man03-17} && J^{ij}=p^i \partial_{p^j} - p^j \partial_{p^i} + M^{ij}\,, \hspace{1.4cm} M^{ij} = \alpha^i \alphab^j - \alpha^j \alphab^i\,;
\\
&& \hspace{-2cm}  \hbox{\it Dynamical generators}:
\nonumber\\
\label{01092018-man03-18} && P^- =  -\frac{p^i p^i + m^2}{2\beta}\,, \qquad J^{-i} = - \partial_\beta p^i + \partial_{p^i} P^- + \frac{1}{\beta}(M^{ij} p^j + M^i)\,,
\\
\label{01092018-man03-19} && \hspace{1cm} \beta\equiv p^+\,,\qquad \partial_\beta\equiv \partial/\partial \beta\,, \quad
\partial_{p^i}\equiv \partial/\partial p^i\,,
\eeq
where, in \rf{01092018-man03-19}, we explain our notation.
Operator $M^{ij}$ \rf{01092018-man03-17} stands for a spin operator of the $so(d-2)$ algebra. Commutation relations for the operators $M^{ij}$, $M^i$ take the form
\beq
\label{01092018-man03-20} & [M^{ij},M^{kl}] =  \delta^{jk} M^{il} + 3\hbox{ terms}\,, \qquad  [M^i,M^{jk}] = \delta^{ij} M^k - \delta^{ik} M^j \,, &
\\
\label{01092018-man03-21} & [M^i, M^j ] = - m^2 M^{ij}\,. &
\eeq
Realization of the operator $M^i$ on space of various ket-vectors is given by

\vspace{-0.8cm}
\beq
&&\hspace{-2.8cm}  \hbox{\it Continuous-spin field (massless, $m^2=0$, and massive, $m^2< 0$)}:
\nonumber\\
\label{01092018-man03-22} && M^i = g \bar\alpha^i + A^i \gb\,, \hspace{2cm} A^i  =  \alpha^i - \alpha^2 \frac{1}{2N_\alpha+d-2}\alphab^i\,,
\\
\label{01092018-man03-23} && g =  g_\upsilon \upsilonb\,, \qquad  \gb  = - \upsilon g_\upsilon\,,\qquad  g_\upsilon  = \Bigl(\frac{F_\upsilon}{(N_\upsilon + 1) (2N_\upsilon + d - 2)} \Bigr)^{1/2} \,,
\\
\label{01092018-man03-24} && F_\upsilon  =    \kappa^2 -  N_\upsilon (N_\upsilon + d - 3) m^2\,;
\eeq
\noindent
\beq
\label{01092018-man03-25} && \hspace{-6cm} \hbox{\it Arbitrary spin massive field, $m^2> 0$}:\hspace{1cm} M^i = m (\zeta\alphab^i - \alpha^i \zetab);
\\
\label{01092018-man03-26} && \hspace{-6cm} \hbox{\it Arbitrary spin massless field, $m=0$}: \hspace{1cm} M^i = 0\,.
\eeq

\vspace{-0.2cm}
For massive/massless continuous-spin fields, eigenvalues of the 2nd- and 4th-order Casimir operators of the Poincar\'e algebra are given by $C_2=m^2$, $C_4=\kappa^2$, while for arbitrary spin-$s$ massive/massless fields one has the relations $C_2=m^2$, $C_4=m^2s(s+d-3)$. Note that, in terms of the operators $M^i$, $M^{ij}$ \rf{01092018-man03-20},\rf{01092018-man03-21}, the 4th-order Casimir operator of the Poincar\'e algebra takes the form $C_4 = -  M^i M^i - \half m^2 M^{ij}M^{ij}$.

Now we are ready to present a field theoretical realization of the Poincar\'e algebra generators. This is to say that, at quadratic order in fields, a field theoretical realization of the generators  \rf{01092018-man03-04},\rf{01092018-man03-05} takes the form
\be \label{01092018-man03-27}
G_\smpt  =\int \beta d^{d-1}p\, \langle\phi(p)| G |\phi(p)\rangle\,, \qquad
d^{d-1}p \equiv  d\beta d^{d-2}p\,,
\ee
where $G$ stands for the differential operators presented in
\rf{01092018-man03-16}-\rf{01092018-man03-18}, while
$G_\smpt$ stands for the field theoretical generators. Equal time commutator of $G_\smpt$ \rf{01092018-man03-27} with $\phik$ takes the form $[ |\phi\rangle,G_\smpt\,] = G
|\phi\rangle$. Also we recall that the light-cone gauge action takes the form
\be \label{01092018-man03-30}
S = \int dx^+  d^{d-1} p\,\, \langle \phi(p)|{\rm
i}\, \beta
\partial^- |\phi(p)\rangle +\int dx^+ P^-\,,
\ee
where $\partial^-=\partial/\partial x^+$ and $P^-$ is the Hamiltonian. For free fields  fields, the Hamiltonian is obtained from relations \rf{01092018-man03-18},\rf{01092018-man03-27}.

\newsection{ \large Cubic interaction vertices and light-cone gauge dynamical principle} \label{sec-03}

Detailed discussion of the setup we use for studying the light-cone gauge cubic vertices may be found in Sec.3 in Ref.\cite{Metsaev:2017cuz}. Here, to make our presentation self-contained, we just briefly review our result for complete system of equations required to determine the cubic vertices uniquely.

In theories of interacting fields, the dynamical generators $G^\dyn = P^-, J^{-i}$ \rf{01092018-man03-05} can be expanded as
\be \label{01092018-man03-31}
G^\dyn  = \sum_{n=2}^\infty  G_\smpn^\dyn\,,
\ee
where $G_\smpn^\dyn$ stands for a functional that has $n$ powers of ket-vector $\phik$. At quadratic order in fields, contribution to $G^\dyn$ is governed by $G_\smpt$ \rf{01092018-man03-27}, while, at cubic order in fields, contributions to $G^\dyn$ are described by $G_\smp3^\dyn$. Our aim in this section is to describe the complete system of equations which allows us to determine $G_\smp3^\dyn = P_\smp3^-, J_\smp3^{-i}$ uniquely.

We start with the presentation of the expressions for dynamical generators $P_\smp3^-$, $J_\smp3^{-i}$ given by
\beq
\label{01092018-man03-32} && P_\smp3^- =  \int\!\! d\Gamma_3 \,\, \langle \Phi_\smp3 ||p_\smp3^-\rangle \,,
\\
\label{01092018-man03-33} &&  J_\smp3^{-i} = \int \!\! d\Gamma_3\,\,  \langle \Phi_\smp3 | j_\smp3^{-i}\rangle -  \frac{1}{3} \bigl(\sum_{a=1,2,3} \partial_{p_a^i} \langle \Phi_\smp3 | \bigr)|p_\smp3^-\rangle \,,
\eeq
where we use the notation
\beq
\label{01092018-man03-34} && \langle \Phi_\smp3| \equiv \prod_{a=1,2,3} \langle \phi(p_a,\alpha_a)|\,,\qquad\quad \langle \phi(p_a,\alpha_a)| \equiv |\phi(p_a,\alpha_a)\rangle^\dagger\,,
\\
\label{01092018-man03-35} && d\Gamma_3 \equiv (2\pi)^{d-1} \delta^{d-1} \bigl( \sum_{a=1,2,3} p_a \bigr)
\prod_{a=1,2,3} \frac{d^{d-1} p_a}{(2\pi)^{(d-1)/2}} \,, \qquad d^{d-1}p_a = d^{d-2}p_ad\beta_a\,.
\eeq
The ket-vectors $|p_\smp3^-\rangle$ and $|j_\smp3^{-i}\rangle$ appearing in \rf{01092018-man03-32}, \rf{01092018-man03-33}
can be presented as
\be  \label{01092018-man03-36}
|p_\smp3^-\rangle = p_\smp3^-(\Po,\beta_a, \alpha_a)|0\rangle\,, \qquad
|j_\smp3^{-i}\rangle = j_\smp3^{-i}(\Po,\beta_a, \alpha_a)|0\rangle\,,
\ee
where index $a=1,2,3$ labels three
fields entering dynamical generators \rf{01092018-man03-32},\rf{01092018-man03-33}. Quantities  $p_\smp3^-$ and $j_\smp3^{-i}$ in \rf{01092018-man03-36} are referred to as densities. Often the density $p_\smp3^-$ is referred to as cubic interaction vertex. The quantities $\beta_a$ \rf{01092018-man03-36} are three light-cone momenta \rf{01092018-man03-19}, while the quantity $\alpha_a$ \rf{01092018-man03-36} is a shortcut for the oscillators entering ket-vectors\rf{01092018-man03-09}-\rf{01092018-man03-11}: for continuous-spin field, the $\alpha_a$ stands for oscillators $\alpha_a^i$, $\upsilon_a$, while, for massive and massless fields, the $\alpha_a$ stands for oscillators $\alpha_a^i$, $\zeta_a$ and $\alpha_a^i$ respectively. A quantity $\Po$ in \rf{01092018-man03-36} stands for a momentum $\Po^i$ defined by the relations
\be  \label{01092018-man03-37}
\Po^i \equiv \frac{1}{3}\sum_{a=1,2,3} \betach_a p_a^i\,, \qquad
\betach_a\equiv \beta_{a+1}-\beta_{a+2}\,, \quad \beta_a\equiv
\beta_{a+3}\,.
\ee

\noindent {\bf Complete system of equations for cubic vertex}.
The complete system of equations for the cubic vertex $p_\smp3^-$ and the density $j_\smp3^{-i}$ discussed in Ref.\cite{Metsaev:2017cuz} takes the form
\beq
\label{01092018-man03-38} && \Jbf^{+-} | p_\smp3^- \rangle =0 \,, \hspace{5.5cm} \hbox{kinematical } \  J^{+-}-\hbox{ symmetry};
\\
\label{01092018-man03-39} &&  \Jbf^{ij} |p_\smp3^-\rangle = 0\,, \hspace{5.7cm} \hbox{kinematical } \  J^{ij}-\hbox{ symmetries};
\\
\label{01092018-man03-40} && |j_\smp3^{-i}\rangle = - (\Pbf^-)^{-1} \Jbf^{-i\dagger}|p_\smp3^-\rangle\,, \qquad \hspace{2.2cm} \hbox{ dynamical } P^-, J^{-i} \hbox{ symmetries };\qquad
\\
&& \hspace{3cm} \hbox{ Light-cone gauge dynamical principle:}
\nonumber\\
\label{01092018-man03-41} && |p_\smp3^-\rangle \hbox{ and } \ |j_\smp3^{-i}\rangle \hspace{0.5cm} \hbox{ are expandable in } \Po^i;
\\
\label{01092018-man03-42} && |p_\smp3^-\rangle \ne \Pbf^- |V\rangle, \quad |V\rangle \ \hbox{is expandable in } \Po^i; \qquad
\\
\label{01092018-man03-43} && |p_\smp3^-\rangle, |j_\smp3^{-i}\rangle, |V\rangle \quad \hbox{ are finite for } \Pbf^-=0 \,,
\eeq
where operators $\Jbf^{+-}$, $\Jbf^{ij}$, $\Pbf^-$, $\Jbf^{-i\dagger}$ appearing in \rf{01092018-man03-38}-\rf{01092018-man03-43} are given
\beq
\label{01092018-man03-44} && \hspace{-1cm} \Jbf^{+-} =  \Po^j\partial_{\Po^j} + \sum_{a=1,2,3} \beta_a\partial_{\beta_a} \,,
\\
\label{01092018-man03-45} && \hspace{-1cm}  \Jbf^{ij}  =  \Po^i \partial_{\Po^j}  -
\Po^j \partial_{\Po^i} + \Mbf^{ij}\,, \qquad \Mbf^{ij} \equiv \sum_{a=1,2,3} M_a^{ij}\,,
\\
\label{01092018-man03-46} && \hspace{-1cm} \Pbf^-  = \frac{\Po^i \Po^i}{2\beta} - \sum_{a=1,2,3} \frac{m_a^2}{2\beta_a}\,, \hspace{1cm} \beta  \equiv  \beta_1 \beta_2 \beta_3\,,
\\
\label{01092018-man03-47} && \hspace{-1cm}  \Jbf^{-i\dagger}  =   - \frac{\Po^i}{\beta} \No_\beta + \frac{1}{\beta} \Mo^{ij} \Po^j + \sum_{a=1,2,3}  \frac{\check\beta_a }{6\beta_a} m_a^2 \partial_{\Po^i}  + \frac{1}{\beta_a} M_a^{i\dagger}\,,
\\
\label{01092018-man03-49} && \No_\beta    \equiv  \frac{1}{3}\sum_{a=1,2,3} \check\beta_a \beta_a \partial_{\beta_a}\,, \qquad
\Mo^{ij}   \equiv \frac{1}{3}\sum_{a=1,2,3} \check\beta_a M_a^{ij}\,, \qquad \betach_a \equiv \beta_{a+1} - \beta_{a+2}\,.\qquad
\eeq

\subsection{ Equations for parity invariant cubic interaction vertices}

Cubic vertices depend, among other things, on the following variables
\be \label{01092018-man03-49-a1}
\Po^i\,,\quad  \beta_a\,, \quad \alpha_a^i\,, \qquad a=1,2,3\,.
\ee
Note also that, if cubic Hamiltonian $P_\smp3^-$ \rf{01092018-man03-32} involves continuous-spin field $\langle\phi(p_a,\alpha_a)|$, then cubic vertex, besides variables in \rf{01092018-man03-49-a1}, depends on the oscillators $\upsilon_a$ \rf{01092018-man03-32}, while if cubic Hamiltonian $P_\smp3^-$ \rf{01092018-man03-32} involves massive field $\langle\phi(p_a,\alpha_a)|$, then cubic vertex, besides variables in \rf{01092018-man03-49-a1}, depends  on the oscillator $\zeta_a$. Restrictions imposed by $J^{ij}$-symmetries
\rf{01092018-man03-39} imply that the vertex $p_\smp3^-$ depends on invariants of the $so(d-2)$ algebra.  The oscillators $\upsilon_a$, $\zeta_a$, and momenta $\beta_a$ are the invariants of the  $so(d-2)$ algebra. Besides these invariants, in the problem under consideration, the remaining invariants of the $so(d-2)$ algebra can be constructed by using the $\Po^i$, $\alpha_a^i$, the delta-Kroneker $\delta^{ij}$, and the Levi-Civita symbol $\epsilon^{ i_1\ldots i_{d-2} }$. We ignore invariants that involve one Levi-Civita symbol.%
\footnote{ For finite-component fields, recent interesting discussion of parity-odd vertices may be found in Ref.\cite{Kessel:2018ugi}. Generalization of our discussion to the case of light-cone gauge parity-odd vertices is straightforward. We do not discuss parity-odd vertices because their inclusion would make our discussion unwieldy. Note that full actions in light-cone gauge bosonic string field theories are formulated entirely in terms of the parity invariant vertices. We believe therefore that full actions in light-cone gauge bosonic continuous-spin field theories can also be formulated entirely in terms of the parity invariant vertices. For massless fields in 5d flat space, the pattern of the derivation of the light-cone gauge parity-odd cubic vertices may be found in Sec. 8.1 in Ref.\cite{Metsaev:2005ar}.}
This is to say that, in  this paper, vertices that depend on the invariants given by
\be \label{01092018-man03-50}
\Po^i\Po^i,\qquad \alpha_a^i \Po^i,\qquad \alpha_a^i \alpha_b^i\,, \qquad \upsilon_a\,,\qquad  \zeta_a\,, \qquad \beta_a\,,
\ee
are referred to as parity invariant vertices.  If $P_\smp3^-$ \rf{01092018-man03-32} involves continuous-spin field $\langle\phi(p_a,\alpha_a)|$, or massless arbitrary spin field  $\langle\phi(p_a,\alpha_a)|$, then by virtue of the second constraints in \rf{01092018-man03-12},\rf{01092018-man03-14} the invariant $\alpha_a^i\alpha_a^i$ does not contribute to the $P_\smp3^-$, while, if $P_\smp3^-$ \rf{01092018-man03-32}  involves massive field $\langle \phi(p_a,\alpha_a)|$, then by virtue of the second constraint in \rf{01092018-man03-12}, the contribution of $\alpha_a^i\alpha_a^i$, can be replaced by the contribution of $(-\zeta_a^2$). Also note that, by using field redefinitions, one can remove all terms in the vertex $p_\smp3^-$ that  are proportional to $\Po^i\Po^i$ (see Appendix B in Ref.\cite{Metsaev:2005ar}). This implies that, in the vertex $p_\smp3^-$, we can drop down a dependence on the invariant $\Po^i\Po^i$. Taking the above-said into account, we note that  cubic vertices in the list \rf{03092018-man03-01-int}-\rf{13092018-man03-01-int} take the following respective forms:

\vspace{-0.6cm}
\beq
&& \hspace{-2.2cm} \hbox{\bf Cubic vertices with one continuous-spin field:}
\nonumber\\
\label{03092018-man03-01-add} && \hspace{-1cm} (0,s_1)\hbox{-}(0,s_2)\hbox{-}(m_3,\kappa_3)_\smCSF^\vph\,, \hspace{2.4cm}  p_\smp3^-   = p_\smp3^-(\beta_a,B_a, \alpha_{aa+1}\,,\upsilon_3)\,;
\\
\label{04092018-man03-01-add} && \hspace{-1cm} (0,s_1)\hbox{-}(m_2,s_2)\hbox{-}(m_3,\kappa_3)_\smCSF^\vph\,, \hspace{2.1cm} p_\smp3^-   = p_\smp3^-(\beta_a,B_a, \alpha_{aa+1}\,,\zeta_2,\ \upsilon_3)\,;
\\
\label{05092018-man03-01-add} && \hspace{-1cm} (m_1,s_1)\hbox{-}(m_2,s_2)\hbox{-}(m_3,\kappa_3)_\smCSF^\vph\,, \hspace{1.8cm}   p_\smp3^-   = p_\smp3^-(\beta_a,B_a, \alpha_{aa+1}\,,\zeta_1, \zeta_2,\ \upsilon_3);
\\
&& \hspace{-2.2cm} \hbox{\bf Cubic vertices with two continuous-spin fields:}
\nonumber\\
\label{06092018-man03-01-add} && \hspace{-1cm} (m_1,\kappa_1)_\smCSF\hbox{-}(0,\kappa_2)_\smCSF\hbox{-}(0,s_3)\,,
\nonumber\\
\label{07092018-man03-01-add} && \hspace{-1cm} (m,\kappa_1)_\smCSF\hbox{-}(m,\kappa_2)_\smCSF\hbox{-}(0,s_3)\,,
\nonumber\\
\label{08092018-man03-01-add} && \hspace{-1cm} (m_1,\kappa_1)_\smCSF\hbox{-}(m_2,\kappa_2)_\smCSF\hbox{-}(0,s_3)\,, \hspace{1.7cm} p_\smp3^- = p_\smp3^-(\beta_a,B_a\,,\alpha_{aa+1}\,,\upsilon_1\,, \upsilon_2)\,;
\\
&& \hspace{-1cm} (0,\kappa_1)_\smCSF\hbox{-}(m_2,\kappa_2)_\smCSF\hbox{-}(m_3,s_3)\,,
\nonumber\\
\label{09092018-man03-01-add} && \hspace{-1cm} (m_1,\kappa_1)_\smCSF\hbox{-}(m_2,\kappa_2)_\smCSF\hbox{-}(m_3,s_3)\,,
\hspace{1.4cm} p_\smp3^- = p_\smp3^-(\beta_a, B_a\,, \alpha_{aa+1}\,,\upsilon_1\,, \upsilon_2,\ \zeta_3);
\\
&& \hspace{-2.2cm} \hbox{\bf Cubic vertices with three continuous-spin fields:}
\nonumber\\
\label{10092018-man03-01-add} && \hspace{-1cm} (0,\kappa_1)_\smCSF\hbox{-}(0,\kappa_2)_\smCSF\hbox{-}(m_3,\kappa_3)_\smCSF\,, \hspace{1cm}
\nonumber\\
\label{11092018-man03-01-add} && \hspace{-1cm} (m,\kappa_1)_\smCSF\hbox{-}(m,\kappa_2)_\smCSF\hbox{-}(0,\kappa_3)_\smCSF\,,
\nonumber\\
\label{12092018-man03-01-add} && \hspace{-1cm} (m_1,\kappa_1)_\smCSF\hbox{-}(m_2,\kappa_2)_\smCSF\hbox{-}(0,\kappa_3)_\smCSF\,,
\nonumber\\
\label{13092018-man03-01-add} && \hspace{-1cm} (m_1,\kappa_1)_\smCSF\hbox{-}(m_2,\kappa_2)_\smCSF\hbox{-}(m_3,\kappa_3)_\smCSF\,, \hspace{1cm} p_\smp3^- = p_\smp3^-(\beta_a,B_a\,, \alpha_{aa+1}\,,\upsilon_a)\,;
\eeq
where we introduce the notation
\be \label{01092018-man03-51}
B_a \equiv \frac{\alpha_a^i \Po^i}{\beta_a}\,, \qquad \alpha_{ab} \equiv \alpha_a^i\alpha_b^i\,.
\ee
In other words, in place of the invariant $\alpha_a^i \Po^i$ \rf{01092018-man03-50}, we prefer to use the invariant  $B_a$ \rf{01092018-man03-51}.

Using representations for cubic vertices in \rf{03092018-man03-01-add}-\rf{13092018-man03-01-add}, we now present more convenient form of equations for the cubic vertices. Namely, using $\Jbf^{-i}$ \rf{01092018-man03-47}, we find the following relation
\be \label{01092018-man03-52}
\Jbf^{-i\dagger}|p_\smp3^-\rangle  = \Pbf^- \sum_{a=1,2,3} \frac{2\betach_a}{3\beta_a}\alpha_a^i \partial_{B_a} |p_\smp3^-\rangle + \Po^i G_\beta |p_\smp3^-\rangle + \sum_{a=1,2,3} \frac{\alpha_a^i}{\beta_a} G_{a,\Po^2} |p_\smp3^-\rangle\,,
\ee
where explicit form of operators $G_{a,\Po^2}$, $G_\beta$ may be found in Appendices, B,C,D. Using explicit form of operators $G_{a,\Po^2}$, $G_\beta$, we then conclude that equations \rf{01092018-man03-40},\rf{01092018-man03-43},\rf{01092018-man03-52} lead  to the equations
\beq
\label{01092018-man03-53} && G_a |p_\smp3^-\rangle =0\,,\qquad  a=1,2,3;
\\
\label{01092018-man03-54} && G_\beta |p_\smp3^-\rangle = 0\,.
\eeq
In turn, equations \rf{01092018-man03-52}-\rf{01092018-man03-54} and the ones in \rf{01092018-man03-40} imply the following expressions for density $|j_\smp3^{-i}\rangle$   corresponding to the respective vertices with one continuous-spin field \rf{03092018-man03-01-add}-\rf{05092018-man03-01-add}, two continuous-spin field \rf{06092018-man03-01-add},\rf{09092018-man03-01-add} and three continuous-spin field  \rf{11092018-man03-01-add}
\beq
\label{01092018-man03-55} && |j_\smp3^{-i}\rangle = - \sum_{a=1,2,3} \frac{2 \betach_a}{3\beta_a} \alpha_a^i \partial_{B_a} |p_\smp3^-\rangle - \frac{2\beta}{\beta_3^3} \frac{ g_{\upsilon_3} \partial_{\upsilon_3} }{ 2 N_3  + d-2 } \partial_{B_3}^2 |p_\smp3^-\rangle\,,
\\
\label{01092018-man03-56} && |j_\smp3^{-i}\rangle = - \sum_{a=1,2,3} \frac{2 \betach_a}{3\beta_a} \alpha_a^i \partial_{B_a} |p_\smp3^-\rangle - \sum_{a=1,2} \frac{2\beta}{\beta_a^3} \frac{ g_{\upsilon_a} \partial_{\upsilon_a} }{ 2 N_a  + d-2 } \partial_{B_a}^2 |p_\smp3^-\rangle\,,
\\
\label{01092018-man03-57} && |j_\smp3^{-i}\rangle = - \sum_{a=1,2,3} \frac{2 \betach_a}{3\beta_a} \alpha_a^i \partial_{B_a} |p_\smp3^-\rangle - \sum_{a=1,2,3} \frac{2\beta}{\beta_a^3} \frac{ g_{\upsilon_a} \partial_{\upsilon_a} }{ 2 N_a  + d-2 } \partial_{B_a}^2 |p_\smp3^-\rangle\,,
\eeq
where operator $N_a$ is defined in \rf{02092018-man03-06app}. We note also that, in terms of vertices presented in \rf{03092018-man03-01-add}-\rf{13092018-man03-01-add}, equations given in \rf{01092018-man03-38} takes the form
\be \label{01092018-man03-58}
\sum_{a=1,2,3} \beta_a \partial_{\beta_a} p_\smp3^-=0\,.
\ee

Summarizing our discussion in this Section, we note that, the cubic vertices describing interactions of fields in  \rf{03092018-man03-01-add}-\rf{13092018-man03-01-add} should satisfy equations \rf{01092018-man03-53},\rf{01092018-man03-54},\rf{01092018-man03-58}.
The density $|j_\smp3^{-i}\rangle$ corresponding to three groups of cubic vertices in \rf{03092018-man03-01-add}-\rf{13092018-man03-01-add} are given by the respective three relations in \rf{01092018-man03-55}-\rf{01092018-man03-57}. Equations \rf{01092018-man03-53},\rf{01092018-man03-54}, and \rf{01092018-man03-58} constitute the complete system of equations which to the best of our present understanding determine parity invariant vertices uniquely.
Note that, if we consider the light-cone gauge formulation of the Einstein and Yang-Mills theories, then we can verify that Eqs.\rf{01092018-man03-53},\rf{01092018-man03-54}, \rf{01092018-man03-58} admit to fix the cubic interaction vertices unambiguously (see Ref.\cite{Metsaev:2005ar}). Therefore it seems reasonable to use Eqs.\rf{01092018-man03-53},\rf{01092018-man03-54}, \rf{01092018-man03-58} for studying the cubic vertices of the continuous-spin field theory.
We now consider solutions of equations \rf{01092018-man03-53},\rf{01092018-man03-54}, \rf{01092018-man03-58} for vertices  \rf{03092018-man03-01-add}-\rf{13092018-man03-01-add} in turn.

\newsection{ \large Parity invariant cubic vertices for one continuous-spin massive field and two massless fields}\label{sec-one-cont}

In this Section, we discuss parity invariant cubic vertices which involve one massive continuous-spin field and two finite component massless/massive fields. According to our classification, such vertices can be separated into three particular cases given in  \rf{03092018-man03-01-int}-\rf{05092018-man03-01-int}. Let us discuss these particular cases in turn.

\subsection{  One continuous-spin massive field and two massless fields}

We now discuss parity invariant cubic vertices for one continuous-spin massive field and two arbitrary spin massless fields. This is to say that, using the shortcut $(m,\kappa)_\smCSF$ for a continuous-spin massive  field and the shortcut $(0,s)$ for a spin-$s$ massless field, we study cubic vertices for the following three fields:
\beq
\label{03092018-man03-01} && (0,s_1)\hbox{-}(0,s_2)\hbox{-}(m_3,\kappa_3)_\smCSF^\vph \hspace{0.8cm} m_3^2 <  0\,,
\nonumber\\
&& \hbox{\small two massless fields and one continuous-spin massive field.} \hspace{3cm}
\eeq
Relation \rf{03092018-man03-01} tells us that  the spin-$s_1$  and spin-$s_2$  massless fields carry the respective external line indices $a=1$ and $a=2$, while the continuous-spin massive field corresponds to $a=3$.

For fields \rf{03092018-man03-01}, we find the following general solution to cubic vertex $p_\smp3^-$ (see Appendix B)
\beq
\label{03092018-man03-02} p_\smp3^-  & = & U_\upsilon U_\Gamma U_\beta U_B U_W U_B^{-1} V_\Asf^{(7)} \,, \qquad \Asf = +,-\,,
\\
\label{03092018-man03-03} && p_\smp3^- = p_\smp3^-  (\beta_a, B_a, \alpha_{aa+1}\,, \upsilon_3)\,,
\\
\label{03092018-man03-04} && V_\Asf^{(7)} = V_\Asf^{(7)}(B_3, \alpha_{aa+1})\,.
\eeq
In \rf{03092018-man03-02}, we introduce two vertices $V_\Asf^{(7)}$ labelled by the superscript $\Asf=+-$. In \rf{03092018-man03-03} and  \rf{03092018-man03-04}, the arguments of the generic vertex $p_\smp3^-$ and the vertices $V_\Asf^{(7)}$ are shown explicitly. The definition of the arguments $B_a$ and $\alpha_{ab}$ may be found in  \rf{02092018-man03-04app}. Various quantities $U$ appearing in \rf{03092018-man03-02} are differential operators w.r.t. the $B_a$ and $\alpha_{aa+1}$. These quantities will be presented below. For two vertices $V_\Asf^{(7)}$ \rf{03092018-man03-04}, we find the following solution:
\beq
\label{03092018-man03-05} && \hspace{-1cm} V_\pm^{(7)}  =  F(\alphabf_3,\betabf_3,\gammabf_3;\, \frac{1\pm z_3}{2} )V_\pm\,, \qquad V_\pm = V_\pm (\alpha_{12},\alpha_{23},\alpha_{31})\,,
\\
\label{03092018-man03-06} && \alphabf_3 = \nu_3 + \half + \sigma_3 \,, \hspace{1.8cm}   \betabf_3 = \nu_3 + \half - \sigma_3 \,,
\qquad \sigma_3 = \Bigl( \frac{(d-3)^2}{4} + \frac{\kappa_3^2}{m_3^2} \Bigr)^{1/2} \,,\qquad
\\
\label{03092018-man03-07} && \gammabf_3 = \nu_3  + 1\,, \hspace{2.8cm} z_3 =  \frac{2B_3}{\kappa_3} \,,
\eeq
where, in \rf{03092018-man03-05} and below, the $F(\alphabf,\betabf,\gammabf;x)$ stands for the hypergeometric function. For the hypergeometric function, we use notation and convention in Chapter 15 in Ref.\cite{NIST}.
In \rf{03092018-man03-05}, in place of the variable $B_3$, we use new variable $z_3$   \rf{03092018-man03-07}. Operator $\nu_3$ is defined below.

We see that the generic vertex $p_\smp3^-$ \rf{03092018-man03-03} depends on the ten variables, while, the vertices $V_\Asf$ \rf{03092018-man03-05} depend only on the three variables. By definition, the vertices $V_\Asf$ \rf{03092018-man03-05} are expandable in the three variables $\alpha_{12}$, $\alpha_{23}$, $\alpha_{31}$. The general solution \rf{03092018-man03-02} for the vertex $p_\smp3^-$ is expressed  in terms of the operators $U$, $\nu_3$ acting on the vertices $V_\pm$  \rf{03092018-man03-05}.To complete the description of the vertex $p_\smp3^-$ we now provide expressions for the operators $U$, $\nu_3$. These operators are given by
\beq
\label{03092018-man03-08} && \hspace{-1.5cm} U_\upsilon = \upsilon_3^{N_3}\,, \qquad  \qquad N_3 = N_{B_3} + N_{ \alpha_{31} }  + N_{ \alpha_{23} }\,,
\\
\label{03092018-man03-09} && \hspace{-1.5cm} U_\Gamma   =  \Bigl( \bigl(-\frac{\kappa_3^2}{m_3^2}\bigr)_\vph^{N_3} \frac{2^{N_3} \Gamma( N_3 + \frac{d-2}{2}) }{ \Gamma( N_3 - \lambda_{3+}) \Gamma( N_3 - \lambda_{3-}) \Gamma(N_3 + 1) } \Bigr)^{1/2}\,,
\qquad\lambda_{3\pm} = \frac{d-3}{2} \pm \sigma_3\,,\qquad
\\
\label{03092018-man03-10} && \hspace{-1.5cm} U_\beta   = \exp\bigl(- \frac{\betach_3}{2\beta_3} \kappa_3 \partial_{B_3} \bigr)\,,
\\
\label{03092018-man03-11} && \hspace{-1.5cm}  U_{\partial\alpha}   = \exp\Bigl(- \frac{2B_1B_2}{m_3^2}  \partial_{ \alpha_{12} } + \frac{2B_1}{m_3^2} (B_3+ \half \kappa_3) \partial_{\alpha_{31}} + \frac{2B_2}{m_3^2} (B_3 - \half \kappa_3) \partial_{\alpha_{23}}  \Bigr) \,,\qquad
\\
\label{03092018-man03-12} && \hspace{-1.5cm} U_B = \bigl( - \frac{\kappa_3^2}{m_3^2} + \frac{4}{m_3^2} B_3^2\bigr)^{ -(\nu_3+1)/2 }
\,,
\\
\label{03092018-man03-13} && \hspace{-1.5cm} U_W = \sum_{n=0}^\infty \frac{\Gamma(\nu_3+n)}{4^nn!\Gamma(\nu_3+2n) } W_3^n\,, \qquad W_3 =     2 \alpha_{12} \partial_{\alpha_{31}}\partial_{\alpha_{23}}
\,,
\\
\label{03092018-man03-15} &&  \hspace{-1.5cm}  \nu_3  =   N_{\alpha_{23}} +   N_{\alpha_{31}} + \frac{d-4}{2}\,,
\eeq
where $\Gamma$ \rf{03092018-man03-09},\rf{03092018-man03-13} stands for the Gamma-function.
Quantities $\betach_a$, $N_{B_a}$, $N_{\alpha_{ab}}$, $N_a$ appearing in  \rf{03092018-man03-08}-\rf{03092018-man03-15} are defined in \rf{02092018-man03-04app}-\rf{02092018-man03-06app}.

Expressions above-presented in \rf{03092018-man03-02}-\rf{03092018-man03-15} provide the complete generating form description of cubic interaction vertices for two chains of totally symmetric massless fields \rf{01092018-man03-15} with one continuous-spin massive field. Now our aim is to describe cubic vertices for one continuous-spin massive field and two totally symmetric massless fields with arbitrary but fixed spin-$s_1$ and spin-$s_2$ values. Using the first algebraic constraint for massless spin-$s$ field in \rf{01092018-man03-14} it is easy to see that vertices we are interested in must satisfy the algebraic constraints
\be \label{03092018-man03-16}
( N_{\alpha_a} - s_a )|p_\smp3^-\rangle  = 0\,,\qquad a=1,2.
\ee
Two constraints given in \rf{03092018-man03-16} tell us simply that the cubic vertex $p_\smp3^-$ should be degree-$s_1$ and degree-$s_2$ homogeneous polynomial in the oscillators $\alpha_1^i$ and $\alpha_2^i$ respectively. It is easy to check, that, in terms of the vertices $V_\pm$ \rf{03092018-man03-05}, algebraic constraints \rf{03092018-man03-16} take the form
\be \label{03092018-man03-17}
( N_{ \alpha_{12} } + N_{ \alpha_{31} } - s_1) V_\Asf =0 \,,
\qquad ( N_{ \alpha_{12} } + N_{ \alpha_{23} } - s_2) V_\Asf =0 \,.
\ee
Vertices $V_\Asf$ satisfy one and the same equations \rf{03092018-man03-17}. Therefore to simplify our presentation we drop the superscript $\Asf$ and use a vertex $V$ in place of the vertices $V_\Asf$, i.e., we use $V=V_\Asf$. Doing so, we note that the general solution to constraints \rf{03092018-man03-17} can be presented as
\be \label{03092018-man03-18}
V  =   V(s_1,s_2;\,n) \,, \qquad  V(s_1,s_2;\, n )  =  \alpha_{12}^{n^\vph} \alpha_{23}^{s_2- n^\vph}  \alpha_{31}^{s_1- n^\vph} \,.
\ee
The integer $n$ appearing in \rf{03092018-man03-18} is the freedom of our solution for the vertex $V$. In other words the integer $n$ labels all possible
cubic vertices that can be constructed for three fields in \rf{03092018-man03-01}. In order for vertices \rf{03092018-man03-18} to be sensible, the integer $n$ should satisfy the restrictions
\be \label{03092018-man03-19}
n \geq 0\,, \qquad s_1-n \geq 0\,, \qquad s_2-n \geq 0\,,
\ee
which amount to the requirement that the powers of all variables $\alpha_{12}$,  $\alpha_{23}$,  $\alpha_{31}$ in
\rf{03092018-man03-18} be non--negative. From \rf{03092018-man03-19}, we see that allowed values of $n$ are given by
\be \label{03092018-man03-20}
n = 0,1,\ldots, s_{\min}^\vph,  \qquad s_{\min}^\vph \equiv \min(s_1,s_2)\,.
\ee

Expressions for cubic interaction vertices given in  \rf{03092018-man03-02}-\rf{03092018-man03-15}, \rf{03092018-man03-18} and allowed values for $n$ presented in \rf{03092018-man03-20} provide the complete description and classification of cubic interaction vertices that can be constructed for two spin-$s_1$ and spin-$s_2$ massless fields and one continuous-spin massive field. From \rf{03092018-man03-20}, we find that, given spin values $s_1$ and $s_2$, a number of cubic vertices $V_+$ (or $V_-$) which can be built for the fields in \rf{03092018-man03-01} is given by
\be \label{03092018-man03-21}
{\sf N} = s_{\min}+1\,.
\ee

\noindent {\bf Cubic vertex for continuous-spin massive and massless scalar fields}. For illustration purposes we consider cubic vertex for one continuous-spin massive field and two massless scalar fields. For two scalar fields, we have spin values $s_1=0$, $s_2=0$, i.e., $s_{\min}=0$. Plugging $s_{\min}=0$ in \rf{03092018-man03-20}, we get $n=0$. This implies that there is only one vertex $V_+$ and one vertex $V_-$. Plugging $n=0$ in \rf{03092018-man03-18}, we get $V_\pm=1$. In turn, plugging $V_\pm=1$ in \rf{03092018-man03-05}, we  get the vertices $V_\pm^{(7)}$:
\be \label{03092018-man03-22}
V_\pm^{(7)}  =  F(\lambda_{3+},\lambda_{3-},\frac{d-2}{2}; \half \pm   \frac{B_3}{\kappa_3})\,,   \qquad \lambda_{3\pm} = \frac{d-3}{2}\pm   \Bigl( \frac{(d-3)^2}{4} + \frac{\kappa_3^2}{m_3^2} \Bigr)^{1/2} \,.\qquad
\ee
Finally, plugging $V_\pm^{(7)}$ \rf{03092018-man03-22} into \rf{03092018-man03-02} and using expressions for operators $U$ \rf{03092018-man03-08}-\rf{03092018-man03-15}, we get the full expressions for two cubic interaction vertices $p_\smp3^-$,
\beq
\label{03092018-man03-23} && \hspace{-1cm} p_\smp3^- = U  F(\lambda_{3+},\lambda_{3-},\frac{d-2}{2}; \frac{\upsilon_3 B_3}{\kappa_3}-\frac{\beta_1}{\beta_3})\,,
\\
\label{03092018-man03-24}&& \hspace{-1cm} p_\smp3^- = U  F(\lambda_{3+},\lambda_{3-},\frac{d-2}{2}; -\frac{\upsilon_3 B_3}{\kappa_3} - \frac{\beta_2}{\beta_3})\,,
\\
\label{03092018-man03-26} && \hspace{-1cm}   U   =  \Bigl( \bigl(-\frac{\kappa_3^2}{m_3^2}\bigr)_\vph^{ N_{B_3} } \frac{2^{ N_{B_3} } \Gamma( N_{B_3} + \frac{d-2}{2}) }{ \Gamma( N_{B_3} - \lambda_{3+}) \Gamma( N_{B_3}- \lambda_{3-}) \Gamma(N_{B_3} + 1) } \Bigr)^{1/2}\,,\qquad N_{B_3} = B_3\partial_{B_3}\,,\qquad
\eeq
where $\lambda_{3\pm}$ are given in \rf{03092018-man03-22}.

\subsection{  One continuous-spin massive field, one massless field, and one massive field}

We discuss parity invariant cubic vertices for one continuous-spin massive field, one arbitrary spin massless field, and one arbitrary spin massive field. This is to say that, using the shortcut $(m,\kappa)_\smCSF$ for a continuous-spin massive  field and the shortcut $(m,s)$ for a mass-$m$ and spin-$s$ massive field, we study cubic vertices for the following three fields:
\beq
\label{04092018-man03-01} && (0,s_1)\hbox{-}(m_2,s_2)\hbox{-}(m_3,\kappa_3)_\smCSF^\vph \hspace{0.8cm} \hspace{1cm} m_2^2 > 0\,, \qquad  m_3^2 <  0\,,
\nonumber\\
&&\hbox{\small one massless field, one massive field, and one continuous-spin massive field.}\quad
\eeq
Relation \rf{04092018-man03-01} tells us that  spin-$s_1$ massless and spin-$s_2$  massive fields carry the respective external line indices $a=1$ and $a=2$, while the continuous-spin massive field corresponds to $a=3$.

For fields \rf{04092018-man03-01}, we find the following general solution to cubic vertex $p_\smp3^-$ (see Appendix B):
\beq
\label{04092018-man03-02} p_\smp3^-  & = & U_\upsilon U_\Gamma U_\beta  U_\zeta U_{\partial B} U_{\partial \alpha} U_B U_W U_B^{-1} V_\Asf^{(9)} \,, \qquad \Asf = +,-\,,
\\
\label{04092018-man03-03} && p_\smp3^- = p_\smp3^-  (\beta_a, B_a, \alpha_{aa+1}\,, \zeta_2\,, \upsilon_3)\,,
\\
\label{04092018-man03-04} && V_\Asf^{(9)} = V_\Asf^{(9)}(B_2,B_3, \alpha_{aa+1})\,.
\eeq
In \rf{04092018-man03-02}, we introduce two vertices $V_\Asf^{(9)}$ labelled by the superscripts $\Asf = +,-$. In \rf{04092018-man03-03} and  \rf{04092018-man03-04}, the arguments of the generic vertex $p_\smp3^-$ and the vertices $V_\Asf^{(9)}$ are shown explicitly. The definition of the arguments $B_a$ and $\alpha_{ab}$ may be found in \rf{02092018-man03-04app}. Various quantities $U$ appearing in \rf{04092018-man03-02} are differential operators w.r.t. the $B_a$ and $\alpha_{aa+1}$. These quantities will be presented below. For two vertices $V_\pm^{(9)}$ \rf{04092018-man03-04}, we find the following solution:
\beq
\label{04092018-man03-05} && \hspace{-1cm} V_\pm^{(9)}  =  F(\alphabf_3,\betabf_3,\gammabf_3;\, \frac{1\pm z_3}{2} )V_\pm\,, \qquad V_\pm = V_\pm (B_2,\alpha_{12},\alpha_{23},\alpha_{31})\,,
\\
\label{04092018-man03-06} && \alphabf_3 = \nu_3 + \half + \sigma_3 \,, \hspace{1.8cm}   \betabf_3 = \nu_3 + \half - \sigma_3 \,,
\qquad \sigma_3 = \Bigl( \frac{(d-3)^2}{4} + \frac{\kappa_3^2}{m_3^2} \Bigr)^{1/2} \,,\qquad
\\
\label{04092018-man03-07} && \gammabf_3 = \nu_3  + 1\,, \hspace{2.8cm} z_3 =  \frac{2m_3^2 B_3}{ \kappa_3(m_3^2-m_2^2) }\,,
\eeq
where the $F(\alphabf,\betabf,\gammabf;x)$ stands for the hypergeometric function. In \rf{04092018-man03-05}, in place of the variable $B_3$, we use new variable $z_3$   \rf{04092018-man03-07}. A quantity $\nu_3$ is defined below.

From \rf{04092018-man03-03}, we see that the generic vertex $p_\smp3^-$ depends on the eleven variables, while, from \rf{04092018-man03-05}, we learn that the vertices $V_\Asf$ depend only on the four variables. Note also that, by definition, the vertices $V_\Asf$ \rf{04092018-man03-05} are expandable in the variables $B_2$, $\alpha_{12}$, $\alpha_{23}$, $\alpha_{31}$. From \rf{04092018-man03-02}, we see that the general solution for the vertex $p_\smp3^-$ is expressed  in terms of the operators $U$, $\nu_3$ and the vertices $V_\Asf$  \rf{04092018-man03-05}. In order  to complete the description of the vertex $p_\smp3^-$ we should provide expressions for the operators $U$, $\nu_3$. These operators are given by
\beq
\label{04092018-man03-08} && \hspace{-1.5cm} U_\upsilon = \upsilon_3^{N_3}\,, \qquad  \qquad N_3 = N_{B_3} + N_{ \alpha_{31} }  + N_{ \alpha_{23} }\,,
\\
\label{04092018-man03-09} && \hspace{-1.5cm} U_\Gamma   =  \Bigl( \bigl(-\frac{\kappa_3^2}{m_3^2}\bigr)_\vph^{N_3} \frac{2^{N_3} \Gamma( N_3 + \frac{d-2}{2}) }{ \Gamma( N_3 - \lambda_{3+}) \Gamma( N_3 - \lambda_{3-}) \Gamma(N_3 + 1) } \Bigr)^{1/2}\,,
\qquad\lambda_{3\pm} = \frac{d-3}{2} \pm \sigma_3 \,,\qquad
\\
\label{04092018-man03-10} && \hspace{-1.5cm} U_\beta   = \exp\bigl(- \frac{\betach_2}{2\beta_2} \zeta_2 m_2 \partial_{B_2} - \frac{\betach_3}{2\beta_3} \kappa_3 \partial_{B_3} \bigr)\,,
\\
&& \hspace{-1.5cm} U_\zeta  = \exp\Bigl(-\frac{\zeta_2}{m_2} B_1\partial_{\alpha_{12}} + \frac{\zeta_2}{m_2}( B_3 - \half \kappa_3 ) \partial_{\alpha_{23}} - \frac{m_3^2}{2m_2}\zeta_2 \partial_{B_2} \Bigr)\,,
\\
&& \hspace{-1.5cm} U_{\partial B} = \exp\bigl( \frac{m_2^2}{2m_3^2}\kappa_3  \partial_{ B_3}  \bigr)
\\
\label{04092018-man03-11} && \hspace{-1.5cm} U_{\partial\alpha} = \exp\Bigl(  \frac{\kappa_3 }{m_3^2} B_1  \partial_{ \alpha_{31} } -  \frac{\kappa_3 }{m_3^2}  B_2 \partial_{ \alpha_{23} }
\nonumber\\
&& + \frac{2}{m_2^2-m_3^2}  B_1 B_2 \partial_{\alpha_{12}} + \frac{2(m_2^2+ m_3^2)}{(m_2^2- m_3^2)^2}  B_2 B_3 \partial_{\alpha_{23}} - \frac{2}{m_2^2 - m_3^2}  B_3 B_1 \partial_{\alpha_{31}} \Bigr)\,,
\\
\label{04092018-man03-12} && \hspace{-1.5cm} U_B = \bigl( - \frac{\kappa_3^2}{m_3^2 } + \frac{4m_3^2 }{(m_2^2- m_3^2)^2} B_3^2 \bigr)^{ -(\nu_3+1)/2 }
\,,
\\
\label{04092018-man03-13} && \hspace{-1.5cm} U_W = \sum_{n=0}^\infty \frac{\Gamma(\nu_3 + n)}{4^nn!\Gamma(\nu_3 +2n) } W_3^n\,,  \hspace{1cm} W_3 =     2 \alpha_{12} \partial_{\alpha_{31}} \partial_{\alpha_{23}}  - \frac{4m_2^2}{(m_2^2- m_3^2)^2} B_2^2\partial_{\alpha_{23}}^2  \,,
\\
\label{04092018-man03-15} &&  \hspace{-1.5cm}  \nu_3  =   N_{\alpha_{23}} +   N_{\alpha_{31}} + \frac{d-4}{2}\,,
\eeq
where quantities $\betach_a$, $N_{B_a}$, $N_{\alpha_{ab}}$, $N_a$ appearing in  \rf{04092018-man03-08}-\rf{04092018-man03-15} are defined in \rf{02092018-man03-04app}-\rf{02092018-man03-06app}.

Expressions above-presented in \rf{04092018-man03-02}-\rf{04092018-man03-15} provide the complete generating form description of cubic vertices for coupling of one continuous-spin massive field to arbitrary spin massless and massive fields. Namely, these vertices describe an coupling of one continuous-spin massive field to two chains of massless and massive fields \rf{01092018-man03-15}. Now our aim is to describe cubic vertices for coupling of one continuous-spin massive field to massless and massive fields having the respective arbitrary but fixed spin-$s_1$ and spin-$s_2$ values. Using the first algebraic constraints in \rf{01092018-man03-13},\rf{01092018-man03-14} it is easy to see that vertices we are interested in must satisfy the algebraic constraints
\be \label{04092018-man03-16}
( N_{\alpha_1} - s_1 )|p_\smp3^-\rangle  = 0\,,\qquad ( N_{\alpha_2} + N_{\zeta_2} - s_2 )|p_\smp3^-\rangle  = 0\,.
\ee
First constraint in \rf{04092018-man03-16} tells us that the $p_\smp3^-$ should be degree-$s_1$ homogeneous polynomial in the $\alpha_1^i$, while, from the second constraint in \rf{04092018-man03-16}, we learn that the $p_\smp3^-$ should be degree-$s_2$ homogeneous polynomial in the $\alpha_2^i$, $\zeta_2$. In terms of the $V_\pm$ \rf{04092018-man03-05}, constraints \rf{04092018-man03-16} take the form
\be \label{04092018-man03-17}
( N_{ \alpha_{12} } + N_{ \alpha_{31} } - s_1) V_\pm =0 \,,
\qquad ( N_{B_2} + N_{ \alpha_{12} } + N_{ \alpha_{23} } - s_2) V_\pm =0 \,.
\ee
Vertices $V_\pm$ satisfy one and the same equations \rf{04092018-man03-17}. Therefore to simplify our presentation we drop the superscripts $\pm$ and use a vertex $V$ in place of the vertices $V_\pm$, i.e., we use $V=V_\pm$. Doing so, we note that the general solution to constraints \rf{04092018-man03-17} can be presented
\be \label{04092018-man03-18}
V  =   V(s_1,s_2;\,n,k) \,, \qquad  V(s_1,s_2;\, n,k)  =  B_2^{s_2-s_1 + k -n}\alpha_{12}^{s_1-k} \alpha_{23}^n  \alpha_{31}^k \,.
\ee
The integers $n$, $k$ appearing in \rf{04092018-man03-18} are the freedom of our solution for the vertex $V$. In other words, the integers $n$, $k$ label all possible
cubic vertices that can be constructed for three fields in \rf{04092018-man03-01}. In order for vertices \rf{04092018-man03-18} to be sensible, the integers $n$, $k$ should satisfy the restrictions
\be \label{04092018-man03-19}
0 \leq k \leq s_1\,, \qquad 0 \leq n \leq s_2 - s_1+k\,,
\ee
which amount to the requirement that the powers of all variables $B_2$, $\alpha_{12}$,  $\alpha_{23}$,  $\alpha_{31}$ in \rf{04092018-man03-18} be non--negative.

Expressions for cubic interaction vertices given in  \rf{04092018-man03-02}-\rf{04092018-man03-15}, \rf{04092018-man03-18} and values of $n$, $k$ which satisfy restrictions \rf{04092018-man03-19} provide the complete description and classification of cubic interaction vertices that can be constructed for one spin-$s_1$ massless field, one spin-$s_2$ massive field and one continuous-spin massive field.

\subsection{ One continuous-spin massive field and two massive fields}

In this section, we discuss parity invariant cubic vertices for one continuous-spin massive field and two arbitrary spin massive fields. This is to say that, using the shortcut $(m,\kappa)_\smCSF$ for a continuous-spin massive  field and the shortcut $(m,s)$ for a spin-$s$ massive field, we study cubic vertices for the following three fields:
\beq
\label{05092018-man03-01} && (m_1,s_1)\hbox{-}(m_2,s_2)\hbox{-}(m_3,\kappa_3)_\smCSF^\vph \hspace{0.8cm} \hbox{\small two massive fields and one continuous-spin massive field}\quad
\nonumber\\
&&   m_1^2 > 0\,, \quad   m_2^2 > 0\,, \quad  m_3^2 <  0\,.
\eeq
Relation \rf{05092018-man03-01} tells us that  the spin-$s_1$  and spin-$s_2$  massive fields carry the respective external line indices $a=1,2$, while the continuous-spin massive field corresponds to $a=3$.

For fields \rf{05092018-man03-01}, we find the following general solution to cubic vertex $p_\smp3^-$ (see Appendix B)
\beq
\label{05092018-man03-02} p_\smp3^-  & = & U_\upsilon U_\Gamma U_\beta U_\zeta U_{\partial B} U_{\partial \alpha} U_B U_W U_B^{-1} V_\Asf^{(7)} \,, \qquad \Asf = +,-\,,
\\
\label{05092018-man03-03} && p_\smp3^- = p_\smp3^-  (\beta_a, B_a, \alpha_{aa+1}\,, \zeta_1,\zeta_2, \upsilon_3)\,,
\\
\label{05092018-man03-04} && V_\Asf^{(7)} = V_\Asf^{(7)}(B_a, \alpha_{aa+1})\,.
\eeq
In \rf{05092018-man03-02}, we introduce two vertices $V_\Asf^{(7)}$ labelled by the superscripts $\Asf=\pm$. In \rf{05092018-man03-03} and  \rf{05092018-man03-04}, the arguments of the generic vertex $p_\smp3^-$ and the vertices $V_\Asf^{(7)}$ are shown explicitly. The definition of the arguments $B_a$ and $\alpha_{ab}$ may be found in \rf{02092018-man03-04app}. Various quantities $U$ appearing in \rf{05092018-man03-02} are differential operators w.r.t. the $B_a$ and $\alpha_{aa+1}$. These operators are given below. For two vertices $V_\Asf^{(7)}$ \rf{05092018-man03-04}, we find the following solution:
\beq
\label{05092018-man03-05} && \hspace{-1cm} V_\pm^{(7)}  =  F(\alphabf_3,\betabf_3,\gammabf_3;\, \frac{1\pm z_3}{2} )V_\pm\,, \qquad V_\pm = V_\pm (B_1, B_2,\alpha_{12},\alpha_{23},\alpha_{31})\,,
\\
\label{05092018-man03-06} && \alphabf_3 = \nu_3 + \half + \sigma_3 \,, \hspace{1.8cm}   \betabf_3 = \nu_3 + \half - \sigma_3 \,,
\qquad \sigma_3 = \Bigl( \frac{(d-3)^2}{4} + \frac{\kappa_3^2}{m_3^2} \Bigr)^{1/2} \,,\qquad
\\
\label{05092018-man03-07} && \gammabf_3 = \nu_3  + 1\,, \hspace{2.8cm} z_3 =  -\frac{2 m_3^2 }{\kappa_3\sqrt{D}} B_3 \,,
\eeq
where the $F(\alphabf,\betabf,\gammabf;x)$ stands for the hypergeometric function. In \rf{05092018-man03-05}, in place of the variable $B_3$, we use new variable $z_3$ \rf{05092018-man03-07}. Quantities $D$, $\nu_3$ are defined below.

From \rf{05092018-man03-03}, we see that the generic vertex $p_\smp3^-$ depends on the twelve variables, while, from \rf{05092018-man03-05}, we learn that the vertices $V_\pm$ depend only on the five variables. Note also that, by definition, the vertices $V_\pm$ \rf{05092018-man03-05} are expandable in the five variables $B_1$, $B_2$, $\alpha_{12}$, $\alpha_{23}$, $\alpha_{31}$. From \rf{05092018-man03-02}, we see that the general solution for the vertex $p_\smp3^-$ is expressed  in terms of the operators $U$, $\nu_3$ and the vertices $V_\pm$  \rf{05092018-man03-05}. In order to complete the description of the vertex $p_\smp3^-$ we should provide expressions for the operators $U$, $\nu_3$. These operators are given by
\beq
\label{05092018-man03-08} && \hspace{-1.5cm} U_\upsilon = \upsilon_3^{N_3}\,, \qquad  \qquad N_3 = N_{B_3} + N_{ \alpha_{31} }  + N_{ \alpha_{23} }\,,
\\
\label{05092018-man03-09} && \hspace{-1.5cm} U_\Gamma   =  \Bigl( \bigl(-\frac{\kappa_3^2}{m_3^2}\bigr)_\vph^{N_3} \frac{2^{N_3} \Gamma( N_3 + \frac{d-2}{2}) }{ \Gamma( N_3 - \lambda_{3+}) \Gamma( N_3 - \lambda_{3-}) \Gamma(N_3 + 1) } \Bigr)^{1/2}\,,
\qquad\lambda_{3\pm} = \frac{d-3}{2} \pm \sigma_3 \,,\qquad
\\
\label{05092018-man03-10} && \hspace{-1.5cm} U_\beta   = \exp\bigl( - \frac{\betach_1}{2\beta_1} \zeta_1 m_1 \partial_{B_1} - \frac{\betach_2}{2\beta_2} \zeta_2 m_2 \partial_{B_2} - \frac{\betach_3}{2\beta_3} \kappa_3 \partial_{B_3} \bigr)\,,
\\
&& \hspace{-1.5cm} U_\zeta = \exp\Bigl( - \frac{\zeta_1}{2m_1} (m_2^2 - m_3^2)\partial_{B_1} - \frac{\zeta_2}{2m_2} (m_3^2 - m_1^2)\partial_{B_2} - \frac{\zeta_1}{m_1}(B_3 + \half \kappa_3)\partial_{\alpha_{31}}
\nonumber\\
&& \,\, + \,\, (\frac{\zeta_1}{m_1} B_2 - \frac{\zeta_2}{m_2} B_1 + \frac{m_3^2 - m_1^2 - m_2^2}{2m_1m_2} \zeta_1 \zeta_2 )\partial_{\alpha_{12}}
+ \frac{\zeta_2}{m_2}(B_3 - \half \kappa_3)\partial_{\alpha_{23}}\Bigr)\,,
\\
&& \hspace{-1.5cm} U_{\partial B} = \exp\bigl( \frac{m_2^2 -m_1^2}{2m_3^2}\kappa_3  \partial_{ B_3}  \bigr)\,,
\\
\label{05092018-man03-11} && \hspace{-1.5cm} U_{\partial\alpha} = \exp\Bigl( \frac{\kappa_3 }{m_3^2} B_1  \partial_{ \alpha_{31} } -  \frac{\kappa_3 }{m_3^2}  B_2 \partial_{ \alpha_{23} } +   \sum_{a=1,2,3} \frac{2h_{aa+1}}{D}  B_a B_{a+1} \partial_{ \alpha_{aa+1} } )\,,
\\
\label{05092018-man03-12} && \hspace{-1.5cm} U_B = \bigl( - \frac{\kappa_3^2}{m_3^2} + \frac{4m_3^2}{D} B_3^2\bigr)^{ -(\nu_3+1)/2 } \,,
\\
\label{05092018-man03-13} && \hspace{-1.5cm} U_W = \sum_{n=0}^\infty \frac{\Gamma(\nu_3+n)}{4^nn!\Gamma(\nu_3+2n) } W_3^n\,,
\quad
\\
\label{05092018-man03-15} &&  \hspace{-1.5cm}  \nu_3  =   N_{\alpha_{23}} +   N_{\alpha_{31}} + \frac{d-4}{2}\,,\qquad W_3 =   - \frac{4m_1^2}{D} B_1^2\partial_{\alpha_{31}}^2  - \frac{4m_2^2}{D} B_2^2\partial_{\alpha_{23}}^2  + 2 \alpha_{12} \partial_{\alpha_{31}}\partial_{\alpha_{23}}
\,,\quad
\\
\label{05092018-man03-15-a1} &&  \hspace{-1.5cm} h_{aa+1} = m_a^2 + m_{a+1}^2 - m_{a+2}^2\,,
\\
\label{05092018-man03-15-a2} &&  \hspace{-1.5cm}  D = m_1^4 + m_2^4 +m_3^4 - 2(m_1^2m_2^2 + m_2^2 m_3^2 + m_3^2 m_1^2)\,,
\eeq
where quantities $\betach_a$, $N_{B_a}$, $N_{\alpha_{ab}}$, $N_a$ appearing in  \rf{05092018-man03-08}-\rf{05092018-man03-15} are defined in \rf{02092018-man03-04app}-\rf{02092018-man03-06app}.

Expressions above-presented in \rf{05092018-man03-02}-\rf{05092018-man03-15-a2} provide the complete generating form description of cubic interaction vertices for coupling of one continuous-spin massive field to two chains of totally symmetric massive fields \rf{01092018-man03-15}. Now our aim is to describe cubic vertices for coupling of one continuous-spin massive field to two totally symmetric massive fields having arbitrary but fixed spin-$s_1$ and spin-$s_2$ values. Using the first algebraic constraint in \rf{01092018-man03-13} it is easy to see that vertices we are interested in must satisfy the algebraic constraints
\be \label{05092018-man03-16}
( N_{\alpha_a} + N_{\zeta_a} - s_a )|p_\smp3^-\rangle  = 0\,,\qquad  a=1,2\,.
\ee
Two constraints given in \rf{05092018-man03-16} tell us simply that the cubic vertex $p_\smp3^-$ should be degree-$s_1$ and degree-$s_2$ homogeneous polynomial in the oscillators $\alpha_1^i$, $\zeta_1$ and $\alpha_2^i$, $\zeta_2$ respectively. It is easy to check, that, in terms of the vertices $V_\pm$ \rf{05092018-man03-05}, algebraic constraints \rf{05092018-man03-16} take the form
\be \label{05092018-man03-17}
( N_{B_1} + N_{ \alpha_{12} } + N_{ \alpha_{31} } - s_1) V_\Asf =0 \,,
\qquad ( N_{B_2} + N_{ \alpha_{12} } + N_{ \alpha_{23} } - s_2) V_\Asf =0 \,.
\ee
Vertices $V_\Asf$ satisfy one and the same equations \rf{05092018-man03-17}. Therefore to simplify our presentation we drop the superscript $\Asf$ and use a vertex $V$ in place of the vertices $V_\Asf$, i.e., we use $V=V_\Asf$. Doing so, we note that the general solution to constraints \rf{05092018-man03-17} can be presented as
\be \label{05092018-man03-18}
V  =   V(s_1,s_2;\,n_1,n_2,n_3) \,, \qquad  V(s_1,s_2;\, n_1,n_2,n_3)  =  B_1^{s_1-n_2-n_3} B_2^{s_2 - n_1 -n_3} \alpha_{12}^{n_3} \alpha_{23}^{n_1}  \alpha_{31}^{n_2} \,.
\ee
The integers $n_1$, $n_2$, $n_3$ appearing in \rf{05092018-man03-18} are the freedom of our solution for the vertex $V$. In other words, the three integers $n_a$ label all possible
cubic vertices that can be constructed for three fields in \rf{05092018-man03-01}. In order for vertices \rf{05092018-man03-18} to be sensible, the integers $n_a$ should satisfy the restrictions
\be \label{05092018-man03-19}
n_2 + n_3 \leq s_1\,, \qquad n_1+n_3 \leq s_2\,, \qquad n_a \geq 0\,,\qquad a=1,2,3\,,
\ee
which amount to the requirement that the powers of $B_1$, $B_2$, $\alpha_{aa+1}$ in \rf{05092018-man03-18} be non--negative.
Expressions for cubic vertices given in  \rf{05092018-man03-02}-\rf{05092018-man03-15}, \rf{05092018-man03-18} and restrictions on  values of $n_a$ presented in \rf{05092018-man03-19} provide the complete description and classification of cubic interaction vertices that can be constructed for two spin-$s_1$ and spin-$s_2$ massive fields and one continuous-spin massive field.

\newsection{ \large Parity invariant cubic vertices for two  continuous-spin massive/massless fields and one arbitrary spin massive/massless field}\label{sec-two-cont}

In this Section, we discuss parity invariant cubic vertices which involve two continuous-spin massive/massless fields and one finite component massive/massless field. According to our classification, such vertices can be separated into five particular cases given in  \rf{06092018-man03-01-int}-\rf{09092018-man03-01-int}. Let us discuss these particular cases in turn.

\subsection{ \large One continuous-spin massive field, one continuous-spin massless field and one arbitrary spin massless field}

We start with discussion of vertices involving one continuous-spin massive field, one continuous-spin massless spin and one arbitrary spin massless field. This is to say that, using the shortcuts $(m,\kappa)_\smCSF$ and $(0,\kappa)_\smCSF$ for the respective  continuous-spin massive and massless fields and the shortcut $(0,s)$ for a spin-$s$ massless field, we study cubic vertices for the following three fields:
\beq
\label{06092018-man03-01} && \hspace{-1cm} (m_1,\kappa_1)_\smCSF\hbox{-}(0,\kappa_2)_\smCSF\hbox{-}(0,s_3)\,, \qquad    m_1^2 < 0\,,
\nonumber\\
&& \hspace{-1cm} \hbox{\small one continuous-spin massive field, one continuous-spin massless field, and one massless field}\quad
\eeq
Relation \rf{06092018-man03-01} tells us that massive and massless continuous-spin fields carry the respective external line indices $a=1$ and $a=2$, while the massless spin-$s_3$ field corresponds to $a=3$.

For fields \rf{06092018-man03-01}, we find the following general solution to cubic vertex $p_\smp3^-$ (see Appendix C)
\beq
\label{06092018-man03-02} p_\smp3^-  & = & U_\upsilon U_\Gamma U_\beta U_{\partial B}  U_{\partial \alpha} U_B U_{JB}U_W U_B^{-1} V_{\Asf\Bsf}^{(8)} \,, \qquad \Asf,\Bsf = +,-\,,
\\
\label{06092018-man03-03} && p_\smp3^- = p_\smp3^-  (\beta_a, B_a, \alpha_{aa+1}\,, \upsilon_1,\upsilon_2)\,,
\\
\label{06092018-man03-04} && V_{\Asf\Bsf}^{(8)} = V_{\Asf\Bsf}^{(8)}(B_1,B_2,\alpha_{aa+1})\,.
\eeq
In \rf{06092018-man03-02}, we introduce four vertices $V_{\Asf\Bsf}^{(8)}$ labelled by the superscripts $\Asf,\Bsf$. In \rf{06092018-man03-03} and  \rf{06092018-man03-04}, the arguments of the generic vertex $p_\smp3^-$ and the vertices $V_{\Asf\Bsf}^{(8)}$ are shown explicitly. The definition of the arguments $B_a$ and $\alpha_{ab}$ may be found in \rf{02092018-man03-04app}. Various operators $U$ appearing in \rf{06092018-man03-02} are defined below. For the four vertices $V_{\Asf\Bsf}^{(8)}$ \rf{06092018-man03-04}, we find the following solution:
\beq
\label{06092018-man03-05} && \hspace{-1cm} V_{\Asf\Bsf}^{(8)}  = F_\Asf J_\Bsf V_{\Asf\Bsf}\,, \qquad V_{\Asf\Bsf} = V_{\Asf\Bsf}(\alpha_{12},\alpha_{23},\alpha_{31})\,, \qquad  \Asf,\Bsf = +,-\,,
\\
\label{06092018-man03-05-a1} && F_\pm = F(\alphabf_1,\betabf_1,\gammabf_1;\, \frac{1\pm z_1}{2})\,, \qquad J_+ = I_{\nu_2}(\sqrt{z_2})\,, \hspace{1cm} J_- = K_{\nu_2} (\sqrt{z_2})\,,
\\
\label{06092018-man03-06} && \alphabf_1 = \nu_1 + \half + \sigma_1 \,, \hspace{1.8cm}   \betabf_1 = \nu_1 + \half - \sigma_1 \,,
\qquad \sigma_1 = \Bigl( \frac{(d-3)^2}{4} + \frac{\kappa_1^2}{m_1^2} \Bigr)^{1/2} \,,\qquad
\\
\label{06092018-man03-07} && \gammabf_1 = \nu_1  + 1\,, \hspace{2.8cm} z_1 =  \frac{2B_1}{\kappa_1}  \,,\hspace{2cm} z_2 = \frac{4\kappa_2}{m_1^2} B_2\,,
\eeq
where $F(\alphabf,\betabf,\gammabf;x)$ stands for the hypergeometric function, while $I_\nu$ and $K_\nu$ are the modified Bessel functions.
In \rf{06092018-man03-05-a1}, in place of the variables $B_1$ and $B_2$ we use the respective  variables $z_1$ and $z_2$  \rf{06092018-man03-07}. Operators $\nu_1$, $\nu_2$ are defined below.

From \rf{06092018-man03-03}, we see that the generic vertex $p_\smp3^-$ depends on the eleven variables, while, from \rf{06092018-man03-05}, we learn that the vertices $V_{\Asf\Bsf}$ depend only on the three variables. Note also that, by definition, the vertices $V_{\Asf\Bsf}$ \rf{06092018-man03-05} are expandable in the three variables $\alpha_{12}$, $\alpha_{23}$, $\alpha_{31}$. We now
present the explicit expressions for the operators $U$, $\nu_1$, $\nu_2$. These operators are given by
\beq
\label{06092018-man03-08} && \hspace{-1.5cm} U_\upsilon =  \upsilon_1^{N_1}\upsilon_2^{N_2}\,, \qquad N_1 = N_{B_1} + N_{\alpha_{12}} + N_{\alpha_{31}}\,,  \quad N_2 = N_{B_2} + N_{\alpha_{12}} + N_{\alpha_{23}}\,,
\\
\label{06092018-man03-09} && \hspace{-1.5cm} U_\Gamma   =  \Bigl( \bigl(-\frac{\kappa_1^2}{m_1^2}\bigr)_\vph^{N_1} \frac{2^{N_1+N_2} \Gamma( N_1 + \frac{d-2}{2})\Gamma( N_2 + \frac{d-2}{2}) }{ \Gamma( N_1 - \lambda_{1+}) \Gamma( N_1 - \lambda_{1-}) \Gamma(N_1 + 1) \Gamma( N_2 + 1)} \Bigr)^{1/2}\,,
\\
&& \lambda_{1\pm} = \frac{d-3}{2} \pm \sigma_1 \,,\qquad
\\
\label{06092018-man03-10} && \hspace{-1.5cm} U_\beta   = \exp\bigl( - \frac{\betach_1}{2\beta_1} \kappa_1 \partial_{B_1} - \frac{\betach_2}{2\beta_2} \kappa_2 \partial_{B_2} \bigr)\,,
\\
&& \hspace{-1.5cm} U_{\partial B}  = \exp\bigl( \frac{\kappa_2}{2}\partial_{B_2} \bigr)\,,
\\
\label{06092018-man03-11} && \hspace{-1.5cm} U_{\partial\alpha} = \exp\Bigl(-\frac{\kappa_1\kappa_2}{m_1^2}\partial_{\alpha_{12} } + \frac{\kappa_1}{m_1^2} B_2 \partial_{\alpha_{12}} - \frac{\kappa_1}{m_1^2} B_3 \partial_{\alpha_{31}} + \frac{2\kappa_2}{m_1^2} B_1 \partial_{\alpha_{12}}
\nonumber\\
&& \hspace{2.7cm} +\,\, \frac{2B_1B_2}{m_1^2}\partial_{\alpha_{12}} -
\frac{2B_2B_3}{m_1^2}\partial_{\alpha_{23}} + \frac{2B_3 B_1}{m_1^2}\partial_{\alpha_{31}} \Bigr)\,,
\\
\label{06092018-man03-12} && \hspace{-1.5cm} U_B =   \bigl( - \frac{\kappa_1^2}{m_1^2} + \frac{4}{m_1^2} B_1^2 \bigr)_\vph^{ -(\nu_1+1)/2} \,, \qquad U_{JB} = \bigl(\frac{4\kappa_2}{m_1^2}B_2\bigr)^{-\nu_2/2}\,,
\\
\label{06092018-man03-13} && \hspace{-1.5cm} U_W = U_{\nu_1,W_1} U_{\nu_2,W_{23}}\,,
\hspace{3cm} U_{\nu,W} \equiv \sum_{n=0}^\infty \frac{\Gamma(\nu+n)}{4^nn!\Gamma(\nu+2n) } W^n\,,
\\
\label{06092018-man03-14} && \hspace{-1.5cm}  W_1  =  2\alpha_{23} \partial_{\alpha_{12}} \partial_{\alpha_{31}} - \partial_{\alpha_{12}}^2 \,,
\hspace{1.8cm} W_{23}  =  2\alpha_{31}  \partial_{\alpha_{12}} \partial_{\alpha_{23}}\,,
\\
\label{06092018-man03-15} &&  \hspace{-1.5cm}  \nu_1  =   N_{\alpha_{12}} +   N_{\alpha_{31}} + \frac{d-4}{2}\,,\hspace{2.1cm} \nu_2  =   N_{\alpha_{12}} +   N_{\alpha_{23}} + \frac{d-4}{2}\,,
\eeq
where quantities $\betach_a$, $N_{B_a}$, $N_{\alpha_{ab}}$, $N_a$ appearing in  \rf{06092018-man03-08}-\rf{06092018-man03-15} are defined in \rf{02092018-man03-04app}-\rf{02092018-man03-06app}.

Expressions \rf{06092018-man03-02}-\rf{06092018-man03-15} provide the complete generating form description of cubic vertices for coupling of  continuous-spin massless and massive fields to arbitrary spin massless field. Namely, these vertices describe coupling of two continuous-spin fields to chain of massless fields \rf{01092018-man03-15}. Now our aim is to describe cubic vertices for coupling of continuous-spin massless and massive fields to massless field with arbitrary but fixed spin-$s_3$ value. Using the first algebraic constraint in \rf{01092018-man03-14}, it is easy to see that vertices we are interested in must satisfy the algebraic constraint
\be \label{06092018-man03-16}
( N_{\alpha_3}  - s_3 )|p_\smp3^-\rangle  = 0\,,
\ee
which implies that the cubic vertex $p_\smp3^-$ should be degree-$s_3$ homogeneous polynomial in the $\alpha_3^i$. It is easy to check, that, in terms of the vertices $V_{\Asf\Bsf}$ \rf{06092018-man03-05}, algebraic constraint \rf{06092018-man03-16} takes the form
\be \label{06092018-man03-17}
( N_{ \alpha_{23} } + N_{ \alpha_{31} } - s_3) V =0 \,, \qquad V = V_{\Asf\Bsf}\,,
\ee
where to simplify the notation we drop the superscripts $\Asf\Bsf$ in $V_{\Asf\Bsf}$. Obviously, general solution to constraints \rf{06092018-man03-17} can be presented as
\beq
\label{06092018-man03-18} && V  =   V(s_3;\,n,k) \,, \qquad  V(s_3;\, n,k )  =  \alpha_{12}^n \alpha_{23}^k  \alpha_{31}^{s_3-k} \,,
\\
\label{06092018-man03-19} && n\geq 0\,, \qquad k\geq 0\,, \quad s_3-k \geq 0\,.
\eeq
The integers $n$, $k$ appearing in \rf{06092018-man03-18} are the freedom of our solution for the vertex $V$, i..e, the integers $n$, $k$ label all possible
cubic vertices that can be constructed for three fields in \rf{06092018-man03-01}. In order for vertices \rf{06092018-man03-18} to be sensible, the integers should satisfy the restrictions in \rf{06092018-man03-19},  which amount to the requirement that the powers of all variables $\alpha_{12}$,  $\alpha_{23}$,  $\alpha_{31}$ in
\rf{06092018-man03-18} be non--negative. From \rf{06092018-man03-19}, we see that allowed values of $n$, $k$ are given by
\be \label{06092018-man03-20}
k = 0,1,\ldots, s_3\,,  \qquad n=0,1,\ldots, \infty\,.
\ee

Expressions for cubic  vertices given in  \rf{06092018-man03-02}-\rf{06092018-man03-15}, \rf{06092018-man03-18} and allowed values for $n$ and $k$ presented in \rf{06092018-man03-20} provide the complete description and classification of cubic interaction vertices that can be constructed for one continuous-spin massive field, one continuous-spin massless field, and one spin-$s_3$ massless field.

\subsection{ Two continuous-spin massive fields with the same mass values and one massless field}

In this section, we discuss parity invariant cubic vertices for two  continuous-spin massive fields having the same masses and one arbitrary spin massless field. This is to say that, using the shortcut $(m,\kappa)_\smCSF$ for a continuous-spin massive  field and the shortcut $(0,s)$ for a spin-$s$ massless field, we study cubic vertices for the following three fields:
\beq
\label{07092018-man03-01} && \hspace{-1cm} (m_1,\kappa_1)_\smCSF\hbox{-}(m_2,\kappa_2)_\smCSF\hbox{-}(0,s_3)\,, \qquad   m_1=m\,, \quad m_2=m\,,\quad  m^2 < 0\,,
\nonumber\\
&& \hspace{-1cm} \hbox{\small two continuous-spin massive fields with the same masses and one massless field.}\quad
\eeq
Relation \rf{07092018-man03-01} tells us that  the massive continuous-spin fields carry the respective external line indices $a=1,2$, while the massless spin-$s_3$ field corresponds to $a=3$.

For fields \rf{07092018-man03-01}, we find the following general solution to cubic vertex $p_\smp3^-$ (see Appendix C)
\beq
\label{07092018-man03-02} p_\smp3^-  & = & U_\upsilon U_\Gamma U_\beta U_{\partial B}  U_{\partial \alpha} U_e U_W  V_{\Asf\Bsf}^{(7)} \,, \qquad \Asf,\Bsf = 1,2,\ldots,6\,,
\\
\label{07092018-man03-03} && p_\smp3^- = p_\smp3^-  (\beta_a, B_a, \alpha_{aa+1}\,, \upsilon_1,\upsilon_2)\,,
\\
\label{07092018-man03-04} && V_{\Asf\Bsf}^{(7)} = V_{\Asf\Bsf}^{(7)}(B_a,Z,\alpha_{12})\,,
\\
&& Z = ( B_1 - \half \kappa_1  )\alpha_{23} + ( B_2 + \half \kappa_2) \alpha_{31} \,.
\eeq
In \rf{07092018-man03-02}, we introduce vertices $V_{\Asf\Bsf}^{(7)}$ labelled by the superscripts $\Asf,\Bsf$. In \rf{07092018-man03-03} and  \rf{07092018-man03-04}, the arguments of the generic vertex $p_\smp3^-$ and the vertices $V_{\Asf\Bsf}^{(7)}$ are shown explicitly. The definition of the arguments $B_a$ and $\alpha_{ab}$ may be found in \rf{02092018-man03-04app}. Operators $U$ appearing in \rf{07092018-man03-02} are defined below. For the vertices $V_{\Asf\Bsf}^{(7)}$ \rf{07092018-man03-04}, we find the following solution:
\beq
\label{07092018-man03-05} && \hspace{-1.2cm} V_{\Asf\Bsf}^{(6)}  = E_{1\Asf} E_{2\Bsf}V_{\Asf\Bsf}\,, \hspace{0.6cm} V_{\Asf\Bsf} = V_{\Asf\Bsf} (B_3,Z,\alpha_{12})\,, \qquad \Asf,\Bsf = 1,2,\ldots, 6\,,
\\
&& \hspace{-1.3cm} E_{a1} = B_a^\rho\,, \hspace{2.3cm} E_{a2} = B_a^\rho \ln B_a\,, \hspace{2.1cm} \hbox{ for } \quad \kappa_a^2 = -\frac{(d-3)^2}{4} m^2\,,
\\
&& \hspace{-1.3cm} E_{a3} = B_a^{\rho_{a+}}, \hspace{2cm} E_{a4} =  B_a^{\rho_{a-}}\,, \hspace{2.7cm} \hbox{ for } \quad  \kappa_a^2 < -\frac{(d-3)^2}{4} m^2\,,
\\
&& \hspace{-1.3cm} E_{a5} = B_a^{\rho_{a+}} + B_a^{\rho_{a-}}\,, \hspace{0.5cm} E_{a6} = \irm (B_a^{\rho_{a+}} - B_a^{\rho_{a-}})\,, \hspace{1cm} \hbox{ for } \quad \kappa_a^2 >  -\frac{(d-3)^2}{4}m^2\,,\qquad
\\
&& \hspace{-1.3cm}  \rho  =  -\omega_Z^\vph\,, \hspace{1cm} \rho_{a\pm}  =  - \omega_Z^\vph \pm \Bigl( \frac{(d-3)^2}{4} + \frac{\kappa_a^2}{m^2} \Bigr)^{1/2} \,,\qquad
\omega_Z^\vph = N_{ \alpha_{12} } + N_Z + \frac{d-3}{2}\,,\quad
\eeq
where $a=1,2$, $N_Z= Z\partial_Z$. The generic vertex $p_\smp3^-$ \rf{07092018-man03-03} depends on eleven variables, while, the vertices $V_{\Asf\Bsf}$  \rf{07092018-man03-05} depend only on three variables. By definition, the vertices $V_{\Asf\Bsf}$ \rf{07092018-man03-05} are expandable in the three variables $B_3$, $Z$, $\alpha_{12}$. In order  to complete the description of the vertex $p_\smp3^-$ we now  present the explicit form of the operators $U$. These operators are given by
\beq
\label{07092018-man03-08} && \hspace{-1.5cm} U_\upsilon =  \upsilon_1^{N_1}\upsilon_2^{N_2}\,, \qquad N_1 = N_{B_1} + N_{\alpha_{12}} + N_{\alpha_{31}}\,,  \quad N_2 = N_{B_2} + N_{\alpha_{12}} + N_{\alpha_{23}}\,,
\\
\label{07092018-man03-09} && \hspace{-1.5cm} U_\Gamma   =  \prod_{a=1,2} \Bigl( \bigl( - \frac{\kappa_a^2}{m^2} \bigr)_\vph^{N_a} \frac{2^{N_a}\Gamma( N_a + \frac{d-2}{2}) }{ \Gamma( N_a - \lambda_{a+}) \Gamma( N_a - \lambda_{a-}) \Gamma(N_a + 1) } \Bigr)^{1/2}\,,
\\
&& \lambda_{a\pm} = \frac{d-3}{2} \pm \sigma_a \,,\qquad
\\
\label{07092018-man03-10} && \hspace{-1.5cm} U_\beta   = \exp\bigl( - \frac{\betach_1}{2\beta_1} \kappa_1 \partial_{B_1} - \frac{\betach_2}{2\beta_2} \kappa_2 \partial_{B_2} \bigr)\,,
\\
&& \hspace{-1.5cm} U_{ \partial B } = \exp( -\frac{\kappa_1}{2}\partial_{B_1}  + \frac{\kappa_2}{2}\partial_{B_2} \bigr)\,,
%
%
\\
\label{07092018-man03-11} && \hspace{-1.5cm} U_{\partial\alpha} =
\exp\Bigl( - \frac{\kappa_1\kappa_2}{m^2} \partial_{\alpha_{12}} - \frac{\kappa_2}{m^2} B_1 \partial_{\alpha_{12}}  + \frac{\kappa_1}{m^2} B_2 \partial_{\alpha_{12}}    - \frac{\kappa_1}{m^2} B_2 B_3 \partial_Z + \frac{\kappa_2}{m^2} B_3B_1 \partial_Z
\nonumber\\
\hspace{-1.7cm} && +\,\, \frac{1}{2m^2} B_1B_2 \partial_{\alpha_{12}} - \frac{1}{m^2} B_1 B_2 B_3\partial_Z \Bigr)\,,
\\
\label{07092018-man03-12} && \hspace{-1.5cm}  U_e = \bigl(-\frac{m^2}{\kappa_1^2}\bigr)^{\omega_1/2}
\bigl(-\frac{m^2}{\kappa_2^2}\bigr)^{\omega_2/2}\,,
\\
\label{07092018-man03-13} && \hspace{-1.5cm} U_W = U_{\omega_1,W_1} U_{\omega_2,W_{21}}\,,  \qquad U_{\omega,W}= \sum_{n=0}^\infty \frac{\Gamma(\omega+n)}{4^nn!\Gamma(\omega+2n) } W^n\,,
\\
\label{07092018-man03-14} && \hspace{-1.5cm} W_1   =   \frac{1}{m^2}  B_2^2 B_3^2\partial_Z^2 - \partial_{\alpha_{12}}^2   - 2 B_2\partial_{\alpha_{12}} \partial_{B_1}    \,,
\hspace{1cm} W_{2 1}  =    \frac{1}{m^2}  B_1^2 B_3^2\partial_Z^2   - 2 B_1\partial_{\alpha_{12}} \partial_{B_2}   \,,
\\
\label{07092018-man03-15} &&  \hspace{-1.5cm} \omega_1^\vph = N_{B_1}  + N_{ \alpha_{12} } + N_Z + \frac{d-3}{2}\,, \qquad \omega_2^\vph = N_{B_2} + N_{ \alpha_{12} } + N_Z + \frac{d-3}{2}\,,\quad
N_Z = Z\partial_Z \,, \quad
\eeq
where quantities $\betach_a$, $N_{B_a}$, $N_{\alpha_{ab}}$, $N_a$ appearing in  \rf{07092018-man03-08}-\rf{07092018-man03-15} are defined in \rf{02092018-man03-04app}-\rf{02092018-man03-06app}.

Expressions above-presented in \rf{07092018-man03-02}-\rf{07092018-man03-15} provide the complete generating form description of cubic vertices for coupling of two continuous-spin massive fields to chain of massless fields \rf{01092018-man03-15}. Now our aim is to describe cubic vertices for coupling of two continuous-spin massive fields to  massless field having arbitrary but fixed spin-$s_3$ value. Using the first  algebraic constraint in \rf{01092018-man03-14}, it is easy to see that vertices we are interested in must satisfy the algebraic constraint
\be \label{07092018-man03-16}
( N_{\alpha_3}  - s_3 )|p_\smp3^-\rangle  = 0\,,
\ee
which implies that the vertex $p_\smp3^-$ should be degree-$s_3$ homogeneous polynomial in the oscillators $\alpha_3^i$. In terms of the vertices $V_{\Asf\Bsf}$ \rf{07092018-man03-05}, algebraic constraint \rf{07092018-man03-16} takes the form
\be \label{07092018-man03-17}
( N_{B_3 } + N_Z - s_3) V =0 \,, \qquad V \equiv V_{\Asf\Bsf}\,,
\ee
where to simplify the notation we drop the superscripts $\Asf\Bsf$ in $V_{\Asf\Bsf}$.  General solution to constraint \rf{07092018-man03-17} can be presented as
\beq
\label{07092018-man03-18} && V  =   V(s_3;\,n,k) \,, \qquad  V(s_3;\, n,k )  =   B_3^{s_3-k} Z^k   \alpha_{12}^n \,,
\\
\label{07092018-man03-19} && 0\leq k \leq s_3\,, \qquad n \geq 0\,.
\eeq
The integers $n$ and $k$ appearing in \rf{07092018-man03-18} are the freedom of our solution for the vertex $V$. In other words, these integers label all possible
cubic vertices that can be constructed for three fields in \rf{07092018-man03-01}. In order for vertices \rf{07092018-man03-18} to be sensible, the integers $n$, $k$ should satisfy the restrictions in \rf{07092018-man03-19} which amount to the requirement that the powers of all variables $B_3$, $Z$ and $\alpha_{12}$ in \rf{07092018-man03-18} be non--negative. From \rf{07092018-man03-19}, we see that allowed values of $n$, $k$ are given by
\be \label{07092018-man03-20}
k = 0,1,\ldots, s_3^\vph,  \qquad n=0,1,\ldots, \infty\,.
\ee

Expressions for cubic interaction vertices given in  \rf{07092018-man03-02}-\rf{07092018-man03-15}, \rf{07092018-man03-18} and values for $n$, $k$ presented in \rf{07092018-man03-20} provide the complete description and classification of cubic vertices that can be constructed for two continuous-spin massive fields and one spin-$s_3$ massless field \rf{07092018-man03-01}.

\subsection{ Two continuous-spin massive field with different masses and one massless field}

In this section, we discuss parity invariant cubic vertices for two continuous-spin massive fields having different masses and one arbitrary spin massless field. This is to say that, using the shortcut $(m,\kappa)_\smCSF$ for a continuous-spin massive  field and the shortcut $(0,s)$ for a spin-$s$ massless field, we study cubic vertices for the following three fields:
\beq
\label{08092018-man03-01} && \hspace{-1cm} (m_1,\kappa_1)_\smCSF\hbox{-}(m_2,\kappa_2)_\smCSF\hbox{-}(0,s_3)\,, \qquad    m_1^2 < 0\,,\qquad    m_2^2 < 0\,,\qquad m_1\ne m_2\,,
\nonumber\\
&& \hspace{-1cm} \hbox{\small two continuous-spin massive fields with different masses and one massless field.}\quad
\eeq
Relation \rf{08092018-man03-01} tells us that two continuous-spin massive fields carry the external line indices $a=1,2$, while the spin-$s_3$ massless field corresponds to $a=3$.

For fields \rf{08092018-man03-01}, we find the following general solution to cubic vertex $p_\smp3^-$ (see Appendix C)
\beq
\label{08092018-man03-02} p_\smp3^-  & = & U_\upsilon U_\Gamma U_\beta U_{\partial B}  U_{\partial \alpha} U_B U_W U_B^{-1} V_{\Asf\Bsf}^{(8)} \,, \qquad \Asf,\Bsf = +,-\,,
\\
\label{08092018-man03-03} && p_\smp3^- = p_\smp3^-  (\beta_a, B_a, \alpha_{aa+1}\,, \upsilon_1,\upsilon_2)\,,
\\
\label{08092018-man03-04} && V_{\Asf\Bsf}^{(8)} = V_{\Asf\Bsf}^{(8)}(B_1,B_2,\alpha_{aa+1})\,.
\eeq
In \rf{08092018-man03-02}, we introduce four vertices $V_{\Asf\Bsf}^{(8)}$ labelled by the superscripts $\Asf,\Bsf=\pm$. In \rf{08092018-man03-03} and  \rf{08092018-man03-04}, the arguments of the generic vertex $p_\smp3^-$ and the vertices $V_{\Asf\Bsf}^{(8)}$ are shown explicitly. The definition of the arguments $B_a$ and $\alpha_{ab}$ may be found in \rf{02092018-man03-04app}. Various quantities $U$ appearing in \rf{08092018-man03-02} are differential operators w.r.t. the $B_a$ and $\alpha_{aa+1}$. These quantities will be presented below. For four vertices $V_{\Asf\Bsf}^{(8)}$ \rf{08092018-man03-04}, we find the following solution:
\beq
\label{08092018-man03-05} && \hspace{-0.8cm} V_{\Asf\Bsf}^{(6)}  = F_{1\Asf} F_{2\Bsf} V\,, \qquad V_{\Asf\Bsf} = V_{\Asf\Bsf} (\alpha_{12},\alpha_{23},\alpha_{31})\,, \qquad \Asf,\Bsf = +,-\,,
\\
\label{08092018-man03-06} && F_{a\pm} = F(\alphabf_a,\betabf_a,\gammabf_a;\, \frac{1\pm z_a}{2})\,,  \qquad \alphabf_a = \nu_a + \half + \sigma_a \,, \qquad   \betabf_a = \nu_a + \half - \sigma_a \,,
\\
\label{08092018-man03-07}  && \gammabf_a = \nu_a  + 1\,, \quad \ z_a =  \frac{2m_a^2 B_a}{\kappa_a(m_1^2-m_2^2)}\,, \quad \ \sigma_a = \Bigl( \frac{(d-3)^2}{4} + \frac{\kappa_a^2}{m_a^2} \Bigr)^{1/2} \,,\quad a=1,2\,,\qquad
\eeq
where the $F(\alphabf,\betabf,\gammabf;x)$ stands for the hypergeometric function.
In \rf{08092018-man03-05}, in place of the variables $B_1$, $B_2$, we use new variables $z_1$, $z_2$ defined in  \rf{08092018-man03-07}. Operators $\nu_a$ are given below.

The generic vertex $p_\smp3^-$ \rf{08092018-man03-03} depends on the eleven variables, while,  the vertices $V_{\Asf\Bsf}$  \rf{08092018-man03-05} depend only on the three variables. Note also that, by definition, the vertices $V_{\Asf\Bsf}$ \rf{08092018-man03-05} are expandable in the three variables $\alpha_{12}$, $\alpha_{23}$, $\alpha_{31}$.  In order  to complete the description of the vertex $p_\smp3^-$ we now provide expressions for the operators $U$, $\nu_1$, $\nu_2$. These operators are given by
\beq
\label{08092018-man03-08} && \hspace{-1.2cm} U_\upsilon =  \upsilon_1^{N_1}\upsilon_2^{N_2}\,, \qquad N_1 = N_{B_1} + N_{\alpha_{12}} + N_{\alpha_{31}}\,,  \quad N_2 = N_{B_2} + N_{\alpha_{12}} + N_{\alpha_{23}}\,,
\\
\label{08092018-man03-09} && \hspace{-1.2cm} U_\Gamma   =  \prod_{a=1,2} \Bigl( \bigl( - \frac{\kappa_a^2}{m_a^2} \bigr)_\vph^{N_a} \frac{2^{N_a}\Gamma( N_a + \frac{d-2}{2}) }{ \Gamma( N_a - \lambda_{a+}) \Gamma( N_a - \lambda_{a-}) \Gamma(N_a + 1) } \Bigr)^{1/2}\,,\quad \lambda_{a\pm} = \frac{d-3}{2} \pm \sigma_a\,, \qquad
\\
\label{08092018-man03-10} && \hspace{-1.2cm} U_\beta   = \exp\bigl( - \frac{\betach_1}{2\beta_1} \kappa_1 \partial_{B_1} - \frac{\betach_2}{2\beta_2} \kappa_2 \partial_{B_2} \bigr)\,,
\\
&& \hspace{-1.2cm} U_{\partial B}  = \exp\bigl( - \frac{m_2^2}{2m_1^2} \kappa_1 \partial_{B_1}  + \frac{m_1^2}{2m_2^2} \kappa_2 \partial_{B_2} \bigr)\,,
\\
\label{08092018-man03-11} && \hspace{-1.2cm} U_{\partial\alpha}  = \exp\Bigl(- \frac{\kappa_1\kappa_2(m_1^2+m_2^2)}{ 2m_1^2 m_2^2 }\partial_{\alpha_{12} } + \frac{\kappa_1}{m_1^2} (B_2\partial_{\alpha_{12}} - B_3\partial_{\alpha_{31}}) + \frac{\kappa_2}{m_2^2} ( B_3 \partial_{\alpha_{23}} - B_1\partial_{\alpha_{12}})
\nonumber\\
&& \hspace{0.7cm} +\,\,  \frac{2(m_1^2+m_2^2)}{(m_1^2-m_2^2)^2} B_1 B_2 \partial_{\alpha_{12}} - \frac{2}{m_1^2-m_2^2} B_2 B_3 \partial_{\alpha_{23}} + \frac{2}{m_1^2-m_2^2}  B_3 B_1 \partial_{\alpha_{31}} \Bigr)\,,
\\
\label{08092018-man03-12} && \hspace{-1.2cm} U_B =   \prod_{a=1,2}\bigl( - \frac{\kappa_a^2}{m_a^2} +  \frac{4 m_a^2 }{(m_1^2-m_2^2)^2} B_a^2 \bigr)_\vph^{ -(\nu_a+1)/2}  \,,
\\
\label{08092018-man03-13} && \hspace{-1.2cm} U_W = U_{\nu_1,W_1} U_{\nu_2,W_{21}}\,,\hspace{3cm}
U_{\nu,W}= \sum_{n=0}^\infty \frac{\Gamma(\nu+n)}{4^nn!\Gamma(\nu+2n) } W^n\,,
\\
\label{08092018-man03-14} && \hspace{-1.2cm}  W_1  =  2\alpha_{23} \partial_{\alpha_{12}} \partial_{\alpha_{31}} - \partial_{\alpha_{12}}^2\,,
\hspace{2cm} W_{21}  =  2\alpha_{31}  \partial_{\alpha_{12}} \partial_{\alpha_{23}}\,,
\\
\label{08092018-man03-15} &&  \hspace{-1.2cm}  \nu_1  =   N_{\alpha_{12}} +   N_{\alpha_{31}} + \frac{d-4}{2}\,,\hspace{2cm} \nu_2  =   N_{\alpha_{12}} +   N_{\alpha_{23}} + \frac{d-4}{2}\,,
\eeq
where quantities $\betach_a$, $N_{B_a}$, $N_{\alpha_{ab}}$, $N_a$ appearing in  \rf{08092018-man03-08}-\rf{08092018-man03-15} are defined in \rf{02092018-man03-04app}-\rf{02092018-man03-06app}.

Expressions \rf{08092018-man03-02}-\rf{08092018-man03-15} provide the complete generating form description of cubic vertices for coupling of two continuous-spin massive fields having different masses to chain of massless fields \rf{01092018-man03-15}. Now our aim is to describe cubic vertices for coupling of two continuous-spin massive fields having different masses to massless field having arbitrary but fixed spin-$s_3$ value. Using the first algebraic constraint in \rf{01092018-man03-14}, it is easy to see that vertices we are interested in must satisfy the algebraic constraint
\be \label{08092018-man03-16}
( N_{\alpha_3}  - s_3 )|p_\smp3^-\rangle  = 0\,,
\ee
which implies that the cubic vertex $p_\smp3^-$ should be degree-$s_3$ homogeneous polynomial in the oscillators $\alpha_3^i$. In terms of the vertices $V_{\Asf\Bsf}$ \rf{08092018-man03-05}, algebraic constraint \rf{08092018-man03-16} takes the form
\be \label{08092018-man03-17}
( N_{ \alpha_{23} } + N_{ \alpha_{31} } - s_3) V =0\,, \qquad V = V_{\Asf\Bsf}\,.
\ee
General solution to constraint \rf{08092018-man03-17} can be presented as
\beq
\label{08092018-man03-18} && V  =   V(s_3;\,n,k) \,, \qquad  V(s_3;\, n,k )  =  \alpha_{12}^n \alpha_{23}^k  \alpha_{31}^{s_3-k} \,,
\\
\label{08092018-man03-19} && 0\leq k \leq s_3\,, \qquad n\geq 0\,.
\eeq
The integers $n$, $k$ appearing in \rf{08092018-man03-18} are the freedom of our solution for the vertex $V$. In other words, these integers label all possible
cubic vertices that can be constructed for three fields in \rf{08092018-man03-01}.
In order for vertices \rf{08092018-man03-18} to be sensible, the integers $n$, $k$ should satisfy the restrictions in \rf{08092018-man03-19} which amount to the requirement that the powers of all variables $\alpha_{12}$,  $\alpha_{23}$,  $\alpha_{31}$ in \rf{08092018-man03-18} be non--negative. From \rf{08092018-man03-19}, we see that allowed values of $n$, $k$ are given by
\be \label{08092018-man03-20}
k = 0,1,\ldots, s_3^\vph,  \qquad n=0,1,\ldots, \infty\,.
\ee

Relations \rf{08092018-man03-02}-\rf{08092018-man03-15}, \rf{08092018-man03-18}, and  \rf{08092018-man03-20} provide the complete description and classification of cubic vertices that can be constructed for  fields in \rf{08092018-man03-01}.

\subsection{   One continuous-spin massless field, one continuous-spin massive field, and one arbitrary spin massive field}

In this Section, we discuss parity invariant cubic vertices for one continuous-spin massless field, one continuous-spin massive field, and one arbitrary spin massive field. This is to say that, using the shortcut $(m,\kappa)_\smCSF$ for a continuous-spin mass-$m$ field and the shortcut $(m,s)$ for the mass-$m$ and spin-$s$ field, we study cubic vertices for the following three fields:
\beq
\label{08092018-man03-01-q} && \hspace{-1cm} (0,\kappa_1)_\smCSF\hbox{-}(m_2,\kappa_2)_\smCSF\hbox{-}(m_3,s_3)\,, \qquad       m_2^2 < 0\,,\qquad m_3^2 > 0\,,
\nonumber\\
&& \hspace{-1cm} \hbox{\small one continuous-spin massless field, one continuous-spin massive field, and one massive field.}\quad
\eeq
Relation \rf{08092018-man03-01-q} tells us that two continuous-spin massless and massive fields carry the external line indices $a=1,2$, while the spin-$s_3$ massive field corresponds to $a=3$.

For fields \rf{08092018-man03-01-q}, we find the following general solution to cubic vertex $p_\smp3^-$ (see Appendix C)
\beq
\label{08092018-man03-02-q} p_\smp3^-  & = & U_\upsilon U_\Gamma U_\beta U_\zeta U_{\partial B}  U_{\partial \alpha} U_{JB}  U_B U_W U_B^{-1} V_{\Asf\Bsf}^{(8)} \,, \qquad \Asf,\Bsf = +,-\,,
\\
\label{08092018-man03-03-q} && p_\smp3^- = p_\smp3^-  (\beta_a, B_a, \alpha_{aa+1}\,, \upsilon_1,\upsilon_2,\zeta_3)\,,
\\
\label{08092018-man03-04-q} && V_{\Asf\Bsf}^{(8)} = V_{\Asf\Bsf}^{(8)}(B_a,\alpha_{aa+1})\,.
\eeq
In \rf{08092018-man03-02-q}, we introduce four vertices $V_{\Asf\Bsf}^{(8)}$ labelled by the superscripts $\Asf\Bsf = \pm$. In \rf{08092018-man03-03-q} and  \rf{08092018-man03-04-q}, the arguments of the generic vertex $p_\smp3^-$ and the vertices $V_{\Asf\Bsf}^{(8)}$ are shown explicitly. The definition of the arguments $B_a$ and $\alpha_{ab}$ may be found in \rf{02092018-man03-04app}. Various quantities $U$ appearing in \rf{08092018-man03-02-q} are differential operators w.r.t. the $B_a$ and $\alpha_{aa+1}$. These quantities will be presented below. For four vertices $V_{\Asf\Bsf}^{(8)}$ \rf{08092018-man03-04-q}, we find the following solution:
\beq
\label{08092018-man03-05-q} && \hspace{-0.8cm} V_{\Asf\Bsf}^{(8)}  = J_{1\Asf} F_{2\Bsf} V\,, \qquad V_{\Asf\Bsf} = V_{\Asf\Bsf} (B_3,\alpha_{12},\alpha_{23},\alpha_{31})\,,
\\
\label{08092018-man03-05-q-a1} && J_{1+} = I_{\nu_1}(\sqrt{z_1})\,, \hspace{1cm} J_{1-} = K_{\nu_1} (\sqrt{z_1})\,, \quad z_1 =  -\frac{4\kappa_1B_1}{m_2^2- m_3^2}\,,
\\
\label{08092018-man03-06-q} && F_{2\pm} = F(\alphabf_2,\betabf_2,\gammabf_2;\, \frac{1\pm z_2}{2})\,,  \qquad \alphabf_2 = \nu_2 + \half + \sigma_2 \,, \qquad   \betabf_2 = \nu_2 + \half - \sigma_2 \,,\qquad
\\
\label{08092018-man03-07-q}  && \gammabf_2 = \nu_2  + 1\,, \qquad z_2 =  \frac{2m_2^2 B_2}{\kappa_2(m_2^2-m_3^2)}\,, \qquad \sigma_2 = \Bigl( \frac{(d-3)^2}{4} + \frac{\kappa_2^2}{m_2^2} \Bigr)^{1/2} \,,\qquad
\eeq
where $F(\alphabf,\betabf,\gammabf;x)$ is the hypergeometric function,  while $I_\nu$, $K_\nu$ are the modified Bessel functions.
In \rf{08092018-man03-05-q}, in place of $B_1$, $B_2$, we use $z_1$ \rf{08092018-man03-05-q-a1} and $z_2$ \rf{08092018-man03-07-q}. Operators $\nu_a$ are given below.

The generic vertex $p_\smp3^-$ \rf{08092018-man03-03-q} depends on the twelve variables, while, the vertices $V_{\Asf\Bsf}$ \rf{08092018-man03-05-q} depend only on the four variables. Note also that, by definition, the vertices $V_{\Asf\Bsf}$ \rf{08092018-man03-05-q} are expandable in the four variables $B_3$, $\alpha_{12}$, $\alpha_{23}$, $\alpha_{31}$.  In order  to complete the description of the vertex $p_\smp3^-$ we now provide expressions for the operators $U$, $\nu_1$, $\nu_2$. These operators are given by
\beq
\label{08092018-man03-08-q} && \hspace{-1.2cm} U_\upsilon =  \upsilon_1^{N_1}\upsilon_2^{N_2}\,, \qquad N_1 = N_{B_1} + N_{\alpha_{12}} + N_{\alpha_{31}}\,,  \quad N_2 = N_{B_2} + N_{\alpha_{12}} + N_{\alpha_{23}}\,,
\\
\label{11092018-man03-09-q} && \hspace{-1.3cm} U_\Gamma   =  \Bigl(\frac{2^{N_1} \Gamma(N_1 + \frac{d-2}{2})}{\Gamma( N_1 + 1)}  \bigl( - \frac{\kappa_2^2}{m_2^2} \bigr)_\vph^{N_2} \frac{2^{N_2}\Gamma( N_2 + \frac{d-2}{2}) }{ \Gamma( N_2 - \lambda_{2+}) \Gamma( N_2 - \lambda_{2-}) \Gamma(N_2 + 1) } \Bigr)^{1/2}\,,
\nonumber\\
&& \lambda_{2\pm} = \frac{d-3}{2} \pm \sigma_2\,, \qquad
\\
\label{08092018-man03-10-q} && \hspace{-1.2cm} U_\beta   = \exp\bigl( - \frac{\betach_1}{2\beta_1} \kappa_1 \partial_{B_1} - \frac{\betach_2}{2\beta_2} \kappa_2 \partial_{B_2} - \frac{\betach_3}{2\beta_3} m_3 \zeta_3 \partial_{B_3} \bigr)\,,
\\
&& \hspace{-1.2cm}  U_\zeta = \exp\Bigl(  \frac{\zeta_3}{m_3} ( B_1 - \half \kappa_1  ) \partial_{\alpha_{31}} - \frac{\zeta_3}{m_3} ( B_2 + \half \kappa_2) \partial_{\alpha_{23}} + \frac{m_2^2}{2m_3}\zeta_3\partial_{B_3} \Bigr),
\\
&& \hspace{-1.2cm} U_{\partial B}  = \exp\bigl( - \frac{\kappa_1 (m_2^2 + m_3^2)}{2(m_2^2-m_3^2)} \kappa_1 \partial_{B_1}  - \frac{m_3^2}{2m_2^2} \kappa_2 \partial_{B_2} \bigr)\,,
\\
\label{08092018-man03-11-q} && \hspace{-1.2cm}  U_{\partial\alpha}  = \exp\Bigl(  - \frac{\kappa_2}{q_1} \partial_{\alpha_{12}} - \frac{\kappa_2 B_1}{m_2^2}   \partial_{\alpha_{12}} - \frac{ 2m_2^2}{\kappa_1 q_1^2} B_2 \partial_{\alpha_{12}} +  \frac{2m_3^2}{\kappa_1 q_1^2} B_3 \partial_{\alpha_{31}} +  \frac{ \kappa_2 B_3}{m_2^2} \partial_{\alpha_{23}}
\nonumber\\
&& - \frac{ 2B_1 B_3}{\kappa_1 q_1} \partial_{\alpha_{31}} + \frac{ 2B_1 B_2 }{\kappa_1 q_1} \partial_{\alpha_{12}} +  \frac{ 2(m_2^2+m_3^2) B_2 B_3}{\kappa_1^2 q_1^2} \partial_{\alpha_{23}} \Bigr)\,,\quad q_1 = \frac{m_2^2-m_3^2}{\kappa_1}\,,
\\
\label{08092018-man03-12-q} && \hspace{-1.2cm} U_{JB} = (-\frac{4\kappa_1 B_1}{m_2^2-m_3^2})^{-\nu_1/2}\,, \qquad  U_B =    \bigl( - \frac{\kappa_2^2}{m_2^2} +  \frac{4 m_2^2 }{(m_2^2-m_3^2)^2} B_2^2 \bigr)_\vph^{ -(\nu_2+1)/2}  \,,
\\
\label{08092018-man03-13-q} && \hspace{-1.2cm} U_W = U_{\nu_1,W_1} U_{\nu_2,W_{23}}\,,\hspace{3cm}
U_{\nu,W}= \sum_{n=0}^\infty \frac{\Gamma(\nu+n)}{4^nn!\Gamma(\nu+2n) } W^n\,,
\\
\label{08092018-man03-14-q}  && \hspace{-1.2cm} W_1   =    2 \alpha_{23} \partial_{\alpha_{12}}\partial_{\alpha_{31}}  - \partial_{\alpha_{12}}^2 - \frac{4m_3^2}{(m_2^2-m_3^2)^2} B_3^2 \partial_{\alpha_{31}}^2     \,,
\\
&&  \hspace{-1.2cm}  W_{23}   =    2 \alpha_{31} \partial_{\alpha_{12}}\partial_{\alpha_{23}} - \frac{4m_3^2}{(m_2^2-m_3^2)^2} B_3^2 \partial_{\alpha_{23}}^2   \,,
\\
\label{08092018-man03-15-q} &&  \hspace{-1.2cm}  \nu_1  =   N_{\alpha_{12}} +   N_{\alpha_{31}} + \frac{d-4}{2}\,,\hspace{2cm} \nu_2  =   N_{\alpha_{12}} +   N_{\alpha_{23}} + \frac{d-4}{2}\,,
\eeq
where quantities $\betach_a$, $N_{B_a}$, $N_{\alpha_{ab}}$, $N_a$ appearing in  \rf{08092018-man03-08-q}-\rf{08092018-man03-15-q} are defined in \rf{02092018-man03-04app}-\rf{02092018-man03-06app}.

Expressions above-presented in \rf{08092018-man03-02-q}-\rf{08092018-man03-15-q} provide the complete generating form description of cubic vertices for coupling of two continuous-spin fields to chain of massive fields \rf{01092018-man03-15}. Now our aim is to describe cubic vertices for coupling of two continuous-spin fields to massive field having arbitrary but fixed spin-$s_3$ value. Using the first algebraic constraint \rf{01092018-man03-13}, it is easy to see that vertices we are interested in must satisfy the algebraic constraint
\be \label{08092018-man03-16-q}
( N_{\alpha_3} + N_{\zeta_3}  - s_3 )|p_\smp3^-\rangle  = 0\,,
\ee
which implies that the vertex $p_\smp3^-$ should be degree-$s_3$ homogeneous polynomial in the oscillators $\alpha_3^i$, $\zeta_3$. In terms of the vertices $V_{\Asf\Bsf}$ \rf{08092018-man03-05-q}, algebraic constraint \rf{08092018-man03-16-q} takes the form
\be \label{08092018-man03-17-q}
( N_{B_3} + N_{ \alpha_{23} } + N_{ \alpha_{31} } - s_3) V =0\,, \qquad V = V_{\Asf\Bsf}\,.
\ee
General solution to constraint \rf{08092018-man03-17-q} can be presented as
\beq
\label{08092018-man03-18-q} && V  =   V(s_3;\,n_1,n_2,n_3) \,, \qquad  V(s_3;\, n_1,n_2,n_3 )  = B_3^{s_3-n_1-n_2}  \alpha_{12}^{n_3} \alpha_{23}^{n_1}  \alpha_{31}^{n_2} \,,
\\
\label{08092018-man03-19-q} && n_1 + n_2 \leq s_3\,,\hspace{2cm}  n_a\geq 0\,, \qquad\hbox{ for } \ a=1,2,3\,.
\eeq
The integers $n_1$, $n_2$, $n_3$ appearing in \rf{08092018-man03-18-q} are the freedom of our solution for the vertex $V$. These integers label all possible
cubic vertices that can be constructed for three fields in \rf{08092018-man03-01-q}.
In order for vertices \rf{08092018-man03-18-q} to be sensible, the integers $n_a$ should satisfy the restrictions in \rf{08092018-man03-19-q} which amount to the requirement that the powers of all variables $B_3$, $\alpha_{12}$,  $\alpha_{23}$,  $\alpha_{31}$ in \rf{08092018-man03-18-q} be non--negative.
Relations given in \rf{08092018-man03-02-q}-\rf{08092018-man03-15-q}, \rf{08092018-man03-18-q}, and \rf{08092018-man03-19-q} provide the complete description and classification of cubic interaction vertices that can be constructed for  fields in \rf{08092018-man03-01-q}.

\subsection{  Two continuous-spin massive fields and one massive field}

In this section, we discuss parity invariant cubic vertices for two continuous-spin massive fields and one arbitrary spin massive field. This is to say that, using the shortcut $(m,\kappa)_\smCSF$ for a continuous-spin massive  field and the shortcut $(m,s)$ for mass-$m$ and spin-$s$ massive field, we study cubic vertices for the following three fields:
\beq
\label{09092018-man03-01} && \hspace{-1cm} (m_1,\kappa_1)_\smCSF\hbox{-}(m_2,\kappa_2)_\smCSF\hbox{-}(m_3,s_3)\,, \qquad    m_1^2 < 0\,,\qquad    m_2^2 < 0\,,\qquad m_3^2 > 0 \,,
\nonumber\\
&& \hspace{-1cm} \hbox{\small two continuous-spin massive fields and one massive field.}\quad
\eeq
Relation \rf{09092018-man03-01} tells us that the two continuous-spin massive fields carry the external line indices $a=1,2$, while the spin-$s_3$ massive field corresponds to $a=3$.

For fields \rf{09092018-man03-01}, we find the following general solution to cubic vertex $p_\smp3^-$ (see Appendix C)
\beq
\label{09092018-man03-02} p_\smp3^-  & = & U_\upsilon U_\Gamma U_\beta U_\zeta U_{\partial B}  U_{\partial \alpha} U_B U_W U_B^{-1} V_{\Asf\Bsf}^{(8)} \,, \qquad \Asf,\Bsf = +,-\,,
\\
\label{09092018-man03-03} && p_\smp3^- = p_\smp3^-  (\beta_a, B_a, \alpha_{aa+1}\,, \upsilon_1,\upsilon_2,\zeta_3)\,,
\\
\label{09092018-man03-04} && V_{\Asf\Bsf}^{(8)} = V_{\Asf\Bsf}^{(8)}(B_a,\alpha_{aa+1})\,.
\eeq
In \rf{09092018-man03-02}, we introduce four vertices $V_{\Asf\Bsf}^{(8)}$ labelled by the superscripts $\Asf,\Bsf$. In \rf{09092018-man03-03} and  \rf{09092018-man03-04}, the arguments of the generic vertex $p_\smp3^-$ and the vertices $V_{\Asf\Bsf}^{(8)}$ are shown explicitly. The definition of the arguments $B_a$ and $\alpha_{ab}$ may be found in \rf{02092018-man03-04app}. Various quantities $U$ appearing in \rf{09092018-man03-02} are differential operators w.r.t. the $B_a$ and $\alpha_{aa+1}$. These quantities will be presented below. For four vertices $V_{\Asf\Bsf}^{(8)}$ \rf{09092018-man03-04}, we find the following solution:
\beq
\label{09092018-man03-05} && \hspace{-0.8cm} V_{\Asf\Bsf}^{(6)}  = F_{1\Asf} F_{2\Bsf} V_{\Asf\Bsf}\,, \qquad V_{\Asf\Bsf} = V_{\Asf\Bsf} (B_3\,,\alpha_{12},\alpha_{23},\alpha_{31})\,, \qquad \Asf,\Bsf = +,-\,,
\\
\label{09092018-man03-06} && F_{a\pm} = F(\alphabf_a,\betabf_a,\gammabf_a;\, \frac{1\pm z_a}{2})\,,  \qquad \alphabf_a = \nu_a + \half + \sigma_a \,, \qquad   \betabf_a = \nu_a + \half - \sigma_a \,,
\\
\label{09092018-man03-07}  && \gammabf_a = \nu_a  + 1\,, \qquad z_a =  -\frac{2m_a^2 }{\kappa_a\sqrt{D}}B_a\,, \qquad \sigma_a = \Bigl( \frac{(d-3)^2}{4} + \frac{\kappa_a^2}{m_a^2} \Bigr)^{1/2} \,,\quad a=1,2\,,\qquad
\eeq
where the $F(\alphabf,\betabf,\gammabf;x)$ stands for the hypergeometric function.
In \rf{09092018-man03-05}, in place of the variables $B_a$, we use new variables $z_a$ defined in  \rf{09092018-man03-07}. A quantity $D$ and operators $\nu_a$ are defined below.

The generic vertex $p_\smp3^-$ \rf{09092018-man03-03} depends on the twelve variables, while, the vertices $V_{\Asf\Bsf}$ \rf{09092018-man03-05} depend only on the four variables. Note also that, by definition, the vertices $V_{\Asf\Bsf}$ \rf{09092018-man03-05} are expandable in the four variables $B_3$, $\alpha_{12}$, $\alpha_{23}$, $\alpha_{31}$. To complete the description of the vertex $p_\smp3^-$ we provide expressions for the operators $U$, $\nu_a$. These operators are given by
\beq
\label{09092018-man03-08} && \hspace{-1.5cm} U_\upsilon =  \upsilon_1^{N_1}\upsilon_2^{N_2}\,, \qquad N_1 = N_{B_1} + N_{\alpha_{12}} + N_{\alpha_{31}}\,,  \quad N_2 = N_{B_2} + N_{\alpha_{12}} + N_{\alpha_{23}}\,,
\\
\label{09092018-man03-09} && \hspace{-1.5cm} U_\Gamma   =  \prod_{a=1,2} \Bigl( \bigl( - \frac{\kappa_a^2}{m_a^2} \bigr)_\vph^{N_a} \frac{2^{N_a}\Gamma( N_a + \frac{d-2}{2}) }{ \Gamma( N_a - \lambda_{a+}) \Gamma( N_a - \lambda_{a-}) \Gamma(N_a + 1) } \Bigr)^{1/2}\,,\quad \lambda_{a\pm} = \frac{d-3}{2} \pm \sigma_a\,, \quad
\\
\label{09092018-man03-10} && \hspace{-1.5cm} U_\beta   = \exp\bigl( - \frac{\betach_1}{2\beta_1} \kappa_1 \partial_{B_1} - \frac{\betach_2}{2\beta_2} \kappa_2 \partial_{B_2} - \frac{\betach_3}{2\beta_3} m_3 \zeta_3 \partial_{B_3} \bigr)\,,
\\
&& \hspace{-1.5cm} U_\zeta  = \exp\Bigl( \frac{\zeta_3}{m_3} ( B_1 - \half \kappa_1  ) \partial_{\alpha_{31}} - \frac{\zeta_3}{m_3} ( B_2 + \half \kappa_2) \partial_{\alpha_{23}} - \frac{\zeta_3}{2m_3} \bigl( m_1^2 - m_2^2  \bigr)\partial_{B_3}\Bigr),
\\
&& \hspace{-1.5cm} U_{\partial B}  = \exp\bigl( - \frac{m_2^2}{2m_1^2} \kappa_1 \partial_{B_1}  + \frac{m_1^2}{2m_2^2} \kappa_2 \partial_{B_2} \bigr)\,,
\\
\label{09092018-man03-11} && \hspace{-1.5cm} U_{\partial\alpha}  = \exp\Bigl(- \frac{ \kappa_1\kappa_2 h_{12}}{ 2m_1^2 m_2^2 }\partial_{\alpha_{12} } + \frac{\kappa_1}{ m_1^2} ( B_2\partial_{\alpha_{12}} - B_3\partial_{\alpha_{31}}  ) + \frac{\kappa_2 }{m_2^2} ( B_3 \partial_{\alpha_{23}} - B_1\partial_{\alpha_{12}})
\nonumber\\
&& \hspace{2cm} +\,\,  \frac{2}{D} h_{12} B_1 B_2 \partial_{\alpha_{12}} + \frac{2}{D} h_{23} B_2 B_3 \partial_{\alpha_{23}} + \frac{2}{D} h_{31} B_3 B_1 \partial_{\alpha_{31}}
\Bigr)\,,
\\
\label{09092018-man03-12} && \hspace{-1.5cm} U_B =   \prod_{a=1,2}\bigl( -\frac{\kappa_a^2}{m_a^2} +  \frac{4 m_a^2 }{D} B_a^2 \bigr)_\vph^{ -(\nu_a+1)/2}\,,
\\
\label{09092018-man03-13} && \hspace{-1.5cm} U_W = U_{\nu_1,W_1} U_{\nu_2,W_{21}}\,,
\hspace{2.8cm}  U_{\nu,W}= \sum_{n=0}^\infty \frac{\Gamma(\nu+n)}{4^nn!\Gamma(\nu+2n) } W^n\,,
\\
\label{09092018-man03-14} && \hspace{-1.5cm}  W_1  =          2\alpha_{23} \partial_{\alpha_{12}} \partial_{\alpha_{31}} - \partial_{\alpha_{12}}^2\,,
\hspace{1.8cm} W_{21}  =  2\alpha_{31}  \partial_{\alpha_{12}} \partial_{\alpha_{23}}\,,
\\
\label{09092018-man03-15} &&  \hspace{-1.5cm}  \nu_1  =   N_{\alpha_{12}} +   N_{\alpha_{31}} + \frac{d-4}{2}\,,\hspace{2.1cm} \nu_2  =   N_{\alpha_{12}} +   N_{\alpha_{23}} + \frac{d-4}{2}\,,
\\
\label{09092018-man03-15-a1} &&  \hspace{-1.5cm} h_{aa+1} = m_a^2 + m_{a+1}^2 - m_{a+2}^2\,,
\\
\label{09092018-man03-15-a2} &&  \hspace{-1.5cm}  D = m_1^4 + m_2^4 +m_3^4 - 2(m_1^2m_2^2 + m_2^2 m_3^2 + m_3^2 m_1^2)\,,
\eeq
where quantities $\betach_a$, $N_{B_a}$, $N_{\alpha_{ab}}$, $N_a$ appearing in  \rf{09092018-man03-08}-\rf{09092018-man03-15} are defined in \rf{02092018-man03-04app}-\rf{02092018-man03-06app}.

Expressions above-presented in \rf{09092018-man03-02}-\rf{09092018-man03-15-a2} provide the complete generating form description of cubic vertices for coupling of two continuous-spin massive fields to chain of massive fields \rf{01092018-man03-15}. Now our aim is to describe cubic vertices for coupling of two  continuous-spin massive fields to one massive field having arbitrary but fixed spin-$s_3$ value. Using the first algebraic constraint \rf{01092018-man03-13}, is easy to see that vertices we are interested in must satisfy the algebraic constraint
\be \label{09092018-man03-16}
( N_{\alpha_3} + N_{\zeta_3}  - s_3 )|p_\smp3^-\rangle  = 0\,,
\ee
which implies that the cubic vertex $p_\smp3^-$ should be degree-$s_3$ homogeneous polynomial in the oscillators $\alpha_3^i$, $\zeta_3$. In terms of the vertices $V_{\Asf\Bsf}$ \rf{09092018-man03-05}, algebraic constraint \rf{09092018-man03-16} takes the form
\be \label{09092018-man03-17}
( N_{B_3} + N_{ \alpha_{23} } + N_{ \alpha_{31} } - s_3) V =0 \,,\qquad   V\equiv V_{\Asf\Bsf}\,,
\ee
where to simplify our notation we drop the superscripts $\Asf$, $\Bsf$ and use a vertex $V$ in place of the vertex $V_{\Asf\Bsf}$.
General solution to constraint \rf{09092018-man03-17} can be presented as
\beq
\label{09092018-man03-18} && V  =   V(s_3;\,n_1,n_2,n_3) \,, \qquad  V(s_3;\, n_1,n_2,n_3 )  = B_3^{s_3-n_1-n_2}  \alpha_{12}^{n_3} \alpha_{23}^{n_1}  \alpha_{31}^{n_2} \,,
\\
\label{09092018-man03-19} && n_1 + n_2 \leq s_3\,,\qquad n_a\geq 0\,,\qquad \hbox{ for } a=1,2,3\,.
\eeq
The integers $n_1$, $n_2$, $n_3$ in \rf{09092018-man03-18} are the freedom of our solution for the vertex $V$. In other words, the integers $n_a$ label all possible
cubic vertices that can be constructed for three fields in \rf{09092018-man03-01}. In order for vertices \rf{09092018-man03-18} to be sensible, the integers $n_a$ should satisfy the restrictions \rf{09092018-man03-19} which amount to the requirement that the powers of all variables $B_3$, $\alpha_{12}$,  $\alpha_{23}$,  $\alpha_{31}$ in \rf{09092018-man03-18} be non--negative.  Relations in \rf{09092018-man03-02}-\rf{09092018-man03-15}, \rf{09092018-man03-18}, and \rf{09092018-man03-19} provide the complete description and classification of cubic interaction vertices that can be constructed for  fields in \rf{09092018-man03-01}.

\newsection{ \large Parity invariant cubic vertices for three continuous-spin fields}\label{sec-three-cont}

In this Section, we discuss parity invariant cubic vertices which involve three continuous-spin fields. According to our classification, such vertices can be separated into four particular cases given in  \rf{10092018-man03-01-int}-\rf{13092018-man03-01-int}. Let us discuss these particular cases in turn.

\subsection{ \large Two continuous-spin massless fields and one continuous-spin massive field}

We start with discussion of parity invariant cubic vertices for two continuous-spin massless fields and one continuous-spin massive field. This is to say that, using the shortcut $(m,\kappa)_\smCSF$ for a mass-$m$ continuous-spin field, we study cubic vertices for the following three fields:
\beq
\label{10092018-man03-01} && \hspace{-1cm} (0,\kappa_1)_\smCSF\hbox{-}(0,\kappa_2)_\smCSF\hbox{-}(m_3,\kappa_3)_\smCSF\,, \qquad     m_3^2 < 0 \,,
\nonumber\\
&& \hspace{-1cm} \hbox{\small two continuous-spin massless fields and one continuous-spin massive field.}\quad
\eeq
Relation \rf{10092018-man03-01} tells us that  the massless continuous-spin fields carry the external line indices $a=1,2$, while the massive continuous-spin field corresponds to $a=3$.

For fields \rf{10092018-man03-01}, we find the following general solution to cubic vertex $p_\smp3^-$ (see Appendix D)
\beq
\label{10092018-man03-02} p_\smp3^-  & = & U_\upsilon U_\Gamma U_\beta U_{\partial B}  U_{\partial \alpha} U_{JB} U_B U_W U_B^{-1} V_{\Asf\Bsf\Csf}^{(7)} \,, \qquad \Asf,\Bsf,\Csf = +,-\,,
\\
\label{10092018-man03-03} && p_\smp3^- = p_\smp3^-  (\beta_a, B_a, \alpha_{aa+1}\,, \upsilon_a)\,,
\\
\label{10092018-man03-04} && V_{\Asf\Bsf\Csf}^{(7)} = V_{\Asf\Bsf\Csf}^{(7)}(B_a,\alpha_{aa+1})\,.
\eeq
In \rf{10092018-man03-02}, we introduce eight vertices $V_{\Asf\Bsf\Csf}^{(7)}$ labelled by the superscripts $\Asf,\Bsf,\Csf$. In \rf{10092018-man03-03} and  \rf{10092018-man03-04}, the arguments of the generic vertex $p_\smp3^-$ and the vertices $V_{\Asf\Bsf\Csf}^{(7)}$ are shown explicitly. The definition of the arguments $B_a$ and $\alpha_{ab}$ may be found in \rf{02092018-man03-04app}. Various quantities $U$ appearing in \rf{10092018-man03-02} are differential operators w.r.t. the $B_a$ and $\alpha_{aa+1}$. These quantities will be presented below. For eight vertices $V_{\Asf\Bsf\Csf}^{(7)}$ \rf{10092018-man03-04}, we find the following solution:
\beq
\label{10092018-man03-05} && \hspace{-1.3cm} V_{\Asf\Bsf\Csf}^{(6)}  = J_{1\Asf} J_{2\Bsf} F_{3\Csf} V_{\Asf\Bsf\Csf}\,, \qquad V_{\Asf\Bsf\Csf} = V_{\Asf\Bsf\Csf} (\alpha_{12},\alpha_{23},\alpha_{31})\,,
\\
\label{10092018-man03-05-a1} && \hspace{-1.2cm} J_{a+} = I_{\nu_a}(\sqrt{z_a})\,, \hspace{1cm} J_{a-} = K_{\nu_a} (\sqrt{z_a})\,, \quad z_1 = \frac{4\kappa_1B_1}{m_3^2}\,, \quad z_2 = - \frac{4\kappa_2B_2}{m_3^2}\,, \quad a=1,2\,,\qquad
\\
\label{10092018-man03-06} && \hspace{-1.3cm} F_{3\pm} = F(\alphabf_3,\betabf_3,\gammabf_3;\, \frac{1\pm z_3}{2})\,,  \qquad \alphabf_3 = \nu_3 + \half + \sigma_3 \,, \qquad   \betabf_3 = \nu_3 + \half - \sigma_3 \,,
\\
\label{10092018-man03-07}  &&  \hspace{-1.3cm} \gammabf_3 = \nu_3  + 1\,, \qquad z_3 =  \frac{2B_3 }{\kappa_3}\,, \qquad \sigma_3 = \Bigl( \frac{(d-3)^2}{4} + \frac{\kappa_3^2}{m_3^2} \Bigr)^{1/2} \,,
\eeq
where, $I_\nu(x)$ and $K_\nu(x)$ \rf{10092018-man03-05-a1} are the modified Bessel functions, while the $F(\alphabf,\betabf,\gammabf;x)$ \rf{10092018-man03-06} is the hypergeometric function.
In \rf{10092018-man03-05}, in place of the variables $B_1$, $B_2$, $B_3$, we use new variables $z_1$, $z_2$, $z_3$ defined in \rf{10092018-man03-05-a1} and \rf{10092018-man03-07}. Operators $\nu_a$ are defined below.

The generic vertex $p_\smp3^-$ \rf{10092018-man03-03} depends on the twelve variables, while, the vertices $V_{\Asf\Bsf\Csf}$ \rf{10092018-man03-05} depend only on the three variables. Note also that, by definition, the vertices $V_{\Asf\Bsf\Csf}$ \rf{10092018-man03-05} are expandable in the three variables $\alpha_{12}$, $\alpha_{23}$, $\alpha_{31}$. To complete the description of the vertex $p_\smp3^-$ we provide expressions for the operators $U$, $\nu_a$. These operators are given by
\beq
\label{10092018-man03-08} && \hspace{-1.5cm} U_\upsilon =  \upsilon_1^{N_1}\upsilon_2^{N_2}\upsilon_3^{N_3}\,, \qquad N_a = N_{B_a} + N_{\alpha_{aa+1}} + N_{\alpha_{a+2a}}\,,
\\
\label{10092018-man03-09} && \hspace{-1.5cm} U_\Gamma   =   \Bigl( \bigl( - \frac{\kappa_3^2}{m_3^2} \bigr)_\vph^{N_3} \frac{2^{N_3}\Gamma( N_3 + \frac{d-2}{2}) }{ \Gamma( N_3 - \lambda_{3+}) \Gamma( N_3 - \lambda_{3-}) \Gamma(N_3 + 1) }
\prod_{a=1,2} \frac{2^{N_a} \Gamma(N_a + \frac{d-2}{2})}{\Gamma( N_a + 1)} \Bigr)^{1/2}\,,
\\
&& \lambda_{3\pm} = \frac{d-3}{2} \pm \sigma_3\,, \qquad
\\
\label{10092018-man03-10} && \hspace{-1.5cm} U_\beta   = \exp\bigl( - \ \sum_{a=1,2,3} \frac{\betach_a}{2\beta_a} \kappa_a \partial_{B_a} \bigr)\,,
\\
&& \hspace{-1.5cm} U_{\partial B} = \exp\bigl( \frac{\kappa_1}{2}\partial_{B_1} - \frac{\kappa_2}{2}\partial_{B_2} \bigr)\,,
\\
\label{10092018-man03-11} && \hspace{-1.5cm} U_{\partial \alpha} =   \exp\Bigl( \frac{1}{m_3^2} \bigl( 2 B_1B_3 + \kappa_3 B_1 + 2 \kappa_1 B_3 - \kappa_1\kappa_3) \partial_{\alpha_{31}}
\nonumber\\
& + & \frac{1}{m_3^2} \bigl(2B_2B_3  - \kappa_3B_2 - 2 \kappa_2 B_3 - \kappa_2\kappa_3 ) \partial_{\alpha_{23}  }  - \frac{2}{m_3^2} (B_1B_2-\kappa_1\kappa_2)  \partial_{\alpha_{12}} \Bigr),\qquad
\\
\label{10092018-man03-12} && \hspace{-1.5cm} U_{JB} =   \bigl(\frac{4\kappa_1}{m_3^2}B_1\bigr)_\vph^{-\nu_1/2} \bigl(-\frac{4\kappa_2}{m_3^2}B_2\bigr)_\vph^{-\nu_2/2}\,,  \qquad U_B = \bigl( -\frac{\kappa_3^2}{m_3^2} + \frac{4}{m_3^2} B_3^2 \bigr)_\vph^{ -(\nu_3+1)/2}\,,
\\
\label{10092018-man03-13} && \hspace{-1.5cm} U_W = U_{\nu_1,W_1} U_{\nu_2,W_{23}} U_{\nu_3,W_{312}}\,, \hspace{1.8cm}  U_{\nu,W} \equiv \sum_{n=0}^\infty \frac{\Gamma(\nu+n)}{4^nn!\Gamma(\nu+2n) } W^n\,,
\\
\label{10092018-man03-14} && \hspace{-1.5cm}  W_1   =  2  \alpha_{23} \partial_{\alpha_{12}} \partial_{\alpha_{31}} -  \partial_{\alpha_{12}}^2   -    \partial_{\alpha_{31}}^2\,,\qquad
W_{23} =  2 \alpha_{31} \partial_{\alpha_{12}} \partial_{\alpha_{23}}  -    \partial_{\alpha_{23}}^2\,,
\\
\label{10092018-man03-15} && \hspace{-1.5cm} W_{312} =  2 \alpha_{12} \partial_{\alpha_{23}} \partial_{\alpha_{31}}\,, \hspace{3cm}  \nu_a  =   N_{\alpha_{aa+1}} +   N_{\alpha_{a+2a}} + \frac{d-4}{2}\,,
\eeq
where quantities $\betach_a$, $N_{B_a}$, $N_{\alpha_{ab}}$, $N_a$ appearing in  \rf{10092018-man03-08}-\rf{10092018-man03-15} are defined in \rf{02092018-man03-04app}-\rf{02092018-man03-06app}.

As we have already said, the vertices $V_{\Asf\Bsf\Csf}$ \rf{10092018-man03-05} should be expandable in the three variables $\alpha_{12}$, $\alpha_{23}$, $\alpha_{31}$. Using the simplified notation $V = V_{\Asf\Bsf\Csf}$, we then note that a general representative of the vertex $V$ can be chosen to be
\beq
\label{10092018-man03-18} && V  =   V(n_1,n_2,n_3) \,, \qquad  V(n_1,n_2,n_3) =   \alpha_{12}^{n_3} \alpha_{23}^{n_1}  \alpha_{31}^{n_2} \,,
\\
\label{10092018-man03-19} && n_a \geq 0, \qquad a=1,2,3\,.
\eeq
The three integers $n_a$ in \rf{10092018-man03-18} are the freedom of our solution for the vertex $V$. In other words, the integers $n_a$ label all possible
cubic vertices that can be constructed for three fields in \rf{10092018-man03-01}. In order for vertices \rf{10092018-man03-18} to be sensible, the integers $n_a$ should satisfy the restrictions \rf{10092018-man03-19} which amount to the requirement that the powers of all variables $\alpha_{12}$,  $\alpha_{23}$,  $\alpha_{31}$ in \rf{10092018-man03-18} be non--negative.  Relations given in \rf{10092018-man03-02}-\rf{10092018-man03-15}, \rf{10092018-man03-18}, and \rf{10092018-man03-19} provide the complete description and classification of cubic interaction vertices that can be constructed for  fields in \rf{10092018-man03-01}.

\subsection{  One continuous-spin massless field and two continuous-spin massive fields with equal masses}

In this section, we discuss parity invariant cubic vertices for two continuous-spin massive fields having equal masses and one continuous-spin massless field. This is to say that, using the shortcut $(m,\kappa)_\smCSF$ for a continuous-spin field, we study cubic vertices for the following three fields:
\beq
\label{11092018-man03-01} && \hspace{-1cm} (m_1,\kappa_1)_\smCSF\hbox{-}(m_2,\kappa_2)_\smCSF\hbox{-}(0,\kappa_3)_\smCSF\,, \qquad     m_1=m\,, \qquad m_2=m\,, \qquad m^2 < 0 \,,
\nonumber\\
&& \hspace{-1cm} \hbox{\small two continuous-spin massive fields with equal masses and one continuous-spin massless field.}\quad
\eeq
Relation \rf{11092018-man03-01} tells us that  the massive  continuous-spin fields having equal masses carry the external line indices $a=1,2$, while the continuous-spin massless field corresponds to $a=3$.

For fields \rf{11092018-man03-01}, we find the following general solution to cubic vertex $p_\smp3^-$ (see Appendix D)
\beq
\label{11092018-man03-02} p_\smp3^-  & = & U_\upsilon U_\Gamma U_\beta U_{\partial B}  U_{\partial \alpha} U_e U_W V_{\Asf\Bsf\Csf}^{(6)} \,, \qquad \Asf,\Bsf=1,\ldots,6\,,\quad \Csf = +,-\,,
\\
\label{11092018-man03-03} && p_\smp3^- = p_\smp3^-  (\beta_a, B_a, \alpha_{aa+1}\,, \upsilon_a)\,,
\\
\label{11092018-man03-04} && V_{\Asf\Bsf\Csf}^{(6)} = V_{\Asf\Bsf\Csf}^{(6)}(B_a,\alpha_{aa+1})\,.
\eeq
In \rf{11092018-man03-02}, we introduce vertices $V_{\Asf\Bsf\Csf}^{(6)}$ labelled by the superscripts $\Asf,\Bsf,\Csf$. In \rf{11092018-man03-03} and  \rf{11092018-man03-04}, the arguments of the generic vertex $p_\smp3^-$ and the vertices $V_{\Asf\Bsf\Csf}^{(6)}$ are shown explicitly. The definition of the arguments $B_a$ and $\alpha_{ab}$ may be found in \rf{02092018-man03-04app}. Various quantities $U$ appearing in \rf{11092018-man03-02} are differential operators w.r.t. the $B_a$ and $\alpha_{aa+1}$. These quantities will be presented below. For the vertices $V_{\Asf\Bsf\Csf}^{(6)}$ \rf{11092018-man03-04}, we find the following solution:
\beq
\label{11092018-man03-05} && \hspace{-1.4cm} V_{\Asf\Bsf\Csf}^{(6)}  = E_{1\Asf} E_{2\Bsf} T_{3\Csf} V_{\Asf\Bsf\Csf}\,, \quad V_{\Asf\Bsf\Csf} = V_{\Asf\Bsf\Csf} (\alpha_{12},\alpha_{23},\alpha_{31})\,,\quad \Asf,\Bsf = 1,\ldots, 6\,,\quad \Csf=\pm\,,\qquad
\\
\label{11092018-man03-06} && \hspace{-1.3cm} E_{a1} = B_a^{\rho_a}\,, \hspace{2cm} E_{a2} = B_a^{\rho_a} \ln B_a\,, \hspace{2cm} \hbox{ for } \quad \kappa_a^2 = -\frac{(d-3)^2}{4} m^2\,,
\\
\label{11092018-man03-06-a1} && \hspace{-1.3cm} E_{a3} = B_a^{\rho_{a+}}, \hspace{2cm} E_{a4} =  B_a^{\rho_{a-}}\,, \hspace{2.7cm} \hbox{ for } \quad  \kappa_a^2 < -\frac{(d-3)^2}{4} m^2\,,
\\
\label{11092018-man03-06-a2} && \hspace{-1.3cm} E_{a5} = B_a^{\rho_{a+}} + B_a^{\rho_{a-}}\,, \hspace{0.5cm} E_{a6} = \irm (B_a^{\rho_{a+}} - B_a^{\rho_{a-}})\,, \hspace{1cm} \hbox{ for } \quad \kappa_a^2 >  -\frac{(d-3)^2}{4}m^2\,,\qquad
\\
\label{11092018-man03-06-a3} && \hspace{-1.3cm}  \rho_a  =  -\nu_a^\vph - \half \,, \hspace{1.4cm} \rho_{a\pm}  =  - \nu_a^\vph - \half \pm \sigma_a\,,\hspace{1.4cm} \sigma_a= \Bigl( \frac{(d-3)^2}{4} + \frac{\kappa_a^2}{m^2} \Bigr)^{1/2}\,,
\\
&& \hspace{-1.3cm}  T_{3+} = \cos\bigl( \Omega_3 B_3)\,, \qquad T_{3-} = \frac{\sin\bigl( \Omega_3 B_3)}{\Omega_3}\,,
\eeq
where $a=1,2$, while operators $\nu_a$ and $\Omega_3$ are defined below. Note that that, for $\kappa_a^2$ in \rf{11092018-man03-06-a2}, the $\sigma_a$ is purely imaginary.

The generic vertex $p_\smp3^-$ \rf{11092018-man03-03} depends on the twelve variables, while, the vertices $V_{\Asf\Bsf\Csf}$ \rf{11092018-man03-05} depend only on the three variables. Note also that, by definition, the vertices $V_{\Asf\Bsf\Csf}$ \rf{11092018-man03-05} are expandable in the three variables $\alpha_{12}$, $\alpha_{23}$, $\alpha_{31}$. To complete the description of the vertex $p_\smp3^-$ \rf{11092018-man03-02} we should provide expressions for the operators $U$, $\nu_a$, $\Omega_3$. These operators are given by
\beq
\label{11092018-man03-08} && \hspace{-1.3cm} U_\upsilon =  \upsilon_1^{N_1}\upsilon_2^{N_2}\upsilon_3^{N_3}\,, \qquad N_a = N_{B_a} + N_{\alpha_{aa+1}} + N_{\alpha_{a+2a}}\,,
\\
\label{11092018-man03-09} && \hspace{-1.3cm} U_\Gamma   =  \Bigl(\frac{2^{N_3} \Gamma(N_3 + \frac{d-2}{2})}{\Gamma( N_3 + 1)} \prod_{a=1,2}  \bigl( - \frac{\kappa_a^2}{m^2} \bigr)_\vph^{N_a} \frac{2^{N_a}\Gamma( N_a + \frac{d-2}{2}) }{ \Gamma( N_a - \lambda_{a+}) \Gamma( N_a - \lambda_{a-}) \Gamma(N_a + 1) } \Bigr)^{1/2}\,,
\\
&& \lambda_{a\pm} = \frac{d-3}{2} \pm \sigma_a\,, \qquad
\\
\label{11092018-man03-10} && \hspace{-1.3cm} U_\beta   = \exp\bigl( - \ \sum_{a=1,2,3} \frac{\betach_a}{2\beta_a} \kappa_a \partial_{B_a} \bigr)\,,
\\
&& \hspace{-1.3cm} U_{\partial B} = \exp\bigl( - \frac{\kappa_1}{2}\partial_{B_1} + \frac{\kappa_2}{2}\partial_{B_2} \bigr)\,,
\\
\label{11092018-man03-11} && \hspace{-1.3cm} U_{\partial\alpha} = \exp\Bigl(  (B_3^2+\frac{1}{4}\kappa_3^2) \frac{(  B_2 \partial_{\alpha_{23}} - B_1\partial_{\alpha_{31}} ) }{2\kappa_3 m^2 } -  \frac{B_3B_1}{2m^2 }\partial_{\alpha_{31}}
-\frac{B_2B_3}{2m^2 }\partial_{\alpha_{23}} + \frac{B_1 B_2}{2m^2 }\partial_{\alpha_{12}}
\nonumber\\
&&  \hspace{-1.4cm} - \frac{\kappa_1 B_3}{m^2 }\partial_{\alpha_{31}} + \frac{\kappa_2 B_3}{ m^2 } \partial_{\alpha_{23}} - \frac{\kappa_2 B_1}{ m^2 }\partial_{\alpha_{12}} + \frac{\kappa_1B_2}{ m^2 } \partial_{\alpha_{12}}   - \frac{\kappa_1\kappa_2}{m^2 } \partial_{\alpha_{12}} - \frac{\kappa_2\kappa_3}{m^2 } \partial_{\alpha_{23}}- \frac{\kappa_3\kappa_1}{m^2 } \partial_{\alpha_{31}} \Bigr)\qquad
\\
\label{11092018-man03-12} && \hspace{-1.3cm} U_e = \bigl(-\frac{m^2}{\kappa_1^2}\bigr)_\vph^{\omega_1/2}  \bigl(-\frac{m^2}{\kappa_2^2}\bigr)_\vph^{\omega_2/2}
\\
\label{11092018-man03-13} && \hspace{-1.3cm} U_W = U_{\omega_1,W_1} U_{\omega_2,W_{23}} \,, \hspace{1.8cm}  U_{\omega,W} \equiv \sum_{n=0}^\infty \frac{\Gamma(\omega+n)}{4^nn!\Gamma(\omega+2n) } W^n\,,
\\
\label{11092018-man03-14} && \hspace{-1.3cm} W_1  = - 2 B_2 \partial_{B_1} \partial_{\alpha_{12}} +  2\kappa_3  \partial_{B_1} \partial_{\alpha_{31} }   -  \partial_{\alpha_{12} }^2  +   2\alpha_{23} \partial_{\alpha_{12}} \partial_{\alpha_{31} }\,,\qquad
\\
&& \hspace{-1.3cm}  W_{23}   =     -  2 B_1\partial_{B_2} \partial_{\alpha_{12}} -2 \kappa_3 \partial_{B_2} \partial_{\alpha_{23} }   +   2 \alpha_{31} \partial_{\alpha_{12}}\partial_{\alpha_{23} }\,,
\\
&& \hspace{-1.3cm}  W_{321}  =    \frac{1}{ \kappa_3} \{\nu_3 , B_2\partial_{\alpha_{23}}  - B_1 \partial_{\alpha_{31}} \}    +   2 \alpha_{12} \partial_{\alpha_{31}} \partial_{\alpha_{23}}\,,
\\
&& \hspace{-1.3cm}  \omega_a =  N_{B_a}+  \nu_a + \half \,, \qquad \nu_a  = N_{\alpha_{aa+1}} +   N_{\alpha_{a+2a}} + \frac{d-4}{2}\,,
\\
\label{11092018-man03-15}  && \hspace{-1.3cm}  \Omega_3 = \bigl(- \frac{1 + W_{321}}{m^2}\bigr)^{1/2}\,,
\eeq
where quantities $\betach_a$, $N_{B_a}$, $N_{\alpha_{ab}}$, $N_a$ appearing in  \rf{11092018-man03-08}-\rf{11092018-man03-15} are defined in \rf{02092018-man03-04app}-\rf{02092018-man03-06app}.

Using simplified notation for the vertices $V = V_{\Asf\Bsf\Csf}$, we note that the general expression for the $V$ can be presented as
\beq
\label{11092018-man03-18} && V  =   V(n_1,n_2,n_3) \,, \qquad  V(n_1,n_2,n_3) =   \alpha_{12}^{n_3} \alpha_{23}^{n_1}  \alpha_{31}^{n_2} \,,
\\
\label{11092018-man03-19} && n_a \geq 0, \qquad a=1,2,3\,.
\eeq
The three integers $n_a$ in \rf{11092018-man03-18} are the freedom of our solution for the vertex $V$. In other words these integers label all possible
cubic vertices that can be constructed for three fields in \rf{11092018-man03-01}. In order for vertices \rf{11092018-man03-18} to be sensible, the $n_a$ should satisfy the restrictions \rf{11092018-man03-19} which amount to the requirement that the powers of all variables $\alpha_{12}$,  $\alpha_{23}$,  $\alpha_{31}$ in \rf{11092018-man03-18} be non--negative.  Relations given in \rf{11092018-man03-02}-\rf{11092018-man03-15}, \rf{11092018-man03-18}, and \rf{11092018-man03-19} provide the complete description and classification of cubic interaction vertices that can be constructed for  fields in \rf{11092018-man03-01}.

\subsection{  One continuous-spin massless field and two continuous-spin massive fields with nonequal masses}

In this section, we discuss parity invariant cubic vertices for two continuous-spin massive fields having different masses and one continuous-spin massless field. This is to say that, using the shortcut $(m,\kappa)_\smCSF$ for a mass-$m$ continuous-spin field, we study cubic vertices for the following three fields:
\beq
\label{12092018-man03-01} && \hspace{-1.2cm} (m_1,\kappa_1)_\smCSF\hbox{-}(m_2,\kappa_2)_\smCSF\hbox{-}(0,\kappa_3)_\smCSF\,, \qquad     m_1^2<0\,, \qquad m_2^2<0\,, \qquad m_1^2 \ne m_2^2 \,,
\nonumber\\
&& \hspace{-1.2cm} \hbox{\small two continuous-spin massive fields with nonequal masses and one continuous-spin massless field.}\quad
\eeq
Relation \rf{12092018-man03-01} tells us that  the massive  continuous-spin fields carry the external line indices $a=1,2$, while the continuous-spin massless field corresponds to $a=3$.

For fields \rf{12092018-man03-01}, we find the following general solution to cubic vertex $p_\smp3^-$ (see Appendix D)
\beq
\label{12092018-man03-02} p_\smp3^-  & = & U_\upsilon U_\Gamma U_\beta U_{\partial B}  U_{\partial \alpha} U_{B} U_{JB} U_W U_B^{-1} V_{\Asf\Bsf\Csf}^{(7)} \,, \qquad \Asf,\Bsf,\Csf= +,-\,,
\\
\label{12092018-man03-03} && p_\smp3^- = p_\smp3^-  (\beta_a, B_a, \alpha_{aa+1}\,, \upsilon_a)\,,
\\
\label{12092018-man03-04} && V_{\Asf\Bsf\Csf}^{(7)} = V_{\Asf\Bsf\Csf}^{(7)}(B_a,\alpha_{aa+1})\,.
\eeq
In \rf{12092018-man03-02}, we introduce eight vertices $V_{\Asf\Bsf\Csf}^{(7)}$ labelled by the superscripts $\Asf,\Bsf,\Csf=\pm$. In \rf{12092018-man03-03} and  \rf{12092018-man03-04}, the arguments of the generic vertex $p_\smp3^-$ and the vertices $V_{\Asf\Bsf\Csf}^{(7)}$ are shown explicitly. The definition of the arguments $B_a$ and $\alpha_{ab}$ may be found in \rf{02092018-man03-04app}. Various quantities $U$ appearing in \rf{12092018-man03-02} are differential operators w.r.t. the $B_a$ and $\alpha_{aa+1}$. These quantities will be presented below. For eight vertices $V_{\Asf\Bsf\Csf}^{(7)}$ \rf{12092018-man03-04}, we find the following solution:
\beq
\label{12092018-man03-05} && \hspace{-1.4cm} V_{\Asf\Bsf\Csf}^{(6)}  = F_{1\Asf} F_{2\Bsf} J_{3\Csf} V_{\Asf\Bsf\Csf}\,, \qquad V_{\Asf\Bsf\Csf} = V_{\Asf\Bsf\Csf} (\alpha_{12},\alpha_{23},\alpha_{31})\,,  \qquad \Asf,\Bsf,\Csf= +,-\,,
\\
\label{12092018-man03-06} && \hspace{-1.4cm} F_{a\pm} = F(\alphabf_a,\betabf_a,\gammabf_a;\, \frac{1\pm z_a}{2})\,,  \qquad \alphabf_a = \nu_a + \half + \sigma_a \,, \qquad   \betabf_a = \nu_a + \half - \sigma_a \,,
\\
\label{12092018-man03-07}  &&  \hspace{-1.4cm} \gammabf_a = \nu_a  + 1\,, \qquad z_a =  \frac{2m_a^2B_a }{\kappa_a(m_1^2-m_2^2)}\,, \qquad \sigma_a = \Bigl(\frac{(d-3)^2}{4} + \frac{\kappa_a^2}{m_a^2} \Bigr)^{1/2} \,, \quad a=1,2\,,\qquad
\\
\label{12092018-man03-07-a1} && \hspace{-1.4cm} J_{3+} = I_{\nu_3}(\sqrt{z_3})\,, \hspace{1cm} J_{3-} = K_{\nu_3} (\sqrt{z_3})\,, \quad z_3 =  - \frac{4\kappa_3B_3}{m_1^2 - m_2^2}\,,
\eeq
where $F(\alphabf,\betabf,\gammabf;x)$ stands for the hypergeometric function,  while $I_\nu(x)$ and $K_\nu(x)$ are the modified Bessel functions.
In \rf{12092018-man03-05}, in place of the variables $B_a$, we use new variables $z_a$ defined in  \rf{12092018-man03-07},\rf{12092018-man03-07-a1}. Operators $\nu_a$ are defined below.

From \rf{12092018-man03-03}, we see that the generic vertex $p_\smp3^-$ depends on the twelve variables, while, from \rf{12092018-man03-05}, we learn that the vertices $V_{\Asf\Bsf\Csf}$ depend only on the three variables. Note also that, by definition, the vertices $V_{\Asf\Bsf\Csf}$ \rf{12092018-man03-05} are expandable in the three variables $\alpha_{12}$, $\alpha_{23}$, $\alpha_{31}$. To complete our description of the vertex $p_\smp3^-$ \rf{12092018-man03-02} we should provide expressions for the operators $U$, $\nu_a$. These operators are given by
\beq
\label{12092018-man03-08} && \hspace{-1.5cm} U_\upsilon =  \upsilon_1^{N_1}\upsilon_2^{N_2}\upsilon_3^{N_3}\,, \qquad N_a = N_{B_a} + N_{\alpha_{aa+1}} + N_{\alpha_{a+2a}}\,,
\\
\label{12092018-man03-09} && \hspace{-1.5cm} U_\Gamma   =  \Bigl(\frac{2^{N_3} \Gamma(N_3 + \frac{d-2}{2})}{\Gamma( N_3 + 1)} \prod_{a=1,2}  \bigl( - \frac{\kappa_a^2}{m_a^2} \bigr)_\vph^{N_a} \frac{2^{N_a}\Gamma( N_a + \frac{d-2}{2}) }{ \Gamma( N_a - \lambda_{a+}) \Gamma( N_a - \lambda_{a-}) \Gamma(N_a + 1) } \Bigr)^{1/2}\,,
\\
&& \lambda_{a\pm} = \frac{d-3}{2} \pm \sigma_a\,, \qquad
\\
\label{12092018-man03-10} && \hspace{-1.5cm} U_\beta   = \exp\bigl( - \ \sum_{a=1,2,3} \frac{\betach_a}{2\beta_a} \kappa_a \partial_{B_a} \bigr)\,,
\\
&& \hspace{-1.5cm} U_{\partial B} = \exp\bigl(- \frac{m_1^2}{2m_2^2}\kappa_1\partial_{B_1} + \frac{m_1^2}{2m_2^2}\kappa_2\partial_{B_2} - \frac{m_1^2+m_2^2}{2(m_1^2-m_2^2)}\kappa_3\partial_{B_3} \bigr)\,,
\\
\label{12092018-man03-11} && \hspace{-1.5cm} U_{\partial \alpha} = \exp\Bigl( \bigl(  \frac{2(m_1^2 +m_2^2)}{(m_1^2 - m_2^2)^2}B_1 B_2  - \frac{\kappa_2 B_1}{m_2^2} + \frac{ \kappa_1 B_2}{ m_1^2} - \frac{\kappa_1\kappa_2 (m_1^2 +m_2^2)}{2 m_1^2 m_2^2} \bigr) \partial_{\alpha_{12}}
\nonumber\\
&&  \hspace{-1.2cm} + \bigl( \frac{ 2 B_1B_3 }{\kappa_3 q_3}   - \frac{ \kappa_1 B_3   }{m_1^2} - \frac{2m_1^2}{\kappa_3 q_3^2} B_1  - \frac{\kappa_1}{q_3}\bigr) \partial_{ \alpha_{31} } - \bigl( \frac{ 2B_2 B_3 }{\kappa_3 q_3} -  \frac{ \kappa_2 B_3   }{ m_2^2}  - \frac{2m_2^2}{\kappa_3 q_3^2} B_2  - \frac{\kappa_2}{q_3}\bigr)         \partial_{ \alpha_{23} } \Bigr),\qquad
\\
&& \hspace{-0.8cm} q_3 = (m_1^2-m_2^2)/\kappa_3\,,
\\
\label{12092018-man03-12} && \hspace{-1.5cm} U_B =   \prod_{a=1,2}  \bigl( -\frac{\kappa_a^2}{m_a^2} + \frac{4 m_a^2 }{(m_1^2-m_2^2)^2} B_a^2 \bigr)_\vph^{ -(\nu_a+1)/2}\,, \qquad U_{JB} = \bigl( - \frac{4\kappa_3}{m_1^2-m_2^2}B_3\bigr)_\vph^{-\nu_3/2}\,,
\\
\label{12092018-man03-13} && \hspace{-1.5cm} U_W = U_{\nu_1,W_1} U_{\nu_2,W_{23}} U_{\nu_3,W_{312}}\,, \hspace{1.8cm}  U_{\nu,W} \equiv \sum_{n=0}^\infty \frac{\Gamma(\nu+n)}{4^nn!\Gamma(\nu+2n) } W^n\,,
\\
\label{12092018-man03-14} && \hspace{-1.5cm}  W_1   =  2  \alpha_{23} \partial_{\alpha_{12}} \partial_{\alpha_{31}} -  \partial_{\alpha_{12}}^2   -    \partial_{\alpha_{31}}^2\,,\qquad
W_{23} =  2 \alpha_{31} \partial_{\alpha_{12}} \partial_{\alpha_{23}}  -    \partial_{\alpha_{23}}^2\,,
\\
\label{12092018-man03-15} && \hspace{-1.5cm} W_{312} =  2 \alpha_{12} \partial_{\alpha_{23}} \partial_{\alpha_{31}}\,, \hspace{3cm}  \nu_a  =   N_{\alpha_{aa+1}} +   N_{\alpha_{a+2a}} + \frac{d-4}{2}\,,
\eeq
where quantities $\betach_a$, $N_{B_a}$, $N_{\alpha_{ab}}$, $N_a$ appearing in  \rf{12092018-man03-08}-\rf{12092018-man03-15} are defined in \rf{02092018-man03-04app}-\rf{02092018-man03-06app}.

Using the notation for the vertices $V = V_{\Asf\Bsf\Csf}$, we note that the general $V$ can be presented as
\beq
\label{12092018-man03-18} && V  =   V(n_1,n_2,n_3) \,, \qquad  V(n_1,n_2,n_3) =   \alpha_{12}^{n_3} \alpha_{23}^{n_1}  \alpha_{31}^{n_2} \,,
\\
\label{12092018-man03-19} && n_a \geq 0, \qquad a=1,2,3\,.
\eeq
The three integers $n_a$ in \rf{12092018-man03-18} are the freedom of our solution for the vertex $V$. These integers label all possible
cubic vertices that can be constructed for three fields in \rf{12092018-man03-01}. In order for vertices \rf{12092018-man03-18} to be sensible, the integers $n_a$ should satisfy the restrictions \rf{12092018-man03-19} which amount to the requirement that the powers of all variables $\alpha_{12}$,  $\alpha_{23}$,  $\alpha_{31}$ in \rf{12092018-man03-18} be non--negative.  Relations given in \rf{12092018-man03-02}-\rf{12092018-man03-15}, \rf{12092018-man03-18}, and \rf{12092018-man03-19} provide the complete description and classification of cubic interaction vertices that can be constructed for  fields in \rf{12092018-man03-01}.

\subsection{   Three continuous-spin massive fields}

Finally, we discuss parity invariant cubic vertices for three continuous-spin massive fields. Using shortcut $(m,\kappa)_\smCSF$ for a continuous-spin massive field, we study cubic vertices for the following three fields:
\beq
\label{13092018-man03-01} && \hspace{-1cm} (m_1,\kappa_1)_\smCSF\hbox{-}(m_2,\kappa_2)_\smCSF\hbox{-}(m_3,\kappa_3)_\smCSF\,, \qquad    m_1^2 < 0\,,\qquad    m_2^2 < 0\,,\qquad m_3^2 < 0 \,,
\nonumber\\
&& \hspace{-1cm} \hbox{\small three continuous-spin massive fields.}\quad
\eeq
Relation \rf{13092018-man03-01} tells us that the mass-$m_a$ continuous-spin massive field  carry the external line index $a$, where $a=1,2,3$.

For fields \rf{13092018-man03-01}, we find the following general solution to cubic vertex $p_\smp3^-$ (see Appendix D)
\beq
\label{13092018-man03-02} p_\smp3^-  & = & U_\upsilon U_\Gamma U_\beta U_{\partial B}  U_{\partial \alpha} U_B U_W U_B^{-1} V_{\Asf\Bsf\Csf}^{(7)} \,,   \qquad \Asf,\Bsf,\Csf = +,-\,,
\\
\label{13092018-man03-03} && p_\smp3^- = p_\smp3^-  (\beta_a, B_a, \alpha_{aa+1}\,, \upsilon_a)\,,
\\
\label{13092018-man03-04} && V_{\Asf\Bsf\Csf}^{(7)} = V_{\Asf\Bsf\Csf}^{(7)}(B_a,\alpha_{aa+1})\,.
\eeq
In \rf{13092018-man03-02}, we introduce eight vertices $V_{\Asf\Bsf\Csf}^{(7)}$ labelled by the superscripts $\Asf,\Bsf,\Csf$. In \rf{13092018-man03-03} and  \rf{13092018-man03-04}, the arguments of the generic vertex $p_\smp3^-$ and the vertices $V_{\Asf\Bsf\Csf}^{(7)}$ are shown explicitly. The definition of the arguments $B_a$ and $\alpha_{ab}$ may be found in \rf{02092018-man03-04app}. Various quantities $U$ appearing in \rf{13092018-man03-02} are differential operators w.r.t. the $B_a$ and $\alpha_{aa+1}$. These operators are given below. For eight vertices $V_{\Asf\Bsf\Csf}^{(7)}$ \rf{13092018-man03-04}, we find the following solution:
\beq
\label{13092018-man03-05} && \hspace{-0.8cm} V_{\Asf\Bsf\Csf}^{(7)}  = F_{1\Asf} F_{2\Bsf} F_{3\Csf} V_{\Asf\Bsf\Csf}\,, \qquad V_{\Asf\Bsf\Csf} = V_{\Asf\Bsf\Csf} (\alpha_{12},\alpha_{23},\alpha_{31})\,, \qquad \Asf,\Bsf,\Csf = +,-\,,
\\
\label{13092018-man03-06} && F_{a\pm} = F(\alphabf_a,\betabf_a,\gammabf_a;\, \frac{1\pm z_a}{2})\,,  \qquad \alphabf_a = \nu_a + \half + \sigma_a \,, \qquad   \betabf_a = \nu_a + \half - \sigma_a \,,
\\
\label{13092018-man03-07}  && \gammabf_a = \nu_a  + 1\,, \qquad z_a =  -\frac{2m_a^2 B_a }{\kappa_a\sqrt{D}}\,, \qquad \sigma_a = \Bigl( \frac{(d-3)^2}{4} + \frac{\kappa_a^2}{m_a^2} \Bigr)^{1/2} \,,\quad a=1,2,3\,,\qquad
\eeq
where the $F(\alphabf,\betabf,\gammabf;x)$ is the hypergeometric function.
In \rf{13092018-man03-05}, in place of the variables $B_a$, we use new variables $z_a$ defined in \rf{13092018-man03-07}. Quantities $\nu_a$ and $D$ are  defined below in \rf{13092018-man03-15}, \rf{13092018-man03-15-a2}.

The generic vertex $p_\smp3^-$ \rf{13092018-man03-03} depends on the twelve variables, while, the vertices $V_{\Asf\Bsf\Csf}$ \rf{13092018-man03-05} depend only on the three variables. Note also that, by definition, the vertices $V_{\Asf\Bsf\Csf}$ \rf{13092018-man03-05} are expandable in the three variables $\alpha_{12}$, $\alpha_{23}$, $\alpha_{31}$. To complete our description of the vertex $p_\smp3^-$ \rf{13092018-man03-03} we should provide expressions for the operators $U$, $\nu_a$. These operators are given by
\beq
\label{13092018-man03-08} && \hspace{-1.1cm} U_\upsilon =  \upsilon_1^{N_1}\upsilon_2^{N_2}\upsilon_3^{N_3}\,, \qquad N_a = N_{B_a} + N_{\alpha_{aa+1}} + N_{\alpha_{a+2a}}\,,
\\
\label{13092018-man03-09} && \hspace{-1.1cm} U_\Gamma   =  \prod_{a=1,2,3} \Bigl( \bigl( - \frac{\kappa_a^2}{m_a^2} \bigr)_\vph^{N_a} \frac{2^{N_a}\Gamma( N_a + \frac{d-2}{2}) }{ \Gamma( N_a - \lambda_{a+}) \Gamma( N_a - \lambda_{a-}) \Gamma(N_a + 1) } \Bigr)^{1/2}\,,\quad \lambda_{a\pm} = \frac{d-3}{2} \pm \sigma_a\,, \qquad
\\
\label{13092018-man03-10} && \hspace{-1.1cm} U_\beta   = \exp\bigl( - \sum_{a=1,2,3} \frac{\betach_a}{2\beta_a} \kappa_a \partial_{B_a}   \bigr)\,,
\\
&& \hspace{-1.2cm} U_{\partial B} = \exp\bigl(- \sum_{a=1,2,3}\frac{m_a^2-m_{a+2}^2}{2m_a^2}\kappa_a\partial_{B_a} \bigr)\,,
\\
\label{13092018-man03-11} && \hspace{-1.1cm}
U_{\partial\alpha}  =  \exp\Bigl(\sum_{a=1,2,3} \Bigl(\frac{ \kappa_{a+1} B_{a+2}}{m_{a+1}^2 } - \frac{ \kappa_{a+2} B_{a+1}}{m_{a+2}^2 }  - \frac{h_{a+1a+2} \kappa_{a+1}\kappa_{a+2}}{ 2 m_{a+1}^2 m_{a+2}^2 } \Bigr)\partial_{\alpha_{a+1a+2} } \Bigr)
\nonumber\\
&&  \hspace{-0.4cm}\times\,\, \exp\Bigl(\sum_{a=1,2,3} \frac{ 2h_{a+1a+2} }{D} B_{a+1}B_{a+2} \partial_{\alpha_{a+1a+2} } \Bigr),
\\
\label{13092018-man03-12} && \hspace{-1.1cm} U_B =  \prod_{a=1,2,3} \bigl( - \frac{\kappa_a^2}{m_a^2} + \frac{4 m_a^2 }{D} B_a^2 \bigr)_\vph^{ -(\nu_a+1)/2} \,,
\\
\label{13092018-man03-13} && \hspace{-1.1cm} U_W = U_{\nu_1,W_1} U_{\nu_2,W_{23}} U_{\nu_3,W_{312}}\,, \hspace{1.8cm}  U_{\nu,W} \equiv \sum_{n=0}^\infty \frac{\Gamma(\nu+n)}{4^nn!\Gamma(\nu+2n) } W^n\,,
\\
\label{13092018-man03-14} && \hspace{-1.1cm}  W_1   =  2  \alpha_{23} \partial_{\alpha_{12}} \partial_{\alpha_{31}} -  \partial_{\alpha_{12}}^2   -    \partial_{\alpha_{31}}^2\,,\qquad
W_{23} =  2 \alpha_{31} \partial_{\alpha_{12}} \partial_{\alpha_{23}}  -    \partial_{\alpha_{23}}^2\,,
\\
\label{13092018-man03-15} && \hspace{-1.1cm} W_{312} =  2 \alpha_{12} \partial_{\alpha_{23}} \partial_{\alpha_{31}}\,, \hspace{3cm}  \nu_a  =   N_{\alpha_{aa+1}} +   N_{\alpha_{a+2a}} + \frac{d-4}{2}\,,
\\
\label{13092018-man03-15-a1} &&  \hspace{-1.1cm}  h_{aa+1} = m_a^2 + m_{a+1}^2 - m_{a+2}^2\,,
\\
\label{13092018-man03-15-a2} &&  \hspace{-1.1cm}  D = m_1^4 + m_2^4 +m_3^4 - 2(m_1^2m_2^2 + m_2^2 m_3^2 + m_3^2 m_1^2)\,,
\eeq
where $\betach_a$, $N_{B_a}$, $N_{\alpha_{ab}}$, $N_a$  are defined in \rf{02092018-man03-04app}-\rf{02092018-man03-06app}.

Using the notation for the vertices $V = V_{\Asf\Bsf\Csf}$, we note that the general representative of the vertex $V$ can be presented as
\beq
\label{13092018-man03-18} && V  =   V(n_1,n_2,n_3) \,, \qquad  V(n_1,n_2,n_3) =   \alpha_{12}^{n_3} \alpha_{23}^{n_1}  \alpha_{31}^{n_2} \,,
\\
\label{13092018-man03-19} && n_a \geq 0, \qquad a=1,2,3\,.
\eeq
The three integers $n_a$ in \rf{13092018-man03-18} are the freedom of our solution for the vertex $V$. These integers label all possible
cubic vertices that can be constructed for three fields in \rf{13092018-man03-01}. In order for vertices \rf{13092018-man03-18} to be sensible, the integers $n_a$ should satisfy the restrictions \rf{13092018-man03-19} which amount to the requirement that the powers of all variables $\alpha_{12}$,  $\alpha_{23}$,  $\alpha_{31}$ in \rf{13092018-man03-18} be non--negative.  Relations given in \rf{13092018-man03-02}-\rf{13092018-man03-15-a2}, \rf{13092018-man03-18}, and \rf{13092018-man03-19} provide the complete description and classification of cubic interaction vertices that can be constructed for  fields in \rf{13092018-man03-01}.

\newsection{ \large Conclusions}\label{sec-07}

In this paper, we used the light-cone gauge approach to construct
the parity invariant cubic vertices for continuous-spin massive/massless fields
and arbitrary spin massive/massless fields.
We investigated three types of the parity invariant cubic vertices:
a) vertices describing coupling of one continuous-spin massive/massless field to two arbitrary spin massive/massless fields; b) vertices describing coupling of two continuous-spin massive/massless fields to one arbitrary spin massive/massless field; c) vertices for self-interacting massive/massless continuous-spin fields. We obtained the complete list of  such cubic vertices. With exception of cubic vertices for massless self-interacting continuous-spin field, results in this paper together with the ones in Ref.\cite{Metsaev:2017cuz} provide the exhaustive solution to the problem of description of all parity invariant cubic vertices for coupling of massive/massless continuous-spin fields to arbitrary spin massive/massless fields as well as vertices for self-interacting massive/massless continuous-spin fields. Results in this paper might have the following applications and generalizations.

\noindent \ibf) Sometimes light-cone gauge formulation turns out to be good starting point for deriving Lorentz covariant formulations.
It is the parity invariant light-cone gauge vertices that turn out to convenient for deriving of Lorentz covariant and BRST gauge invariant formulations \cite{Siegel:1988yz}. For example, all light-cone gauge vertices for massive/massless fields in Ref.\cite{Metsaev:2005ar} have straightforwardly been cast into BRST gauge invariant form in Ref.\cite{Metsaev:2012uy}.%
\footnote{ For Lorentz covariant formulations of massless higher-spin fields in flat space, see also Refs.\cite{Sagnotti:2010at}.}
Various BRST formulations of free continuous-spin fields were discussed in Refs.\cite{Bengtsson:2013vra}-\cite{Buchbinder:2018yoo}. We expect therefore that our results in this paper may be good starting point for deriving Lorentz covariant and BRST gauge invariant formulations of vertices for interacting continuous-spin fields. Various applications of BRST approach for studying interacting finite-component fields  may be found, e.g., in Refs.\cite{Bekaert:2005jf}.

\noindent \iibf) From the perspective of investigation of interrelations between theory of continuous-spin fields and string theory it is important to extend our study to the case of mixed-symmetry fields. During last time, various interesting descriptions of mixed-symmetry fields were obtained in Refs.\cite{Metsaev:2017myp}-\cite{Alkalaev:2018bqe}.
For example, light-cone gauge description in Ref.\cite{Metsaev:2017myp} and methods in this paper provide opportunity for investigation of interacting mixed-symmetry continuous-spin fields. Also, for finite-component mixed-symmetry fields, we mention interesting formulations developed in Refs.\cite{Boulanger:2008up}. Generalization of these formulations to the case continuous-spin fields could also be of some interest.

\noindent \iiibf) In this paper, we investigated cubic vertices for light-cone gauge continuous-spin fields propagating in flat space. Extension of our investigation to the case of continuous-spin fields in AdS space \cite{Metsaev:2016lhs,Metsaev:2017ytk} could be of some interest. In this respect we note that light-cone gauge formulation of free continuous-spin AdS fields was recently obtained in Ref.\cite{Metsaev:2017myp}, while for finite component light-cone gauge AdS fields, the systematic method for building cubic vertices was developed in Refs.\cite{Metsaev:2018xip}. We think therefore that results in Refs.\cite{Metsaev:2017myp,Metsaev:2018xip} provide opportunity for investigation of cubic vertices for continuous-spin AdS fields.
Also we note that, for finite-component interacting AdS fields, many interesting formulations and results were obtained in Refs.\cite{Metsaev:2014sfa}-\cite{Joung:2011ww}, which, upon a generalization,
might be useful for investigation of interaction vertices of continuous-spin AdS fields.
Also we note frame-like formulations of continuous-spin AdS fields obtained in Refs.\cite{Khabarov:2017lth,Ponomarev:2010st} which seem to be convenient for studying interacting AdS fields.

\noindent \ivbf) Generalization of our results to the case of
interacting supersymmetric continuous-spin field theories could be of great interest. As is known, for finite component fields, supersymmetry imposes additional constraints and leads to more simple interacting vertices. We expect therefore that, for continuous-spin field, supersymmetry might simplify a structure of interactions vertices. Supermultiplets for continuous-spin representations are considered in Refs.\cite{Brink:2002zx,Zinoviev:2017rnj}. For finite-component fields, recent study of higher-spin supersymmetric theories can be found in Refs.\cite{Kuzenko:2016qwo,Buchbinder:2018wzq}. Use of twistor-like variables for discussion of supersymmetric theories turns out to be helpful. Recent various interesting applications of twistor-like variables may be found in Refs.\cite{Buchbinder:2018soq,Adamo:2018srx}. Finally, we note that investigation of various algebraic aspects of continuous-spin field theory along the line in Refs.\cite{Basile:2014wua} could also be very interesting.

\medskip

\noindent {\bf Acknowledgments}. This work was supported by Russian Science Foundation grant 14-42-00047.

\setcounter{section}{0}\setcounter{subsection}{0}
\appendix{ \large Notation and conventions  }

Unless otherwise specified, the vector indices of the $so(d-2)$ algebra $i,j,k,l$ run over $1,\ldots ,d-2$. We refer to creation operators $\alpha^i$, $\upsilon$, $\zeta$ and the respective annihilation operators $\alphab^i$, $\upsilonb$, $\zetab$  as oscillators. Our conventions for the commutation relations, the vacuum $|0\rangle $, and hermitian conjugation rules are as follows
\beq
\label{02092018-man03-01app} && [ \bar\alpha^i,\alpha^j] = \delta^{ij}, \quad [\upsilonb,\upsilon]=1,  \quad [\zetab,\zeta]=1,  \quad \alphab^i |0\rangle = 0\,,\quad \upsilonb |0\rangle = 0\,,   \quad \zetab |0\rangle = 0\,,\qquad
\\
\label{02092018-man03-02app} && \alpha^{i \dagger} = \alphab^i\,, \hspace{1.1cm} \upsilon^\dagger = \upsilonb\,, \hspace{1cm} \zeta^\dagger = \zetab\,.
\eeq
Throughout this paper we use the following definitions for momentum $\Po^i$ and quantities $B_a$, $\alpha_{ab}$
\be \label{02092018-man03-04app}
\Po^i \equiv \frac{1}{3}\sum_{a=1,2,3} \betach_a p_a^i\,, \qquad
\betach_a\equiv \beta_{a+1}-\beta_{a+2}\,, \qquad B_a \equiv \frac{\alpha_a^i \Po^i}{\beta_a}\,, \qquad \alpha_{ab} \equiv \alpha_a^i\alpha_b^i\,,
\ee
where $\beta_a\equiv \beta_{a+3}$.
Our notation for the scalar product of the oscillators
and various quantities constructed out of the $B_a$, $\alpha_{ab}$ and  derivatives of the $B_a$, $\alpha_{ab}$  are as follows
\beq
\label{02092018-man03-03app} &&  \hspace{-1.5cm} \alpha^2 \equiv \alpha^i\alpha^i\,,\qquad \alphab^2 \equiv \alphab^i \alphab^i\,,\qquad N_\alpha  \equiv \alpha^i \alphab^i\,, \qquad N_\zeta  \equiv \zeta \zetab\,,\qquad N_\upsilon  \equiv \upsilon \upsilonb\,,
\\
\label{02092018-man03-05app} && \hspace{-1.5cm} N_{B_a} = B_a\partial_{B_a}\,,\qquad \ \ \partial_{B_a} = \partial/\partial B_a\,,
\\
&&  \hspace{-1.5cm} N_{\alpha_{ab}} = \alpha_{ab}\partial_{\alpha_{ab}}\,,\qquad \partial_{\alpha_{ab}} = \partial/\partial \alpha_{ab}\,,
\\
\label{02092018-man03-06app} &&  \hspace{-1.5cm} N_a = N_{B_a} + N_{\alpha_{aa+1}} + N_{\alpha_{a+2a}}\,,
\\
\label{02092018-man03-06app-bb1} && \hspace{-1.5cm} \nu_a =  N_{\alpha_{aa+1}} + N_{\alpha_{a+2a}} + \frac{d-4}{2}\,,
\\
\label{02092018-man03-06app-bb2} &&  \hspace{-1.5cm} g_{v_a}  = \Bigl(\frac{F_a}{(N_a + 1) (2N_a + d - 2)} \Bigr)^{1/2} \,,\qquad  F_a =    \kappa_a^2 -  N_a (N_a + d - 3) m_a^2\,,
\\
\label{02092018-man03-10app} &&  \hspace{-1.5cm} U_{\upsilon_a} = \upsilon_a^{N_a}\,,
\\
\label{02092018-man03-11app} &&  \hspace{-1.5cm} U_{\Gamma_a}   =  \Bigl(\frac{2^{N_a} \Gamma(N_a + \frac{d-2}{2})}{\Gamma( N_a + 1)} \Bigr)^{1/2}\,, \hspace{2.7cm} \hbox{ for massless continuous-spin field}
\\
\label{02092018-man03-12app} &&  \hspace{-1.5cm} U_{\Gamma_a}   =  \Bigl(\bigl( - \frac{\kappa_a^2}{m_a^2} \bigr)_\vph^{N_a} \frac{2^{N_a}\Gamma( N_a + \frac{d-2}{2}) }{ \Gamma( N_a - \lambda_{a+}) \Gamma( N_a - \lambda_{a-}) \Gamma(N_a + 1) } \Bigr)^{1/2}\,,
\nonumber\\
&& \hspace{-0.5cm} \lambda_{a\pm} = \frac{d-3}{2} \pm \Bigl(\frac{(d-3)^2}{4} + \frac{\kappa_a^2}{m_a^2} \Bigr)^{1/2} \,, \hspace{0.5cm} \hbox{ for massive continuous-spin field}
\\
\label{02092018-man03-13app} &&  \hspace{-1.5cm} U_{B_a} = \bigl(\frac{1}{e_a}- \frac{4\kappa_a^2 e_a}{D} B_a^2\bigr)^{ -(\nu_a+1)/2 }\,,
\\
\label{02092018-man03-14app} &&  \hspace{-1.5cm} U_{\nu,W} = \sum_{n=0}^\infty \frac{\Gamma(\nu+n)}{4^nn!\Gamma(\nu+2n) } W^n  \,,
\\
\label{02092018-man03-15app}  &&  \hspace{-1.5cm} U_D = \exp\Bigl( \frac{2}{D} h_{12} B_1 B_2 \partial_{\alpha_{12}} + \frac{2}{D} h_{23} B_2 B_3 \partial_{\alpha_{23}} + \frac{2}{D} h_{31} B_3 B_1 \partial_{\alpha_{31}} \Bigr)\,.
\eeq
Our notation for the quantities constructed out of the masses $m_a$ and continuous-spin parameters $\kappa_a$ are as follows
\beq
\label{02092018-man03-07app} && r_a  =  \half m_{a+1}^2 + \half m_{a+2}^2 - \frac{1}{4} m_a^2\,, \hspace{1cm}  q_a = \frac{m_{a+1}^2- m_{a+2}^2}{\kappa_a}\,,
\\
\label{02092018-man03-08app} && e_a = - \frac{m_a^2}{\kappa_a^2}\,, \hspace{4.3cm} u_a = 1 + \frac{(d-2)(d-4)m_a^2}{4\kappa_a^2}\,,\qquad
\\
\label{02092018-man03-09app} && D = m_1^4 + m_2^4 +m_3^4 - 2(m_1^2m_2^2 + m_2^2 m_3^2 + m_3^2 m_1^2)\,,
\\
\label{02092018-man03-09app-x1} && h_{aa+1} = m_a^2 + m_{a+1}^2 - m_{a+2}^2\,.
\eeq

\appendix{ \large Derivation of vertices $p_\smp3^-$ \rf{03092018-man03-02}, \rf{04092018-man03-02}, \rf{05092018-man03-02} }

We consider three vertices in \rf{03092018-man03-01-add}-\rf{05092018-man03-01-add}. For these vertices, we outline the derivation of the respective solutions in \rf{03092018-man03-02}, \rf{04092018-man03-02},\rf{05092018-man03-02}. We split our derivation in several steps.

\noindent {\bf Realization $G_a$, $G_\beta$ on $p_\smp3^-$ \rf{05092018-man03-01-add} for arbitrary masses}. For $p_\smp3^-$  \rf{05092018-man03-01-add}, we now find $G_\beta$ and $G_{a,\Po^2}$ in \rf{01092018-man03-52}. We use $M_a^i$, $a=1,2$ for massive field \rf{01092018-man03-25} and $M_3^i$ for continuous-spin field \rf{01092018-man03-22}. Plugging such $M_a^i$ into $\Jbf^{-i\dagger}$ \rf{01092018-man03-47}, we cast $\Jbf^{-i\dagger}|p_\smp3^-\rangle$ into the form given in \rf{01092018-man03-52} with the following $G_\beta$ and $G_{a,\Po^2}$:
\beq
\label{14092018-man03-01} &&   \hspace{-1.5cm} G_{1,\Po^2} =   G_1\,, \qquad G_{2,\Po^2} =   G_2\,, \qquad G_{3,\Po^2} =   G_3 +  \Pbf^- \frac{2 \beta}{\beta_3^3} \alpha_3^i \frac{ g_{\upsilon_3} \partial_{\upsilon_3} }{2N_3 + d-2 } \partial_{B_3}^2\,,
\\
\label{14092018-man03-02} && \hspace{-1.5cm} G_1   =    (B_3 - \frac{\beta_1}{\beta_3}  g_{\upsilon_3} \partial_{\upsilon_3} )\partial_{\alpha_{31}}  - ( B_2 + \frac{\beta_1}{\beta_2} m_2 \zeta_2   ) \partial_{\alpha_{12}} +  \half ( \frac{\betach_1}{ \beta_1} m_1^2 +  m_2^2 -m_3^2) \partial_{B_1} + m_1\partial_{\zeta_1}\,,
\\
\label{14092018-man03-03} &&  \hspace{-1.5cm} G_2  =     (B_1 - \frac{\beta_2}{\beta_1}  \zeta_1 m_1)\partial_{\alpha_{12}}
- ( B_3 + \frac{\beta_2}{\beta_3} g_{\upsilon_3} \partial_{\upsilon_3} ) \partial_{\alpha_{23}}  +  \half ( \frac{\betach_2}{ \beta_2} m_2^2 + m_3^2 - m_1^2 ) \partial_{B_2} + m_2\partial_{\zeta_2}\,,
\\
\label{14092018-man03-04} && \hspace{-1.5cm} G_3  =     (B_2 - \frac{\beta_3}{\beta_2} m_2 \zeta_2   )\partial_{\alpha_{2 3}}
-  ( B_1 + \frac{\beta_3}{\beta_1} m_1 \zeta_1 ) \partial_{\alpha_{31}}
+  \half ( \frac{\betach_3}{\beta_3} m_3^2 +  m_1^2 - m_2^2) \partial_{B_3} +  \upsilon_3 g_{\upsilon_3}
\nonumber\\
&& \hspace{-0.9cm} + \,\, \frac{ g_{\upsilon_3} \partial_{\upsilon_3}}{2N_3+d-2} \Bigl( \frac{2\beta_1}{\beta_3} B_1 \partial_{B_3} \partial_{\alpha_{31}} + \frac{2 \beta_2}{\beta_3} B_2 \partial_{B_3} \partial_{\alpha_{23}}
\nonumber\\
&& \hspace{-0.8cm} + \,\, 2 \alpha_{12} \partial_{\alpha_{31}}\partial_{\alpha_{23}}  +   \alpha_{11} \partial_{\alpha_{31}}^2 + \alpha_{22}\partial_{\alpha_{23}}^2
+ \frac{\beta}{\beta_3^2} \sum_{b=1}^3 \frac{m_b^2}{\beta_b} \partial_{B_3}^2 \Bigr)\,,
\\
\label{14092018-man03-05} && \hspace{-1.5cm} G_\beta  =     -  \frac{1}{\beta}  \No_\beta -  \frac{1}{\beta_1^2} m_1\zeta_1  \partial_{B_1} - \frac{1}{\beta_2^2} m_2\zeta_2  \partial_{B_2}  - \frac{1}{\beta_3^2} g_{\upsilon_3} \partial_{\upsilon_3} \partial_{B_3}\,,
\eeq
where $g_{v_a}$ are given in \rf{02092018-man03-06app-bb2}.  Using $G_a$, $G_\beta$ \rf{14092018-man03-02}-\rf{14092018-man03-05}, we now consider equations \rf{01092018-man03-53},\rf{01092018-man03-54}.

\noindent {\bf Vertex $V^{(3)}$}. Multiplying $G_3$ \rf{14092018-man03-04} on the left by $(2N_3+d-2)/\kappa_3$, we use the transformations%
\footnote{ To investigate equations $G_ap_\smp3^-=0$ it is convenient to use equivalence class for the $G_a$. Namely, the $G_a$ and $(2N_a+d-2)G_a$ are considered to be equivalent. For detailed discussion of transformations governed by operators $U_{\upsilon_3}$, $U_{\Gamma_3}$, $U_\beta$ \rf{14092018-man03-06}, see {\bf Steps 2,3,4} in Appendix C in Ref.\cite{Metsaev:2017cuz}.
}
\beq
\label{14092018-man03-06} && \hspace{-1cm} p_\smp3^-  =   U_{\upsilon_3} V^{(1)}\,, \qquad V^{(1)} = U_{\Gamma_3} V^{(2)}\,,  \qquad
V^{(2)}   = U_\beta  V^{(3)}\,,
\\
\label{14092018-man03-07} &&  U_\beta   = \exp\bigl( - \frac{\betach_1}{2\beta_1}m_1\zeta_1 \partial_{B_1} - \frac{\betach_2}{2\beta_2} m_2\zeta_2 \partial_{B_2} - \frac{\betach_3}{2\beta_3} \kappa_3 \partial_{B_3} \bigr)\,,
\eeq
where $ U_{\upsilon_3}$, $U_{\Gamma_3}$ are given in \rf{02092018-man03-10app}, \rf{02092018-man03-12app}. Realization of $G_a$ \rf{14092018-man03-02}-\rf{14092018-man03-04} on $V^{(3)}$ takes the form
\beq
\label{14092018-man03-08} G_1 &  =  &    (B_3 + \half \kappa_3  )\partial_{\alpha_{31}}  - ( B_2 - \half m_2 \zeta_2   ) \partial_{\alpha_{12}} + \half  (m_2^2 - m_3^2)  \partial_{B_1} + m_1\partial_{\zeta_1}\,,
\\
\label{14092018-man03-09} G_2 &  =  &   (B_1 + \half \zeta_1 m_1)\partial_{\alpha_{12}}
- ( B_3 - \half \kappa_3 ) \partial_{\alpha_{23}}
+ \half (m_3^2 - m_1^2) \partial_{B_2} + m_2\partial_{\zeta_2}\,,
\\
\label{14092018-man03-10} G_3 &  =  &  \frac{2N_3+d-2}{\kappa_3 }\Bigl(  (B_2 + \frac{m_2}{2} \zeta_2   )\partial_{\alpha_{23}}
-  ( B_1 - \frac{m_1}{2} \zeta_1 ) \partial_{\alpha_{31}} + \frac{ m_1^2 - m_2^2}{2} \partial_{B_3}\Bigr)\qquad
\nonumber\\
& + &    \frac{F_3}{\kappa_3^2} - (B_1 + \frac{3}{2}m_1\zeta_1) \partial_{B_3} \partial_{\alpha_{31}} - (B_2 - \frac{3}{2}m_2\zeta_2) \partial_{B_3} \partial_{\alpha_{23}}
\nonumber\\
&  +  &  2\alpha_{12} \partial_{\alpha_{31}}\partial_{\alpha_{23}} +   \alpha_{11} \partial_{\alpha_{31}}^2 + \alpha_{22}\partial_{\alpha_{23}}^2
+(\frac{1}{4}m_3^2 - \half m_1^2 -\half  m_2^2) \partial_{B_3}^2  ,\qquad
\\
\label{14092018-man03-11} G_\beta &  =  &    -  \frac{1}{\beta}  \No_\beta \,.
\eeq

\noindent {\bf Dependence of $V^{(3)}$ on $\beta_a$}. We fix dependence of $V^{(3)}$ on $\beta_a$. To this end we use Eqs.\rf{01092018-man03-54},\rf{01092018-man03-58} and relation \rf{14092018-man03-11} to note the following equations for the vertex $V^{(3)}$:
\be \label{14092018-man03-12}
\No_\beta V^{(3)} = 0 \,, \qquad \sum_{a=1,2,3}\beta_a\partial_{\beta_a} V^{(3)} = 0\,.
\ee
Equations \rf{14092018-man03-12} imply that $V^{(3)}$ is independent of $\beta_1$, $\beta_2$, $\beta_3$. Thus, the vertex $V^{(3)}$ is given by
\be \label{14092018-man03-13}
V^{(3)} = V^{(3)}(B_a,\alpha_{aa+1},\zeta_1,\zeta_2)\,.
\ee
We now use results above-obtained to study three vertices  \rf{03092018-man03-01-add}-\rf{05092018-man03-01-add} in turn.
%
%
%
%
%
\be  \label{14092018-man03-14}
\hbox{\bf \large Case \quad $m_1=0$,\quad $m_2=0$,\quad $m_3^2 < 0$. }
\ee
%
%
%
%
%
{\bf Step 1}. This case corresponds to the vertex $p_\smp3^-$ \rf{03092018-man03-01-add}. Setting $m_1=0$, $m_2=0$ and ignoring $\zeta_1$, $\zeta_2$ in \rf{14092018-man03-01}-\rf{14092018-man03-13}, we use vertex $V^{(3)}$ given by
\be \label{14092018-man03-15}
V^{(3)} = V^{(3)}(B_a,\alpha_{aa+1})\,,
\ee
where the realization of operators $G_a$ \rf{14092018-man03-08}-\rf{14092018-man03-10} on the vertex $V_3$ \rf{14092018-man03-15} takes the form%
\footnote{ For massless fields corresponding to $a=1,2$, $\alpha_{11}$-, $\alpha_{22}$-terms in \rf{14092018-man03-04} can be ignored in \rf{14092018-man03-18} by virtue of the second constraint in \rf{01092018-man03-14}.
}
\beq
 \label{14092018-man03-16} && \hspace{-0.7cm} G_1  =     (B_3 + \half \kappa_3  )\partial_{\alpha_{31}}  -  B_2  \partial_{\alpha_{12}} -\half  m_3^2   \partial_{B_1}  \,,
\\
 \label{14092018-man03-17} && \hspace{-0.7cm} G_2  =      B_1  \partial_{\alpha_{12}}
- ( B_3 - \half \kappa_3 ) \partial_{\alpha_{23}} +   \half  m_3^2   \partial_{B_2} \,,
\\
\label{14092018-man03-18} && \hspace{-0.7cm} G_3   =    - r_3 \partial_{B_3}^2  + e_3 (N_{B_3} + \nu_3)(  N_{B_3} + \nu_3 + 1 ) + u_3
\nonumber\\
&& \hspace{-0.7cm} + \frac{2}{\kappa_3}(N_{B_3}+ \nu_3 +1) \bigl(B_2 \partial_{\alpha_{23}} -    B_1   \partial_{\alpha_{31}}\bigr) -  B_1   \partial_{B_3} \partial_{\alpha_{31}} -  B_2   \partial_{B_3} \partial_{\alpha_{23}}  +  2\alpha_{12} \partial_{\alpha_{31}}\partial_{\alpha_{23}}\,, \qquad
\eeq
where, here and below, we use quantities \rf{02092018-man03-07app}-\rf{02092018-man03-09app-x1} taken for masses given in \rf{14092018-man03-14}.

\noindent {\bf Step 2}. We use the transformation
\be\label{14092018-man03-19}
V^{(3)} = U_{\partial\alpha} V^{(4)}\,, \quad U_{\partial\alpha}   = \exp\bigl(- \frac{2B_1B_2}{m_3^2}  \partial_{ \alpha_{12} } + \frac{2B_1}{m_3^2} (B_3+ \half \kappa_3) \partial_{\alpha_{31}} + \frac{2B_2}{m_3^2} (B_3 - \half \kappa_3) \partial_{\alpha_{23}}  \bigr).
\ee
Realization of $G_a$ \rf{14092018-man03-16}-\rf{14092018-man03-18} on $V^{(4)}$ takes the form
\beq
\label{14092018-man03-20} && \hspace{-1.6cm}   G_1 = - \half  m_3^2   \partial_{B_1}\,, \qquad  G_2 = \half  m_3^2   \partial_{B_2}\,,
\\
\label{14092018-man03-21} && \hspace{-1.6cm} G_3   =    - \frac{D}{4\kappa_3^2e_3} \partial_{B_3}^2  + e_3 (N_{B_3} + \nu_3)(  N_{B_3} + \nu_3 + 1 ) + u_3  +   W_3\,, \qquad  W_3 =     2 \alpha_{12} \partial_{\alpha_{31}}\partial_{\alpha_{23}}\,.\quad
\eeq
From \rf{14092018-man03-20} and equations $G_1V^{(4)}=0$, $G_2V^{(4)}=0$, we see that  $V^{(4)}$ is independent of $B_1$, $B_2$,
\be \label{14092018-man03-23}
V^{(4)}= V^{(4)}(B_3,\alpha_{aa+1})\,.
\ee

\noindent {\bf Step 3}. We make the transformations
\be \label{14092018-man03-24}
V^{(4)} = U_{B_3} V^{(5)}\,, \qquad V^{(5)} = U_{\nu_3,W_3} V^{(6)}\,, \qquad
V^{(6)} = U_{B_3}^{-1} V^{(7)}\,,
\ee
where $U_{B_3}$, $U_{\nu_3,W_3}$  are given in \rf{02092018-man03-13app},\rf{02092018-man03-14app}. Realizations of $G_3$ \rf{14092018-man03-21}  on $V^{(5)}$, $V^{(6)}$, $V^{(7)}$ take the forms
\beq
\label{14092018-man03-25} && \hspace{-1.5cm} G_3  =   (B_3^2 - \frac{D}{4\kappa_3^2e_3^2})  \partial_{B_3}^2    +    (1- \frac{4\kappa_3^2e_3^2}{D} B_3^2)^{-1} \bigl( \nu_3^2 - 1 + W_3 \bigr) + \frac{u_3}{e_3} \,, \hspace{1cm} \hbox{for } V^{(5)};
\\
\label{14092018-man03-26} && \hspace{-1.5cm} G_3  =   (B_3^2 - \frac{D}{4\kappa_3^2e_3^2})  \partial_{B_3}^2  +    (1- \frac{4\kappa_3^2e_3^2}{D} B_3^2)^{-1} \bigl( \nu_3^2 - 1 \bigr)  + \frac{u_3}{e_3}  \,, \hspace{2.1cm} \hbox{for } V^{(6)};
\\
\label{14092018-man03-27} && \hspace{-1.5cm} G_3   =    - \frac{D}{4\kappa_3^2e_3} \partial_{B_3}^2  + e_3 (N_{B_3} + \nu_3)(  N_{B_3} + \nu_3 + 1 ) + u_3 \,, \hspace{2.8cm} \hbox{for } V^{(7)}.
\eeq
For $\nu_3$ \rf{02092018-man03-06app-bb1} and $W_3$ \rf{14092018-man03-21}, we note the relation which admits us to get $G_3$ in \rf{14092018-man03-26},
\be
(\nu_3^2+ W_3)U_{\nu_3,W_3}  = U_{\nu_3,W_3} \nu_3^2\,.
\ee
Equation $G_3V^{(7)}=0$ with $G_3$ as in \rf{14092018-man03-27} is the second-order differential equation w.r.t. $B_3$. Two independent solutions of this equation are given in \rf{03092018-man03-05}.
%
%
%
%
%
\be  \label{14092018-man03-27-a1}
\hbox{\bf \large Case \quad $m_1=0$,\quad $m_2^2 > 0$,\quad $m_3^2 < 0$.}
\ee
%
%
%
%
%
{\bf Step 1}. This case corresponds to the vertex $p_\smp3^-$ \rf{04092018-man03-01-add}. Setting $m_1=0$ and ignoring $\zeta_1$ in \rf{14092018-man03-01}-\rf{14092018-man03-13}, we use the vertex $V^{(3)}$ given by
\be  \label{14092018-man03-28}
V^{(3)} = V^{(3)}(B_a,\alpha_{aa+1},\zeta_2)\,,
\ee
where the realization of operators $G_a$ \rf{14092018-man03-08}-\rf{14092018-man03-10} on the vertex $V^{(3)}$ \rf{14092018-man03-28} takes the form%
\footnote{ For massless field corresponding to $a=1$, $\alpha_{11}$-term in \rf{14092018-man03-04} can be ignored in \rf{14092018-man03-31} by virtue of the second constraint in \rf{01092018-man03-14}.}
\beq
\label{14092018-man03-29} G_1 &  =  &    (B_3 + \half \kappa_3  )\partial_{\alpha_{31}}  - ( B_2 - \half m_2 \zeta_2   ) \partial_{\alpha_{12}} +  \half  (m_2^2 - m_3^2)  \partial_{B_1} \,,
\\
\label{14092018-man03-30} G_2 &  =  &   B_1\partial_{\alpha_{12}}
- ( B_3 - \half \kappa_3 ) \partial_{\alpha_{23}} + \half m_3^2  \partial_{B_2} + m_2\partial_{\zeta_2}\,,
\\
\label{14092018-man03-31} G_3 &  =  &  \frac{2N_3+d-2}{\kappa_3 } \Bigl(  (B_2 + \half m_2 \zeta_2   )\partial_{\alpha_{23}}
-   B_1 \partial_{\alpha_{31}} -  \half m_2^2 \partial_{B_3} \Bigr)
\nonumber\\
& + &  \frac{F_3}{\kappa_3^2} - B_1 \partial_{B_3} \partial_{\alpha_{31}} - (B_2 - \frac{3}{2}m_2\zeta_2) \partial_{B_3} \partial_{\alpha_{23}}
\nonumber\\
&  +  &  2\alpha_{12} \partial_{\alpha_{31}}\partial_{\alpha_{23}} + \alpha_{22}\partial_{\alpha_{23}}^2
+(\frac{1}{4}m_3^2  -\half  m_2^2) \partial_{B_3}^2 .
\eeq

\noindent {\bf Step 2}. We use the transformation
\be \label{14092018-man03-32}
V^{(3)} = U_\zeta V^{(4)}\,,\qquad U_\zeta = \exp\Bigl(-\frac{\zeta_2}{m_2} B_1\partial_{\alpha_{12}} + \frac{\zeta_2}{m_2}( B_3 - \half \kappa_3 ) \partial_{\alpha_{23}} - \frac{m_3^2}{2m_2}\zeta_2 \partial_{B_2} \Bigr)\,.
\ee
Realization of $G_a$ \rf{14092018-man03-29}-\rf{14092018-man03-31} on $V^{(4)}$ \rf{14092018-man03-32} takes the form%
\footnote{ For massive field corresponding to $a=2$, operator $G_3$ \rf{14092018-man03-34} involves, besides $\alpha_{22}$-term \rf{14092018-man03-04}, $\zeta_3^2$-term, i.e., we get $(\alpha_{11}+\zeta_2^2)$-term which can be ignored by virtue of the second constraint in \rf{01092018-man03-13}.
}
\beq
\label{14092018-man03-33} && \hspace{-1.4cm} G_1  =     (B_3 + \half \kappa_3  )\partial_{\alpha_{31}}  -  B_2\partial_{\alpha_{12}} +  \half  (m_2^2 - m_3^2)  \partial_{B_1} \,,\qquad G_2 = m_2\partial_{\zeta_2}\,,
\\
\label{14092018-man03-34} && \hspace{-1.4cm}  G_3  =   - r_3 \partial_{B_3}^2  + q_3(N_{B_3} + \nu_3 +1)\partial_{B_3} + e_3 (N_{B_3} + \nu_3)(  N_{B_3} + \nu_3 + 1 ) + u_3
\nonumber\\
&& \hspace{-1.3cm} +\,\,   \frac{2}{\kappa_3}(N_{B_3} + \nu_3 +1) ( B_2\partial_{\alpha_{23}}
-   B_1 \partial_{\alpha_{31}})  - B_1 \partial_{B_3} \partial_{\alpha_{31}} - B_2 \partial_{B_3} \partial_{\alpha_{23}}  + 2 \alpha_{12} \partial_{\alpha_{31}}\partial_{\alpha_{23}}\,,
\eeq
where, here and below, we use quantities \rf{02092018-man03-07app}-\rf{02092018-man03-09app-x1} taken for masses given in \rf{14092018-man03-27-a1}.
From $G_2$ in \rf{14092018-man03-33} and the equation $G_2V^{(4)}=0$, we see that the $V^{(4)}$ is independent of $\zeta_2$,
\be \label{14092018-man03-35}
V^{(4)}= V^{(4)}(B_a,\alpha_{aa+1})\,.
\ee

\noindent {\bf Step 3}. We use the transformations
\be \label{14092018-man03-35-y1}
V^{(4)} = U_{1\partial\alpha} V^{(5)}\,, \qquad
U_{1\partial\alpha} =  \exp\Bigl(  \frac{2B_1B_2}{m_2^2-m_3^2}\partial_{\alpha_{12}} -\frac{2B_1}{m_2^2-m_3^2}(B_3 + \half \kappa_3  )\partial_{\alpha_{31}}  \Bigr).
\ee
Realization $G_1$, $G_3$ \rf{14092018-man03-33},\rf{14092018-man03-34} on $V^{(5)}$ \rf{14092018-man03-35-y1} takes the form
\beq
\label{14092018-man03-36} G_1 & = &   \half  (m_2^2 - m_3^2)  \partial_{B_1}\,,
\\
\label{14092018-man03-37} G_3 &  =  &  - r_3 \partial_{B_3}^2  + q_3(N_{B_3} + \nu_3 +1)\partial_{B_3} + e_3 (N_{B_3} + \nu_3)(  N_{B_3} + \nu_3 + 1 ) + u_3 \qquad
\nonumber\\
& + &  \frac{2}{\kappa_3}(N_{B_3} + \nu_3 +1)  B_2\partial_{\alpha_{23}} -     B_2 \partial_{\alpha_{23}} \partial_{B_3}  + 2 \alpha_{12} \partial_{\alpha_{31}}\partial_{\alpha_{23}}\,.
\eeq
From \rf{14092018-man03-36} and equation $G_1V^{(4)}=0$, we see that  the $V^{(5)}$ is independent of $B_1$,
\be \label{14092018-man03-38}
V^{(5)}= V^{(5)}(B_2,B_3,\alpha_{aa+1})\,.
\ee

\noindent {\bf Step 4}. We use the transformation
\be
\label{14092018-man03-39} V^{(5)} = U_{\partial B}  U_{23\partial \alpha}  V^{(6)}\,, \quad  U_{\partial B} = \exp\bigl( \frac{q_3}{2e_3}  \partial_{ B_3}  \bigr)\,, \quad
U_{23\partial\alpha} = \exp\Bigl( \bigl(\frac{B_2}{\kappa_3 e_3}    + \frac{2h_{23}}{D}  B_2 B_3\bigr) \partial_{\alpha_{23}}  \Bigr)
\ee
and note the relation
\be
\label{14092018-man03-40}   U_{1\partial\alpha} U_{\partial B} U_{23\partial\alpha}
= U_{\partial B} U_{\partial\alpha}\,, \qquad U_{\partial\alpha} = \exp\Bigl( \frac{1}{\kappa_3 e_3}  B_2 \partial_{ \alpha_{23} } - \frac{1}{\kappa_3 e_3}  B_1 \partial_{ \alpha_{31} }\Bigr)U_D\,,
\ee
where $U_D$ is defined in \rf{02092018-man03-15app}. Realization of $G_3$ \rf{14092018-man03-37} on $ V^{(6)}$ takes the form
\beq \label{14092018-man03-41}
&& G_3  =   - \frac{D}{4\kappa_3^2e_3} \partial_{B_3}^2  + e_3 (N_{B_3} + \nu_3)(  N_{B_3} + \nu_3 + 1 ) + u_3 + W_3\,,
\\
\label{14092018-man03-41-b1} &&  W_3 =     2 \alpha_{12} \partial_{\alpha_{31}} \partial_{\alpha_{23}}  - \frac{4m_2^2}{(m_2^2- m_3^2)^2} B_2^2\partial_{\alpha_{23}}^2  \,.
\eeq

\noindent {\bf Step 5}. We make the transformations
\be \label{14092018-man03-42}
V^{(6)} = U_{B_3} V^{(7)}\,, \qquad V^{(7)} = U_{\nu_3,W_3} V^{(8)}\,, \qquad
V^{(8)} = U_{B_3}^{-1} V^{(9)}\,,
\ee
where $U_{B_3}$, $U_{\nu_3,W_3}$  are given in \rf{02092018-man03-13app},\rf{02092018-man03-14app}. Realizations of $G_3$ \rf{14092018-man03-41} on $V^{(7)}$, $V^{(8)}$, $V^{(9)}$ take the forms
\beq
\label{14092018-man03-43} && \hspace{-1.5cm} G_3  =   (B_3^2 - \frac{D}{4\kappa_3^2e_3^2})  \partial_{B_3}^2    +    (1- \frac{4\kappa_3^2e_3^2}{D} B_3^2)^{-1} \bigl( \nu_3^2 - 1 + W_3 \bigr) + \frac{u_3}{e_3} \,, \hspace{1cm} \hbox{for } V^{(7)}\,,
\\
\label{14092018-man03-44} && \hspace{-1.5cm} G_3  =   (B_3^2 - \frac{D}{4\kappa_3^2e_3^2})  \partial_{B_3}^2  +    (1- \frac{4\kappa_3^2e_3^2}{D} B_3^2)^{-1} \bigl( \nu_3^2 - 1 \bigr)  + \frac{u_3}{e_3}  \,, \hspace{2.1cm} \hbox{for } V^{(8)},
\\
\label{14092018-man03-45} && \hspace{-1.5cm} G_3   =    - \frac{D}{4\kappa_3^2e_3} \partial_{B_3}^2  + e_3 (N_{B_3} + \nu_3)(  N_{B_3} + \nu_3 + 1 ) + u_3 \,, \hspace{2.8cm} \hbox{for } V^{(9)}\,.
\eeq
For $\nu_3$ \rf{02092018-man03-06app-bb1} and $W_3$ \rf{14092018-man03-41-b1}, we note the relation which admits us to get $G_3$ in \rf{14092018-man03-44},
\be
(\nu_3^2+ W_3)U_{\nu_3,W_3}  = U_{\nu_3,W_3} \nu_3^2\,.
\ee
Equation $G_3V^{(9)}=0$ with $G_3$ as in \rf{14092018-man03-45} is the second-order differential equation w.r.t. $B_3$. Two independent solutions of this equation are given in \rf{04092018-man03-05}.
%
%
%
%
%
\be  \label{14092018-man03-46}
\hbox{\bf \large Case \quad $m_1^2 > 0$,\quad $m_2^2 > 0$,\quad $m_3^2 < 0$.}
\ee
%
%
%
%
%
{\bf Step 1}. This case corresponds to the vertex $p_\smp3^-$ \rf{05092018-man03-01-add}.  Vertex $V^{(3)}$ is given in \rf{14092018-man03-13}, where the realization of $G_a$ on $V^{(3)}$ is given in \rf{14092018-man03-08}-\rf{14092018-man03-10}.
We make the transformation
\beq
\label{14092018-man03-47} && \hspace{-1.3cm} V^{(3)} = U_\zeta V^{(4)}\,, \qquad  U_\zeta = \exp\Bigl( - \frac{\zeta_1}{2m_1} (m_2^2 - m_3^2)\partial_{B_1} - \frac{\zeta_2}{2m_2} (m_3^2 - m_1^2)\partial_{B_2}
\nonumber\\
&& \hspace{3cm}  + \,\,  (\frac{\zeta_1}{m_1} B_2 - \frac{\zeta_2}{m_2} B_1 + \frac{m_3^2 - m_1^2 - m_2^2}{2m_1m_2} \zeta_1 \zeta_2 )\partial_{\alpha_{12}}
\nonumber\\
&& \hspace{3cm} - \frac{\zeta_1}{m_1}(B_3 + \half \kappa_3)\partial_{\alpha_{31}} + \frac{\zeta_2}{m_2}(B_3 - \half \kappa_3)\partial_{\alpha_{23}}\Bigr)\,.
\eeq
Realization of $G_a$ \rf{14092018-man03-08}-\rf{14092018-man03-10} on $V^{(4)}$ \rf{14092018-man03-47} takes the form
\beq
\label{14092018-man03-49} && \hspace{-1.4cm} G_1 =m_1\partial_{\zeta_1}\,, \qquad G_2 = m_2\partial_{\zeta_2}\,, \qquad %
\\
\label{14092018-man03-50} && \hspace{-1.4cm} G_3  =  - r_3 \partial_{B_3}^2  + q_3(N_{B_3} + \nu_3 +1)\partial_{B_3} + e_3 (N_{B_3} + \nu_3)(  N_{B_3} + \nu_3 + 1 ) + u_3
\nonumber\\
&& \hspace{-1cm} + \,\,   \frac{2}{\kappa_3}(N_{B_3} + \nu_3 +1) \bigl( B_2\partial_{\alpha_{23}} -   B_1   \partial_{\alpha_{31}}\bigr) -  (B_1 \partial_{\alpha_{31}} +  B_2 \partial_{\alpha_{23}}) \partial_{B_3}  + 2 \alpha_{12} \partial_{\alpha_{31}}\partial_{\alpha_{23}}\,,\quad
\eeq
where, here and below, we use quantities defined in \rf{02092018-man03-07app}-\rf{02092018-man03-09app-x1}.
From \rf{14092018-man03-49} and the equations $G_1 V^{(4)}=0$, $G_2 V^{(4)}=0$, we see that the $V^{(4)}$ is independent of $\zeta_1$, $\zeta_2$,
\be \label{14092018-man03-51}
V^{(4)}= V^{(4)}(B_a,\alpha_{aa+1})\,.
\ee

\noindent {\bf Step 2}. We make the transformation
\be \label{14092018-man03-52}
V^{(4)} = U_{\partial B}  U_{\partial \alpha}  V^{(5)}\,, \qquad  U_{\partial B} = \exp\bigl( \frac{q_3}{2e_3}  \partial_{ B_3}  \bigr)\,,
\quad  U_{\partial\alpha}    =  \exp\Bigl( \frac{B_2 \partial_{ \alpha_{23} } - B_1\partial_{ \alpha_{31} }}{\kappa_3 e_3}  \Bigr) U_D\,,
\ee
where $U_D$ is defined in \rf{02092018-man03-15app}. Realization of $G_3$ \rf{14092018-man03-50} on $V^{(5)}$ \rf{14092018-man03-52} takes the form
\beq
\label{14092018-man03-54} && G_3   =   - \frac{D}{4\kappa_3^2e_3} \partial_{B_3}^2  + e_3 (N_{B_3} + \nu_3)(  N_{B_3} + \nu_3 + 1 ) + u_3 +W_3\,,
\nonumber\\
\label{14092018-man03-55} && W_3 =   -  \frac{4m_1^2}{D} B_1^2\partial_{\alpha_{31}}^2  - \frac{4m_2^2}{D} B_2^2\partial_{\alpha_{23}}^2  + 2 \alpha_{12} \partial_{\alpha_{31}}\partial_{\alpha_{23}}\,.
\eeq

\noindent {\bf Step 3}. We make the transformations
\be \label{14092018-man03-56}
V^{(4)} = U_{B_3} V^{(5)}\,, \qquad V^{(5)} = U_{\nu_3,W_3} V^{(6)}\,, \qquad
V^{(6)} = U_{B_3}^{-1} V^{(7)}\,,
\ee
where $U_{B_3}$, $U_{\nu_3,W_3}$  are given in \rf{02092018-man03-13app},\rf{02092018-man03-14app}. Realization of $G_3$ \rf{14092018-man03-54} on $V^{(7)}$ takes the form
\beq
\label{14092018-man03-57} && \hspace{-1.5cm} G_3   =    - \frac{D}{4\kappa_3^2e_3} \partial_{B_3}^2  + e_3 (N_{B_3} + \nu_3)(  N_{B_3} + \nu_3 + 1 ) + u_3 \,.
\eeq
For $\nu_3$ \rf{02092018-man03-06app-bb1} and $W_3$ \rf{14092018-man03-55}, we note the helpful relation,
\be
(\nu_3^2+ W_3)U_{\nu_3,W_3}  = U_{\nu_3,W_3} \nu_3^2\,.
\ee
Equation $G_3V^{(9)}=0$ with $G_3$ as in \rf{14092018-man03-57} is the second-order differential equation w.r.t. $B_3$. Two independent solutions of this equation are given in \rf{05092018-man03-05}.

\appendix{ \large Derivation of $p_\smp3^-$ \rf{06092018-man03-02}, \rf{07092018-man03-02}, \rf{08092018-man03-02}, \rf{08092018-man03-02-q}, \rf{09092018-man03-02} }

We consider five vertices in \rf{06092018-man03-01-add},\rf{09092018-man03-01-add}. For these vertices, we outline the derivation of the respective solutions in \rf{06092018-man03-02}, \rf{07092018-man03-02},\rf{08092018-man03-02},\rf{08092018-man03-02-q},\rf{09092018-man03-02}. We split our derivation in several steps.

\noindent {\bf Realization of $G_a$, $G_\beta$ on $p_\smp3^-$ \rf{09092018-man03-01-add} for arbitrary masses}. First, for $p_\smp3^-$  \rf{09092018-man03-01-add}, we find $G_\beta$ and $G_{a,\Po^2}$ in \rf{01092018-man03-52}. We use $M_3^i$ for massive field \rf{01092018-man03-25} and $M_a^i$ $a=1,2$ for continuous-spin field \rf{01092018-man03-22}. Plugging such $M_a^i$ into $\Jbf^{-i\dagger}$ \rf{01092018-man03-47}, we cast $\Jbf^{-i\dagger}|p_\smp3^-\rangle$ into the form given in \rf{01092018-man03-52} with the following $G_\beta$ and $G_{a,\Po^2}$:
\beq
\label{15092018-man03-01} &&   \hspace{-1cm} G_{a,\Po^2} =   G_a +  \Pbf^- \frac{2 \beta}{\beta_a^3} \alpha_a^i \frac{ g_{\upsilon_a} \partial_{\upsilon_a} }{2N_a + d-2 } \partial_{B_a}^2\,,\qquad a=1,2;
\qquad G_{3,\Po^2} =   G_3\,,
\\
\label{15092018-man03-02}  && \hspace{-1cm}   G_1  =   (B_3 - \frac{\beta_1}{\beta_3}  m_3 \zeta_3 ) \partial_{\alpha_{31}}  - (B_2  \partial_{\alpha_{12}} + \frac{\beta_1}{\beta_2} g_{\upsilon_2} \partial_{\upsilon_2}) \partial_{\alpha_{12}}  + \half \Bigl( \frac{\betach_1}{ \beta_1} m_1^2 +  m_2^2 - m_3^2  \Bigr) \partial_{B_1} + \upsilon_1 g_{\upsilon_1}
\nonumber\\
&& \hspace{-1cm} +\,\,     \frac{ g_{\upsilon_1} \partial_{\upsilon_1}}{2N_1+d-2} \Bigl(\frac{ 2 \beta_2}{\beta_1} B_2 \partial_{B_1} \partial_{\alpha_{12}} + \frac{ 2 \beta_3}{\beta_1} B_3 \partial_{B_1} \partial_{\alpha_{31}}  +   2\alpha_{23} \partial_{\alpha_{12}}\partial_{\alpha_{31}}
\nonumber\\
&&  \hspace{-1cm} + \,\, \alpha_{33} \partial_{\alpha_{31}}^2 + \frac{\beta}{\beta_1^2} \sum_{b=1}^3 \frac{m_b^2}{\beta_b} \partial_{B_1}^2 \Bigr) \,,
\\
\label{15092018-man03-03} && \hspace{-1cm} G_2   =     ( B_1 - \frac{\beta_2}{\beta_1} g_{\upsilon_1} \partial_{\upsilon_1} ) \partial_{\alpha_{12}} -  ( B_3 + \frac{\beta_2}{\beta_3}  m_3 \zeta_3 ) \partial_{\alpha_{23}}  + \half \Bigl( \frac{\betach_2}{ \beta_2} m_2^2 +  m_3^2 - m_1^2  \Bigr)\partial_{B_2} + \upsilon_2 g_{\upsilon_2}
\nonumber\\
&& \hspace{-1cm}  + \,\,       \frac{g_{\upsilon_2} \partial_{\upsilon_2}}{2N_2+d-2} \Bigl(  \frac{ 2 \beta_3}{\beta_2} B_3 \partial_{B_2} \partial_{\alpha_{23}}  + \frac{ 2 \beta_1}{\beta_2} B_1 \partial_{B_2} \partial_{\alpha_{12}}   +  2 \alpha_{31} \partial_{\alpha_{12}}\partial_{\alpha_{23}}
\nonumber\\
&&  \hspace{-1cm} + \,\, \alpha_{33} \partial_{\alpha_{23}}^2
+ \frac{\beta}{\beta_2^2} \sum_{b=1}^3 \frac{m_b^2}{\beta_b} \partial_{B_2}^2 \Bigr)  \,,
\\
\label{15092018-man03-04} && \hspace{-1cm} G_3   =   ( B_2 - \frac{\beta_3}{\beta_2} g_{\upsilon_2} \partial_{\upsilon_2} ) \partial_{\alpha_{23}} -  ( B_1 + \frac{\beta_3}{\beta_1} g_{\upsilon_1} \partial_{\upsilon_1} ) \partial_{\alpha_{31}} +  \half \Bigl( \frac{\betach_3}{ \beta_3} m_3^2 +  m_1^2 - m_2^2  \Bigr) \partial_{B_3} + m_3 \partial_{\zeta_3}\,,\quad
\\
\label{15092018-man03-05} && \hspace{-1cm} G_\beta  =    - \frac{1}{\beta}\No - \frac{1}{\beta_1^2} g_{\upsilon_1} \partial_{\upsilon_1} \partial_{B_1} - \frac{1}{\beta_2^2} g_{\upsilon_2} \partial_{\upsilon_2} \partial_{B_2} - \frac{1}{\beta_3^2}  m_3 \zeta_3 \partial_{B_3} \,,
\eeq
where $g_{v_a}$ are given in \rf{02092018-man03-06app-bb2}.  Using $G_a$, $G_\beta$ \rf{15092018-man03-02}-\rf{15092018-man03-05}, we now consider equations \rf{01092018-man03-53},\rf{01092018-man03-54}.

\noindent {\bf Vertex $V^{(3)}$}. We multiply $G_a$ \rf{15092018-man03-02},\rf{15092018-man03-03}, on the left by $(2N_a+d-2)/\kappa_a$, $a=1,2$ and use the transformations%
\footnote{ For detailed discussion of transformations governed by operators $U_{\upsilon_1}U_{\upsilon_2}$, $U_{\Gamma_1}U_{\Gamma_2}$, $U_\beta$ \rf{15092018-man03-06} see {\bf Steps 2,3,4} in Appendix D in Ref.\cite{Metsaev:2017cuz}.
}
\beq
\label{15092018-man03-06} && \hspace{-1cm} p_\smp3^-  =   U_{\upsilon_1}U_{\upsilon_2} V^{(1)}\,, \qquad V^{(1)} = U_{\Gamma_1}U_{\Gamma_2} V^{(2)}\,,  \qquad V^{(2)}   = U_\beta  V^{(3)}\,,
\\
\label{15092018-man03-07} && U_\beta   = \exp\Bigl( - \frac{\betach_1}{2\beta_1} \kappa_1 \partial_{B_1} - \frac{\betach_2}{2\beta_2} \kappa_2 \partial_{B_2} - \frac{\betach_3}{2\beta_3} m_3\zeta_3 \partial_{B_3} \Bigr)\,,
\eeq
where $U_{\upsilon_a}$, $U_{\Gamma_a}$ are defined in \rf{02092018-man03-10app},\rf{02092018-man03-12app}.
Realization of $G_a$, $G_\beta$ \rf{15092018-man03-02}-\rf{15092018-man03-05} on $V^{(3)}$ takes the form
\beq
\label{15092018-man03-08} \hspace{-0.7cm} G_1 &  =  & \frac{2N_1 + d-2}{ \kappa_1}\Bigl( (B_3 + \half  m_3 \zeta_3 ) \partial_{\alpha_{31}}  - (B_2   - \half \kappa_2) \partial_{\alpha_{12}} + \half \bigl( m_2^2 - m_3^2  \bigr)\partial_{B_1}\Bigr)
\nonumber\\
\hspace{-0.7cm} & + &      \frac{F_1}{\kappa_1^2} - ( B_2 + \frac{3}{2}\kappa_2) \partial_{B_1} \partial_{\alpha_{12}} - (B_3 - \frac{3}{2}m_3\zeta_3) \partial_{B_1} \partial_{\alpha_{31}}  +   2\alpha_{23} \partial_{\alpha_{12}}\partial_{\alpha_{31}}
\nonumber\\
& + & \alpha_{33} \partial_{\alpha_{31}}^2 + \frac{1}{4}(m_1^2 - 2 m_2^2 - 2 m_3^2) \partial_{B_1}^2  \,,
\\
\label{15092018-man03-09} \hspace{-0.7cm} G_2 &  =  & \frac{\kappa_2}{2N_2 +d-2} \Bigl(  ( B_1 + \half \kappa_1 ) \partial_{\alpha_{12}} -  ( B_3 - \half  m_3 \zeta_3 ) \partial_{\alpha_{23}} + \half \bigl( m_3^2 - m_1^2  \bigr) \partial_{B_2}\Bigr)
\nonumber\\
\hspace{-0.7cm} & + &         \frac{ F_2}{\kappa_2^2} - ( B_3 + \frac{3}{2} m_3 \zeta_3) \partial_{B_2} \partial_{\alpha_{23}}  - ( B_1  - \frac{3}{2} \kappa_1) \partial_{B_2} \partial_{\alpha_{12}}   +  2 \alpha_{31} \partial_{\alpha_{12}}\partial_{\alpha_{23}}
\nonumber\\
& + & \alpha_{33} \partial_{\alpha_{23}}^2 + \frac{1}{4}(m_2^2 - 2 m_3^2 - 2 m_1^2) \partial_{B_2}^2    \,,
\\
\label{15092018-man03-10} \hspace{-0.7cm} G_3 &  =  & ( B_2 + \half \kappa_2) \partial_{\alpha_{23}} -  ( B_1 - \half \kappa_1  ) \partial_{\alpha_{31}} + \half \bigl( m_1^2 - m_2^2  \bigr)\partial_{B_3} + m_3 \partial_{\zeta_3}\,,
\\
\label{15092018-man03-11} \hspace{-0.7cm} G_\beta &  =  &   - \frac{1}{\beta}\No  \,,
\eeq
where $F_a$ are given in \rf{02092018-man03-06app-bb2}.

\noindent  {\bf Dependence of $V^{(3)}$ on $\beta_a$}. We fix dependence of $V^{(3)}$ on $\beta_a$. To this end we use Eqs.\rf{01092018-man03-54},\rf{01092018-man03-58} and relation \rf{15092018-man03-11} to note the following equations for the vertex $V^{(3)}$:
\be \label{15092018-man03-12}
\No_\beta V^{(3)} = 0 \,, \qquad \sum_{a=1,2,3}\beta_a\partial_{\beta_a} V^{(3)} = 0\,.
\ee
Equations \rf{15092018-man03-12} imply that $V^{(3)}$ is independent of $\beta_1$, $\beta_2$, $\beta_3$. Thus, the vertex $V^{(3)}$ is given by
\be \label{15092018-man03-13}
V^{(3)} = V^{(3)}(B_a,\alpha_{aa+1},\zeta_3)\,.
\ee
We now use results above-obtained for the study of five vertices in \rf{06092018-man03-01-add},\rf{09092018-man03-01-add} in turn.
%
%
%
%
%
\be  \label{15092018-man03-14}
\hbox{\bf \large Case \quad $m_1^2 < 0$, \quad $m_2 = 0$, \quad $m_3=0$.}
\ee
%
%
%
%
%
{\bf Step 1}. This case corresponds to the first vertex $p_\smp3^-$ \rf{06092018-man03-01-add}. In \rf{15092018-man03-06}, we use operator $U_{\Gamma_2}$ given in \rf{02092018-man03-11app}. Then, in all remaining relations in \rf{15092018-man03-01}-\rf{15092018-man03-13}, we set $m_2=0$, $m_3=0$ and, ignoring $\zeta_3$ in \rf{15092018-man03-01}-\rf{15092018-man03-13}, we  proceed with vertex $V^{(3)}$ given by
\be \label{15092018-man03-15}
V^{(3)} = V^{(3)}(B_a,\alpha_{aa+1})\,,
\ee
where the realization of operators $G_a$ \rf{15092018-man03-08}-\rf{15092018-man03-10} on the vertex $V^{(3)}$ \rf{15092018-man03-15} takes the form%
\footnote{ For massless field corresponding to $a=3$, $\alpha_{33}$-terms in \rf{15092018-man03-02},\rf{15092018-man03-03} can be ignored in \rf{15092018-man03-16},\rf{15092018-man03-17} by virtue of the second constraint in \rf{01092018-man03-14}.}
\beq
\label{15092018-man03-16} \hspace{-0.7cm} G_1 &  =  & \frac{2N_1 + d-2}{ \kappa_1} \Bigl(  B_3  \partial_{\alpha_{31}}  - (B_2   - \half \kappa_2) \partial_{\alpha_{12}}\Bigr) + \frac{F_1}{\kappa_1^2} - ( B_2 + \frac{3}{2}\kappa_2) \partial_{B_1} \partial_{\alpha_{12}}
\nonumber\\
\hspace{-0.7cm} & - &   B_3   \partial_{B_1} \partial_{\alpha_{31}}  + 2\alpha_{23} \partial_{\alpha_{12}}\partial_{\alpha_{31}}
  +  \alpha_{33} \partial_{\alpha_{31}}^2 + \frac{1}{4} m_1^2\partial_{B_1}^2  \,,
\\
\label{15092018-man03-17} \hspace{-0.7cm} G_2 &  =  & \frac{2N_2 +d-2}{\kappa_2} \Bigl(  ( B_1 + \half \kappa_1 ) \partial_{\alpha_{12}} - B_3 \partial_{\alpha_{23}} - \half  m_1^2 \partial_{B_2}\Bigr)
 +    \frac{ F_2}{\kappa_2^2} -  B_3 \partial_{B_2} \partial_{\alpha_{23}}  \qquad
\nonumber\\
\hspace{-0.7cm} & - & ( B_1  - \frac{3}{2} \kappa_1) \partial_{B_2} \partial_{\alpha_{12}}   +  2 \alpha_{31} \partial_{\alpha_{12}}\partial_{\alpha_{23}}
+ \alpha_{33} \partial_{\alpha_{23}}^2 -\half  m_1^2 \partial_{B_2}^2   \,,
\\
\label{15092018-man03-18} \hspace{-0.7cm} G_3 &  =  & ( B_2 + \half \kappa_2) \partial_{\alpha_{23}} -  ( B_1 - \half \kappa_1  ) \partial_{\alpha_{31}} + \half  m_1^2 \partial_{B_3}\,.
\eeq

\noindent {\bf Step 2}. We make the transformation
\be
\label{15092018-man03-19} V^{(3)} = U_{1\partial \alpha} V^{(4)}\,,
\qquad U_{1\partial \alpha} = \exp\Bigl( \frac{2B_3}{m_1^2} ( B_1 - \half \kappa_1  ) \partial_{\alpha_{31}} - \frac{2B_3}{m_1^2}( B_2 + \half \kappa_2) \partial_{\alpha_{23}}\Bigr)\,.
\ee
Realization of $G_a$ \rf{15092018-man03-16}-\rf{15092018-man03-18} on $V^{(4)}$ takes the form
\beq
\label{15092018-man03-20} \hspace{-0.7cm} G_1 &  =  &   - r_1 \partial_{B_1}^2  +  e_1 (N_{B_1} + \nu_1)(  N_{B_1} + \nu_1 + 1 )  + u_1
\nonumber\\
\hspace{-1cm} & - & \frac{2}{\kappa_1} (N_{B_1}+\nu_1 + 1) (B_2   - \half \kappa_2) \partial_{\alpha_{12}} -  ( B_2 + \frac{3}{2}\kappa_2) \partial_{B_1} \partial_{\alpha_{12}}   +   2\alpha_{23} \partial_{\alpha_{12}}\partial_{\alpha_{31}},
\\
\label{15092018-man03-21} \hspace{-0.7cm} G_2 &  =  &   - r_2 \partial_{B_2}^2  + q_2 \bigl(N_{B_2}  + \nu_2 +1 \bigr)  \partial_{B_2}  + u_2
\nonumber\\
\hspace{-1cm} & + & \frac{2}{\kappa_2} (N_{B_2}+\nu_2 + 1) (B_1   + \half \kappa_1) \partial_{\alpha_{12}} -  ( B_1 - \frac{3}{2}\kappa_1) \partial_{B_2} \partial_{\alpha_{12}}   +   2\alpha_{31} \partial_{\alpha_{12}}\partial_{\alpha_{23}},
\\
\label{15092018-man03-22} \hspace{-1cm}  G_3  & =  & \half  m_1^2 \partial_{B_3}\,,
\eeq
where, here and below, we use quantities \rf{02092018-man03-07app}-\rf{02092018-man03-09app-x1} taken for masses given in \rf{15092018-man03-14}.
From \rf{15092018-man03-22} and equation $G_3V^{(4)}=0$, we see that  $V^{(4)}$ is independent of $B_3$,
\be \label{15092018-man03-23}
V^{(4)} = V^{(4)}(B_1,B_2,\alpha_{aa+1})\,.
\ee

\noindent {\bf Step 3}. We make the transformation
\beq
\label{15092018-man03-24} &&  \hspace{-1.5cm} V^{(4)} = U_{\partial B} U_{2\partial \alpha} U_{3\partial \alpha}
V^{(5)}\,, \qquad  U_{\partial B}  = \exp\bigl( \frac{\kappa_2}{2}\partial_{B_2} \bigr)
\nonumber\\
\label{15092018-man03-25} &&   \hspace{-1.5cm} U_{2\partial \alpha} = \exp\Bigl( \frac{\kappa_2}{\kappa_1 e_1} \partial_{\alpha_{12}}  - \frac{B_2 }{\kappa_1 e_1} \partial_{\alpha_{12}}  -   \frac{2B_1B_2 }{\kappa_1^2 e_1}\partial_{\alpha_{12}}\Bigr)\,, \qquad U_{3\partial \alpha} = \exp\bigl(- \frac{2\kappa_2}{\kappa_1^2 e_1} B_1 \partial_{\alpha_{12}} \bigr).\qquad
\eeq
Realization of $G_1$, $G_2$ \rf{15092018-man03-20}, \rf{15092018-man03-21} on $V^{(5)}$ \rf{15092018-man03-24} takes the form
\beq
\label{15092018-man03-26} \hspace{-0.7cm} G_1 &  =  &   - \frac{D}{4\kappa_1^2 e_1} \partial_{B_1}^2   +  e_1 (N_{B_1} + \nu_1)(  N_{B_1} + \nu_1 + 1 )  + u_1 +   \frac{4 }{q_2}B_2 \partial_{\alpha_{12}}^2   +   2\alpha_{23} \partial_{\alpha_{12}} \partial_{\alpha_{31}}, \qquad
\nonumber\\
\label{15092018-man03-27} \hspace{-0.7cm} G_2 &  =  &    q_2 \bigl(N_{B_2}  + \nu_2 +1 \bigr)  \partial_{B_2}  + u_2 + \bigl( \frac{4\kappa_1^2 e_1}{D} B_1^2 - \frac{1}{e_1} \bigr) \partial_{\alpha_{12}}^2   +   2\alpha_{31} \partial_{\alpha_{12}}\partial_{\alpha_{23}}\,.
\eeq
Also, using $U_{\partial \alpha}$ given in \rf{06092018-man03-11}, we note  the relation
\be
U_{1\partial\alpha}  U_{\partial B}  U_{2\partial\alpha} U_{3\partial\alpha}  =  U_{\partial B}U_{\partial \alpha}\,.
\ee

\noindent {\bf Step 4}. We make the transformations
\be
\label{15092018-man03-28} V^{(5)}= U_{B_1} \bigl(-\frac{4}{q_2}B_2\bigr)^{-\nu_2/2} V^{(6)}\,, \qquad V^{(6)}= U_{\nu_1,W_1} U_{\nu_2,W_{23}} V^{(7)}\,,
\qquad V^{(7)}= U_{B_1}^{-1} V^{(8)}\,, \qquad
\ee
where $U_{B_1}$, $U_{\nu,W}$ are given in \rf{02092018-man03-13app}, \rf{02092018-man03-14app} and we use the notation
\beq
\label{15092018-man03-29} && W_1  =  2\alpha_{23} \partial_{\alpha_{12}} \partial_{\alpha_{31}} - \partial_{\alpha_{12}}^2 \,,
\nonumber\\
&& W_2  =    2\alpha_{31}  \partial_{\alpha_{12}} \partial_{\alpha_{23}} - \partial_{\alpha_{12}}^2    \,,
\hspace{2cm} W_{23}  =  2\alpha_{31}  \partial_{\alpha_{12}} \partial_{\alpha_{23}}\,.
\eeq
Realizations of $G_1$, $G_2$ \rf{15092018-man03-26} on $V^{(6)}$,  $V^{(7)}$,  $V^{(8)}$ \rf{15092018-man03-28} take  the forms
\beq
\label{15092018-man03-32} && \hspace{-2cm}  G_1  = (B_1^2 - \frac{D}{4\kappa_1^2e_1^2})  \partial_{B_1}^2    +    (1- \frac{4\kappa_1^2e_1^2}{D} B_1^2)^{-1} \bigl( \nu_1^2 - 1 + W_1 \bigr) + \frac{u_1}{e_1} \,,
\nonumber\\
\label{15092018-man03-33} && \hspace{-2cm}  G_2  =  u_2 +  q_2\bigl(N_{B_2}  + 1 \bigr)\partial_{B_2}    - \frac{ q_2 }{4B_2}(\nu_2^2 + W_2) \,, \hspace{4cm} \hbox{ for } V^{(6)}\,;
\\
\label{15092018-man03-34} && \hspace{-2cm}  G_1  = (B_1^2 - \frac{D}{4\kappa_1^2e_1^2})  \partial_{B_1}^2   +  \bigl( 1- \frac{4\kappa_1^2e_1^2}{D} B_1^2 \bigr)^{-1} \bigl( \nu_1^2 - 1\bigr) + \frac{u_1}{e_1}  \,,
\nonumber\\
\label{15092018-man03-35} && \hspace{-2cm}  G_2  =  u_2 +  q_2\bigl(N_{B_2}  + 1 \bigr)\partial_{B_2}    - \frac{ q_2 }{4B_2} \nu_2^2 \,,  \hspace{5.5cm} \hbox{ for } V^{(7)}\,;
\\
\label{15092018-man03-36} && \hspace{-2cm} G_1  =   - \frac{D}{4\kappa_1^2 e_1} \partial_{B_1}^2  +  e_1 (N_{B_1} + \nu_1)(  N_{B_1} + \nu_1 + 1 )  + u_1\,,
\nonumber\\
\label{15092018-man03-37} && \hspace{-2cm} G_2  =  u_2 +  q_2\bigl(N_{B_2}  + 1 \bigr)\partial_{B_2}    - \frac{ q_2 }{4B_2}\nu_2^2 \,,  \hspace{5.5cm} \hbox{ for } V^{(8)}\,.
\eeq
For $\nu_a$ \rf{02092018-man03-06app-bb1} and $W$-operators \rf{15092018-man03-29}, we note the relations which admit us to get $G_1$, $G_2$ in \rf{15092018-man03-34},
\beq
&& (\nu_1^2+ W_1)U_{\nu_1,W_1}  = U_{\nu_1,W_1} \nu_1^2\,, \hspace{2.5cm} \nu_1^2 U_{\nu_2,W_{23}}  = U_{\nu_2,W_{23}} \nu_1^2\,,
\nonumber\\
&& (\nu_2^2+ W_2)U_{\nu_1,W_1}  = U_{\nu_1,W_1} (\nu_2^2 + W_{23})\,, \hspace{1cm} (\nu_2^2+ W_{23})U_{\nu_2,W_{23}}  = U_{\nu_2,W_{23}} \nu_2^2\,. \qquad
\eeq
Equations $G_1V^{(8)}=0$ and $G_2V^{(8)}=0$ with $G_1$ and $G_2$ as in \rf{15092018-man03-36} constitute a system of two decoupled second-order differential equations w.r.t. $B_1$ and $B_2$. Four independent solutions of these equations are given in \rf{06092018-man03-05}.
%
%
%
%
%
\be  \label{15092018-man03-38}
\hbox{\bf \large Case \quad $m_1 = m$,\quad $m_2 =m$,\quad $m_3=0$,\quad $m^2<  0$.}
\ee
%
%
%
%
%
{\bf Step 1}. This case corresponds to the second vertex $p_\smp3^-$ \rf{06092018-man03-01-add}. In all relations in \rf{15092018-man03-01}-\rf{15092018-man03-13}, we set $m_1=m$, $m_2=m$, $m_3=0$ and, ignoring $\zeta_3$ in \rf{15092018-man03-01}-\rf{15092018-man03-13}, we  use  vertex $V^{(3)}$ given by
\be \label{15092018-man03-39}
V^{(3)} = V^{(3)}(B_a,\alpha_{aa+1})\,,
\ee
where the realization of operators $G_a$ \rf{15092018-man03-08}-\rf{15092018-man03-10} on the vertex $V^{(3)}$ \rf{15092018-man03-39} takes the form%
\footnote{ For massless field corresponding to $a=3$, $\alpha_{33}$-terms in \rf{15092018-man03-02},\rf{15092018-man03-03} can be ignored in \rf{15092018-man03-40},\rf{15092018-man03-41} by virtue of the second constraint in \rf{01092018-man03-14}.}
\beq
\label{15092018-man03-40} && \hspace{-2cm}  G_1  =  \frac{2N_1 + d-2}{ \kappa_1} \Bigl( B_3  \partial_{\alpha_{31}}  - (B_2   - \half \kappa_2) \partial_{\alpha_{12}} + \half  m^2  \partial_{B_1}\Bigr)
\nonumber\\
&& \hspace{-1cm}  + \,\,      \frac{F_1}{\kappa_1^2} - ( B_2 + \frac{3}{2}\kappa_2) \partial_{B_1} \partial_{\alpha_{12}} -  B_3   \partial_{B_1} \partial_{\alpha_{31}}  +   2\alpha_{23} \partial_{\alpha_{12}}\partial_{\alpha_{31}}
- \frac{1}{4} m^2 \partial_{B_1}^2  \,,\qquad
\\
\label{15092018-man03-41} && \hspace{-2cm}  G_2  =  \frac{2N_2 +d-2}{\kappa_2} \Bigl(  ( B_1 + \half \kappa_1 ) \partial_{\alpha_{12}} - B_3 \partial_{\alpha_{23}} - \half  m^2 \partial_{B_2} \Bigr)
\nonumber\\
&& \hspace{-1cm}  + \,\,  \frac{ F_2}{\kappa_2^2} -  B_3 \partial_{B_2} \partial_{\alpha_{23}}  - ( B_1  - \frac{3}{2} \kappa_1) \partial_{B_2} \partial_{\alpha_{12}}   +  2 \alpha_{31} \partial_{\alpha_{12}}\partial_{\alpha_{23}}
- \frac{1}{4}m^2 \partial_{B_2}^2  \,,\qquad
\\
\label{15092018-man03-42} && \hspace{-2cm}  G_3  =  ( B_2 + \half \kappa_2) \partial_{\alpha_{23}} -  ( B_1 - \half \kappa_1  ) \partial_{\alpha_{31}}\,.
\eeq

\noindent {\bf Step 2}. From \rf{15092018-man03-42} and equation $G_3V^{(4)}=0$, we find
\be \label{15092018-man03-43}
V^{(3)} =   V^{(4)}(B_a,\alpha_{12}\,, Z)\,, \qquad Z =   ( B_1 - \half \kappa_1  )\alpha_{23} + ( B_2 + \half \kappa_2) \alpha_{31} \,.
\ee
Realization of $G_1$, $G_2$ \rf{15092018-man03-40},\rf{15092018-man03-41} on $V^{(4)}$ \rf{15092018-man03-43} takes the form
\beq
\label{15092018-man03-44} \hspace{-1.2cm} G_1 &  =  &- r_1 \partial_{B_1}^2  + q_1 (N_{B_1} + \nu_Z^\vph+1) \partial_{B_1} +  e_1 (N_{B_1} + \nu_Z)(  N_{B_1} + \nu_Z^\vph+ 1 ) + u_1
\nonumber\\
\hspace{-1cm} &  &     + \frac{2}{\kappa_1} (N_{B_1} + \nu_Z^\vph+ 1) \Bigl( B_3 (B_2   + \half \kappa_2)\partial_Z- (B_2   - \half \kappa_2) \partial_{\alpha_{12}}  \Bigr)
\nonumber\\
&& - ( B_2 + \frac{3}{2}\kappa_2) \partial_{B_1} \partial_{\alpha_{12}}  - B_3 (B_2 + \half \kappa_2) \partial_{B_1} \partial_Z \,,
\nonumber\\
\label{15092018-man03-45} \hspace{-1.2cm} G_2 &  =  &- r_2 \partial_{B_2}^2  + q_2 (N_{B_2} + \nu_Z^\vph+1) \partial_{B_2} +  e_2 (N_{B_2} + \nu_Z)(  N_{B_2} + \nu_Z^\vph+ 1 ) + u_2
\nonumber\\
\hspace{-1cm} &  &     + \frac{2}{\kappa_2} (N_{B_2} + \nu_Z^\vph + 1) \Bigl( (B_1   + \half \kappa_1) \partial_{\alpha_{12}} -  B_3 (B_1  - \half \kappa_1)\partial_Z \Bigr)
\nonumber\\
&& - ( B_1 - \frac{3}{2}\kappa_1) \partial_{B_2} \partial_{\alpha_{12}}  - B_3(B_1 - \half\kappa_1)\partial_{B_2}\partial_Z  \,,
\\
&& \nu_Z^\vph = N_{ \alpha_{12} } + N_Z + \frac{d-4}{2}\,,\qquad N_Z = Z\partial_Z\,,
\eeq
where, here and below, we use quantities \rf{02092018-man03-07app}-\rf{02092018-man03-09app-x1} taken for masses given in \rf{15092018-man03-38}.

\noindent {\bf Step 3}. We make the transformation
\beq
\label{15092018-man03-46} && \hspace{-2cm} V^{(4)} = U_{ \partial B }U_{ \partial \alpha } V^{(5)}\,, \qquad  U_{ \partial B } = \exp( \frac{q_1}{2e_1}\partial_{B_1}  + \frac{q_2}{2e_2}\partial_{B_2} \bigr)\,,
\nonumber\\
\label{15092018-man03-47} && U_{\partial\alpha}  = \exp\Bigl( - \frac{\kappa_1\kappa_2}{m^2} \partial_{\alpha_{12}} - \frac{\kappa_2}{m^2} B_1 \partial_{\alpha_{12}}  + \frac{\kappa_1}{m^2} B_2 \partial_{\alpha_{12}}  + \frac{1}{2m^2} B_1B_2 \partial_{\alpha_{12}}
\nonumber\\
&&\hspace{2cm}  -\,\,  \frac{\kappa_1}{m^2} B_2 B_3 \partial_Z + \frac{\kappa_2}{m^2} B_3B_1 \partial_Z - \frac{1}{m^2} B_1 B_2 B_3\partial_Z \Bigr).
\eeq
Realization of $G_1$, $G_2$ \rf{15092018-man03-44} on $V^{(5)}$ \rf{15092018-man03-46} takes the form
\beq
\label{15092018-man03-48} \hspace{-1.5cm} G_1 &  =  &  e_1 (N_{B_1} + \nu_Z^\vph)(  N_{B_1} + \nu_Z^\vph + 1 ) + u_1 - \frac{1}{e_2}\partial_{\alpha_{12}}^2   - 2 B_2 \partial_{B_1}  \partial_{\alpha_{12}}   + \frac{1}{m^2}  B_2^2 B_3^2\partial_Z^2 \,,
\nonumber\\
\label{15092018-man03-49} \hspace{-1.5cm} G_2 &  =  &  e_2 (N_{B_2} + \nu_Z^\vph)(  N_{B_2} + \nu_Z^\vph + 1 ) + u_2  -  \frac{1}{e_1}\partial_{\alpha_{12}}^2   - 2 B_1 \partial_{B_2}  \partial_{\alpha_{12}}   + \frac{1}{m^2}  B_1^2 B_3^2\partial_Z^2 \,.
\eeq

\noindent {\bf Step 4}. We make the transformation
\beq
\label{15092018-man03-51} && \hspace{-1cm} V^{(5)} = U_e V^{(6)} \,, \qquad U_e = e_1^{\omega_1/2} e_2^{ \omega_2 /2}
\\
\label{15092018-man03-52} && \hspace{-1cm} \omega_1^\vph = N_{B_1}  + \nu_Z + \half\,, \qquad \omega_2^\vph = N_{B_1}  + \nu_Z + \half\,,
\hspace{1cm} \nu_Z = N_{ \alpha_{12} } + N_Z + \frac{d-4}{2}\,,\qquad
\eeq
where $N_Z=Z\partial_Z$, while $e_a$ are given in \rf{02092018-man03-08app}. Realization of $G_1$, $G_2$ \rf{15092018-man03-48} on $V^{(6)}$ \rf{15092018-man03-51} takes the form
\beq
\label{15092018-man03-54} && \hspace{-1.2cm} G_a    =       (N_{B_a} + \nu_Z^\vph)(  N_{B_a} + \nu_Z^\vph + 1 ) + \frac{u_a}{e_a} + W_a\,,\qquad a=1,2\,,
\\
&& \hspace{-0.5cm} W_1   =    \frac{1}{m^2}  B_2^2 B_3^2\partial_Z^2 - \partial_{\alpha_{12}}^2   - 2 B_2\partial_{\alpha_{12}} \partial_{B_1} \,,
\nonumber\\
\label{15092018-man03-56} && \hspace{-0.5cm} W_2   =  \frac{1}{m^2}  B_1^2 B_3^2\partial_Z^2  - \partial_{\alpha_{12}}^2   - 2 B_1\partial_{\alpha_{12}} \partial_{B_2}\,,
\qquad  W_{2 1}  =   \frac{1}{m^2}  B_1^2 B_3^2\partial_Z^2    - 2 B_1\partial_{\alpha_{12}} \partial_{B_2}\,.\qquad
\eeq
\noindent {\bf Step 5}. We make the transformations
\be \label{15092018-man03-58}
V^{(6)} = U_{\omega_1, W_1} U_{\omega_2,W_{21}} V^{(7)}\,,
\ee
where we use notation in \rf{15092018-man03-52},\rf{15092018-man03-56},\rf{02092018-man03-14app}. Realization of $G_a$ \rf{15092018-man03-54} on $V^{(7)}$ \rf{15092018-man03-58} takes the form
\be \label{15092018-man03-59}
G_a = (N_{B_a} + \nu_Z)(N_{B_a} + \nu_Z + 1 ) + \frac{u_a}{e_a}\,, \qquad a=1,2\,.
\ee
For $\omega_a$ \rf{15092018-man03-52} and $W$-operators \rf{15092018-man03-56}, we note the relations which admit us to get $G_a$ in \rf{15092018-man03-59},
\beq
&& (\omega_1^2+ W_1)U_{\omega_1,W_1}  = U_{\omega_1,W_1} \omega_1^2\,, \hspace{2.5cm} \omega_1^2 U_{\omega_2,W_{21}}  = U_{\omega_2,W_{21}} \omega_1^2\,,
\nonumber\\
&& (\omega_2^2+ W_2)U_{\omega_1,W_1}  = U_{\omega_1,W_1} (\omega_2^2 + W_{21})\,, \hspace{1cm} (\omega_2^2+ W_{21})U_{\omega_2,W_{21}}  = U_{\omega_2,W_{21}} \omega_2^2\,. \qquad
\eeq
Equations $G_1V^{(7)}=0$ and $G_2V^{(7)}=0$ with $G_1$ and $G_2$ as in \rf{15092018-man03-59} constitute a system of two decoupled second-order differential equations w.r.t. $B_1$ and $B_2$. All independent solutions of these equations are given in \rf{07092018-man03-05}.
%
%
%
%
%
\be  \label{15092018-man03-60}
\hbox{\bf \large Case \quad $m_1^2< 0$,\quad $m_2^2< 0$,\quad $m_3=0$,\quad $m_1\ne m_2$. }
\ee
%
%
%
%
%
{\bf Step 1}. This case corresponds to the third vertex $p_\smp3^-$ \rf{06092018-man03-01-add}. In relations \rf{15092018-man03-01}-\rf{15092018-man03-13}, we set $m_3=0$ and, ignoring $\zeta_3$ in \rf{15092018-man03-01}-\rf{15092018-man03-13}, we use vertex $V^{(3)}$ given by
\be \label{15092018-man03-61}
V^{(3)} = V^{(3)}(B_a,\alpha_{aa+1})\,,
\ee
where the realization of operators $G_a$ \rf{15092018-man03-08}-\rf{15092018-man03-10} on the vertex $V^{(3)}$ \rf{15092018-man03-61} takes the form%
\footnote{ For massless field corresponding to $a=3$, $\alpha_{33}$-terms in \rf{15092018-man03-02},\rf{15092018-man03-03} can be ignored in \rf{15092018-man03-62},\rf{15092018-man03-63} by virtue of the second constraint in \rf{01092018-man03-14}.}
\beq
\label{15092018-man03-62} && \hspace{-1.5cm} G_1    = \frac{2N_1 + d-2}{ \kappa_1} \Bigl( B_3  \partial_{\alpha_{31}}  - (B_2   - \half \kappa_2) \partial_{\alpha_{12}} + \frac{m_2^2}{2}  \partial_{B_1}\Bigr) +        \frac{F_1}{\kappa_1^2} - ( B_2 + \frac{3}{2}\kappa_2) \partial_{B_1} \partial_{\alpha_{12}}
\nonumber\\
&& -    B_3   \partial_{B_1} \partial_{\alpha_{31}}  +   2\alpha_{23} \partial_{\alpha_{12}}\partial_{\alpha_{31}}
+   \alpha_{33} \partial_{\alpha_{31}}^2 + \frac{1}{4}(m_1^2 - 2 m_2^2) \partial_{B_1}^2  \,,
\\
\label{15092018-man03-63} &&  \hspace{-1.5cm} G_2    =   \frac{2N_2 +d-2}{\kappa_2} \Bigl( ( B_1 + \half \kappa_1 ) \partial_{\alpha_{12}} - B_3 \partial_{\alpha_{23}} - \frac{m_1^2}{2} \partial_{B_2} \Bigr) + \frac{ F_2}{\kappa_2^2} -  B_3 \partial_{B_2} \partial_{\alpha_{23}}
\nonumber\\
&  - &  ( B_1  - \frac{3}{2} \kappa_1) \partial_{B_2} \partial_{\alpha_{12}}   +  2 \alpha_{31} \partial_{\alpha_{12}}\partial_{\alpha_{23}}
+  \alpha_{33} \partial_{\alpha_{23}}^2 + \frac{1}{4}(m_2^2 - 2 m_1^2) \partial_{B_2}^2 \,,
\\
\label{15092018-man03-64} \hspace{-0.7cm} G_3 &  =  & ( B_2 + \half \kappa_2) \partial_{\alpha_{23}} -  ( B_1 - \half \kappa_1  ) \partial_{\alpha_{31}} + \half \bigl( m_1^2 - m_2^2  \bigr)\partial_{B_3}\,.
\eeq

\noindent {\bf Step 2}. We make the transformation
\be \label{15092018-man03-65}
V^{(3)} = U V^{(4)}\,, \qquad U_{1\partial\alpha} = \exp\Bigl( \frac{2B_3}{m_1^2-m_2^2} ( B_1 - \half \kappa_1  ) \partial_{\alpha_{31}} - \frac{2B_3}{m_1^2-m_2^2}( B_2 + \half \kappa_2) \partial_{\alpha_{23}}\Bigr).
\ee
Realization of $G_1$, $G_2$, $G_3$  \rf{15092018-man03-62}-\rf{15092018-man03-64} on $V^{(4)}$ \rf{15092018-man03-65} takes the form
\beq
\label{15092018-man03-66} \hspace{-0.7cm} G_1 &  =  &   - r_1 \partial_{B_1}^2  + q_1 \bigl(N_{B_1}  + \nu_1 +1 \bigr)  \partial_{B_1} +  e_1 (N_{B_1} + \nu_1)(  N_{B_1} + \nu_1 + 1 )  + u_1
\nonumber\\
\hspace{-1cm} & - & \frac{2}{\kappa_1} (N_{B_1}+\nu_1 + 1) (B_2   - \half \kappa_2) \partial_{\alpha_{12}} -  ( B_2 + \frac{3}{2}\kappa_2) \partial_{B_1} \partial_{\alpha_{12}}   +   2\alpha_{23} \partial_{\alpha_{12}}\partial_{\alpha_{31}}\,,\qquad
\nonumber\\
\label{15092018-man03-67} \hspace{-0.7cm} G_2 &  =  &   - r_2 \partial_{B_2}^2  + q_2 \bigl(N_{B_2}  + \nu_2 +1 \bigr)  \partial_{B_2} +  e_2 (N_{B_2} + \nu_2)(  N_{B_2} + \nu_2 + 2 )  + u_2
\nonumber\\
\hspace{-1cm} & + & \frac{2}{\kappa_2} (N_{B_2}+\nu_2 + 1) (B_1   + \half \kappa_1) \partial_{\alpha_{12}} -  ( B_1 - \frac{3}{2}\kappa_1) \partial_{B_2} \partial_{\alpha_{12}}   +   2\alpha_{31} \partial_{\alpha_{12}}\partial_{\alpha_{23}}\,,
\\
\label{15092018-man03-68} \hspace{-0.7cm} G_3 &  =  &   \half \bigl( m_1^2 - m_2^2  \bigr)\partial_{B_3}\,,
\eeq
where, here and below, we use quantities \rf{02092018-man03-07app}-\rf{02092018-man03-09app-x1} taken for masses given in \rf{15092018-man03-60}.
From \rf{15092018-man03-68} and equation $G_3V^{(4)}=0$, we see that  $V^{(4)}$ is independent of $B_3$,
\be \label{15092018-man03-69}
V^{(4)} = V^{(4)}(B_1,B_2,\alpha_{aa+1})\,.
\ee

\noindent {\bf Step 3}. We make the transformation
\beq
\label{15092018-man03-70} && \hspace{-1.7cm} V^{(4)} = U_{ \partial B } U_{2\partial\alpha}U_{3\partial\alpha}  V^{(5)}\,, \qquad  U_{ \partial B } = \exp( \frac{q_1}{2e_1}\partial_{B_1}  + \frac{q_2}{2e_2}\partial_{B_2} \bigr)\,,
\nonumber\\
\label{15092018-man03-71} && \hspace{-1.7cm} U_{2\partial\alpha} = \exp\Bigl(  \frac{B_1}{ \kappa_2 e_2 }  \partial_{ \alpha_{12} } - \frac{B_2 }{ \kappa_1 e_1 } \partial_{ \alpha_{12} } - \frac{h_{12} }{ 2\kappa_1\kappa_2e_1e_2 }\partial_{\alpha_{12} }\Bigr),
\qquad U_{3\partial\alpha} = \exp\bigl(  \frac{2h_{12}}{D} B_1 B_2 \partial_{ \alpha_{12} } \bigr).\quad
\eeq
Realization of $G_1$, $G_2$ \rf{15092018-man03-66} on $V^{(5)}$ \rf{15092018-man03-70} takes the form
\beq
\label{15092018-man03-72} \hspace{-0.7cm} G_1 &  =  & - \frac{D}{4\kappa_1^2 e_1} \partial_{B_1}^2  +  e_1 (N_{B_1} + \nu_1)(  N_{B_1} + \nu_1 + 1 )  + u_1
\nonumber\\
& + & \bigl( \frac{4\kappa_2^2 e_2}{D} B_2^2  -  \frac{1}{e_2} \bigr) \partial_{\alpha_{12}}^2     +   2\alpha_{23} \partial_{\alpha_{12}}\partial_{\alpha_{31}},\qquad
\nonumber\\
\label{15092018-man03-73} \hspace{-0.7cm} G_2 &  =  & - \frac{D}{4\kappa_2^2 e_2} \partial_{B_2}^2  +  e_2 (N_{B_2} + \nu_2)(  N_{B_2} + \nu_2 + 1 )  + u_2
\nonumber\\
& + & \bigl( \frac{4\kappa_1^2 e_1}{D} B_1^2 -  \frac{1}{e_1} \bigr) \partial_{\alpha_{12}}^2     +   2\alpha_{31} \partial_{\alpha_{12}}\partial_{\alpha_{23}}.
\eeq
Also, using $U_{\partial \alpha}$ in \rf{08092018-man03-11}, we note the relation
\be \label{15092018-man03-74}
U_{1\partial\alpha} U_{\partial B} U_{2\partial\alpha} U_{3\partial\alpha} =  U_{\partial B}  U_{\partial\alpha}\,.
\ee

\noindent {\bf Step 4}. We make the transformation
\be \label{15092018-man03-75}
V^{(5)}  =   U_{B_1} U_{B_2} V^{(6)}\,, \qquad V^{(6)} = U_{\nu_1,W_1} U_{\nu_2,W_{21}} V^{(7)}\,, \qquad V^{(7)} =  U_{B_1}^{-1} U_{B_2}^{-1}V^{(8)}\,,
\ee
where $U_{B_a}$, $U_{\nu,W}$ are defined in \rf{02092018-man03-13app},\rf{02092018-man03-14app} and we use the notation
\beq
&& W_1  =     2\alpha_{23} \partial_{\alpha_{12}} \partial_{\alpha_{31}} - \partial_{\alpha_{12}}^2  \,,
\nonumber\\
\label{15092018-man03-80} && W_2  =    2\alpha_{31}  \partial_{\alpha_{12}} \partial_{\alpha_{23}} - \partial_{\alpha_{12}}^2 \,, \hspace{2cm} W_{21}  =   2\alpha_{31}  \partial_{\alpha_{12}} \partial_{\alpha_{23}}\,.
\eeq
Realizations of $G_1$, $G_2$ \rf{15092018-man03-72} on $V^{(6)}$, $V^{(7)}$, $V^{(8)}$ \rf{15092018-man03-75} take  the forms
\beq
\label{15092018-man03-76}  &&  \hspace{-2cm}  G_a  = (B_a^2 - \frac{D}{4\kappa_a^2e_a^2})  \partial_{B_a}^2   + \frac{u_a}{e_a} +    (1- \frac{4\kappa_a^2e_a^2}{D} B_a^2)^{-1} \bigl( \nu_a^2 - 1 + W_a \bigr) \,,  \hspace{1cm} \hbox{ for } V^{(6)};
\\
\label{15092018-man03-77} && \hspace{-2cm} G_a  = (B_a^2 - \frac{D}{4\kappa_a^2e_a^2})  \partial_{B_a}^2    +    (1- \frac{4\kappa_a^2e_a^2}{D} B_a^2)^{-1} \bigl( \nu_a^2 - 1\bigr) + \frac{u_a}{e_a} \,,  \hspace{2cm} \hbox{ for } V^{(7)};
\\
\label{15092018-man03-78} && \hspace{-2cm} G_a   =    - \frac{D}{4\kappa_a^2e_a} \partial_{B_a}^2  + e_a (N_{B_a} + \nu_a)(  N_{B_a} + \nu_a + 1 ) + u_a \,,  \hspace{2.7cm} \hbox{ for } V^{(8)};
\eeq
where $a=1,2$. For $\nu_a$ \rf{02092018-man03-06app-bb1} and $W$-operators \rf{15092018-man03-80}, we note the relations which admit us to get $G_a$ in \rf{15092018-man03-77},
\beq
&& (\nu_1^2+ W_1)U_{\nu_1,W_1}  = U_{\nu_1,W_1} \nu_1^2\,, \hspace{2.5cm} \nu_1^2 U_{\nu_2,W_{21}}  = U_{\nu_2,W_{21}} \nu_1^2\,,
\nonumber\\
&& (\nu_2^2+ W_2)U_{\nu_1,W_1}  = U_{\nu_1,W_1} (\nu_2^2 + W_{21})\,, \hspace{1cm} (\nu_2^2+ W_{21})U_{\nu_2,W_{21}}  = U_{\nu_2,W_{21}} \nu_2^2\,. \qquad
\eeq
Equations $G_1V^{(8)}=0$ and $G_2V^{(8)}=0$ with $G_1$ and $G_2$ as in \rf{15092018-man03-78} constitute a system of two decoupled second-order differential equations w.r.t. $B_1$ and $B_2$. Four independent solutions of these equations are given in \rf{08092018-man03-05}.

%
%
%
%
%
\be  \label{15092018-man03-60-q}
\hbox{\bf \large Case \quad $m_1=0$,\quad $m_2^2< 0$,\quad $m_3^2>0$. }
\ee
%
%
%
%
%
{\bf Step 1}. This case corresponds to the first vertex $p_\smp3^-$ in \rf{09092018-man03-01-add}. In \rf{15092018-man03-06}, we use operator $U_{\Gamma_1}$ \rf{02092018-man03-11app}. Then, in the remaining relations \rf{15092018-man03-01}-\rf{15092018-man03-13}, we set $m_1=0$ and use vertex $V^{(3)}$ given by
\be \label{15092018-man03-61-q}
V^{(3)} = V^{(3)}(B_a,\alpha_{aa+1},\zeta_3)\,,
\ee
where the realization of operators $G_a$ \rf{15092018-man03-08}-\rf{15092018-man03-10} on the vertex $V^{(3)}$ \rf{15092018-man03-61-q} takes the form
\beq
\label{15092018-man03-08-qq} \hspace{-0.7cm} G_1 &  =  & \frac{2N_1 + d-2}{ \kappa_1}\Bigl( (B_3 + \half  m_3 \zeta_3 ) \partial_{\alpha_{31}}  - (B_2   - \half \kappa_2) \partial_{\alpha_{12}} + \half \bigl( m_2^2 - m_3^2  \bigr)\partial_{B_1}\Bigr)
\nonumber\\
\hspace{-0.7cm} & + &      \frac{F_1}{\kappa_1^2} - ( B_2 + \frac{3}{2}\kappa_2) \partial_{B_1} \partial_{\alpha_{12}} - (B_3 - \frac{3}{2}m_3\zeta_3) \partial_{B_1} \partial_{\alpha_{31}}  +   2\alpha_{23} \partial_{\alpha_{12}}\partial_{\alpha_{31}}
\nonumber\\
& + & \alpha_{33} \partial_{\alpha_{31}}^2  - \half( m_2^2 + m_3^2) \partial_{B_1}^2  \,,
\nonumber\\
\label{15092018-man03-09-qq} \hspace{-0.7cm} G_2 &  =  & \frac{\kappa_2}{2N_2 +d-2} \Bigl(  ( B_1 + \half \kappa_1 ) \partial_{\alpha_{12}} -  ( B_3 - \half  m_3 \zeta_3 ) \partial_{\alpha_{23}} + \half   m_3^2   \partial_{B_2}\Bigr)
\nonumber\\
\hspace{-0.7cm} & + &         \frac{ F_2}{\kappa_2^2} - ( B_3 + \frac{3}{2} m_3 \zeta_3) \partial_{B_2} \partial_{\alpha_{23}}  - ( B_1  - \frac{3}{2} \kappa_1) \partial_{B_2} \partial_{\alpha_{12}}   +  2 \alpha_{31} \partial_{\alpha_{12}}\partial_{\alpha_{23}}
\nonumber\\
& + & \alpha_{33} \partial_{\alpha_{23}}^2 + \frac{1}{4}(m_2^2 - 2 m_3^2 ) \partial_{B_2}^2    \,,
\\
\label{15092018-man03-10-qq} \hspace{-0.7cm} G_3 &  =  & ( B_2 + \half \kappa_2) \partial_{\alpha_{23}} -  ( B_1 - \half \kappa_1  ) \partial_{\alpha_{31}} -  \half  m_2^2  \partial_{B_3} + m_3 \partial_{\zeta_3}\,.
\eeq

\noindent {\bf Step 2}. We make the transformations
\be \label{15092018-man03-65-q}
V^{(3)} = U_\zeta V^{(4)}\,, \quad U_\zeta = \exp\Bigl(  \frac{\zeta_3}{m_3} ( B_1 - \half \kappa_1  ) \partial_{\alpha_{31}} - \frac{\zeta_3}{m_3} ( B_2 + \half \kappa_2) \partial_{\alpha_{23}} + \frac{m_2^2}{2m_3}\zeta_3\partial_{B_3} \Bigr).
\ee
Realization of $G_1$, $G_2$, $G_3$  \rf{15092018-man03-09-qq},\rf{15092018-man03-10-qq} on $V^{(4)}$ \rf{15092018-man03-65-q} takes the form%
\footnote{ For massive field corresponding to $a=3$, operators $G_1$, $G_2$ \rf{15092018-man03-66-q} involve, besides $\alpha_{33}$-term \rf{15092018-man03-09-qq}, $\zeta_3^2$-term, i.e., we get $(\alpha_{33}+\zeta_2^2)$-term which can be ignored by virtue of the second constraint in \rf{01092018-man03-13}.
}

\beq
\label{15092018-man03-66-q} \hspace{-0.7cm} G_1 &  =  &   - r_1 \partial_{B_1}^2  + q_1 \bigl(N_{B_1}  + \nu_1 +1 \bigr)  \partial_{B_1}  + u_1
\nonumber\\
\hspace{-0.7cm} & + &  \frac{2}{\kappa_1}\bigl(N_{B_1}  + \nu_1 +1 \bigr) \bigl(B_3\partial_{\alpha_{31}}  - (B_2   - \half \kappa_2) \partial_{\alpha_{12}}\bigr)
\nonumber\\
& - &    ( B_2 + \frac{3}{2}\kappa_2) \partial_{B_1} \partial_{\alpha_{12}} - B_3 \partial_{B_1} \partial_{\alpha_{31}}  +   2\alpha_{23} \partial_{\alpha_{12}}\partial_{\alpha_{31}} \,,
\nonumber\\
\label{15092018-man03-67-q} \hspace{-0.7cm} G_2 &  =  &   - r_2 \partial_{B_2}^2 + q_2 \bigl(N_{B_2}  + \nu_2 +1 \bigr)  \partial_{B_2} +  e_2 (N_{B_2} + \nu_2)(  N_{B_2} + \nu_2 + 1 )  + u_2
\nonumber\\
\hspace{-0.7cm} & + &  \frac{2}{\kappa_2}\bigl(N_{B_2}  + \nu_2 +1 \bigr) \bigl(( B_1 + \half \kappa_1 ) \partial_{\alpha_{12}} -   B_3 \partial_{\alpha_{23}}\bigr)
\nonumber\\
& - &     B_3 \partial_{B_2} \partial_{\alpha_{23}}  - ( B_1  - \frac{3}{2} \kappa_1) \partial_{B_2} \partial_{\alpha_{12}}   +  2 \alpha_{31} \partial_{\alpha_{12}}\partial_{\alpha_{23}} \,,
\\
\label{15092018-man03-68-q} \hspace{-0.7cm} G_3 &  =  &  m_3\partial_{\zeta_3}\,,
\eeq
where, here and below, we use quantities \rf{02092018-man03-07app}-\rf{02092018-man03-09app-x1} taken for masses given in \rf{15092018-man03-60-q}.
From \rf{15092018-man03-68-q} and equation $G_3V^{(4)}=0$, we see that  $V^{(4)}$ is independent of $\zeta_3$,
\be \label{15092018-man03-69-q}
V^{(4)} = V^{(4)}(B_1,B_2,\alpha_{aa+1})\,.
\ee

\noindent {\bf Step 3}. We make the transformation
\beq
\label{15092018-man03-70-q} && \hspace{-1.7cm} V^{(4)} = U_{ \partial B } U_{ \partial\alpha}   V^{(5)}\,, \qquad  U_{ \partial B } = \exp( - \frac{r_1}{q_1}\partial_{B_1}  + \frac{q_2}{2e_2}\partial_{B_2} \bigr)\,,
\\
\label{15092018-man03-71-q} && \hspace{-1.2cm}  U_{\partial\alpha}  = \exp\Bigl(  - \frac{\kappa_2}{q_1} \partial_{\alpha_{12}} + \frac{B_1}{\kappa_2 e_2}   \partial_{\alpha_{12}} - \frac{ 2m_2^2}{\kappa_1 q_1^2} B_2 \partial_{\alpha_{12}} +  \frac{2m_3^2}{\kappa_1 q_1^2} B_3 \partial_{\alpha_{31}} -  \frac{ B_3}{\kappa_2 e_2} \partial_{\alpha_{23}}
\nonumber\\
&& - \frac{ 2B_1 B_3}{\kappa_1 q_1} \partial_{\alpha_{31}} + \frac{ 2B_1 B_2 }{\kappa_1 q_1} \partial_{\alpha_{12}} +  \frac{ 2h_{23} B_2 B_3}{\kappa_1^2 q_1^2} \partial_{\alpha_{23}} \Bigr)\,.
\eeq
Realization of $G_1$, $G_2$ \rf{15092018-man03-66-q} on $V^{(5)}$ \rf{15092018-man03-70-q} takes the form
\beq
\label{15092018-man03-72-q} \hspace{-0.7cm} G_1 &  =  &  q_1 \bigl(N_{B_1}  + \nu_1 +1 \bigr)  \partial_{B_1}  +  u_1
\nonumber\\
& + &  (\frac{4\kappa_2^2 e_2}{D} B_2^2 - \frac{1}{e_2})\partial_{\alpha_{12}}^2 - \frac{4m_3^2}{D}B_3^2  \partial_{\alpha_{31}}^2 +   2\alpha_{23} \partial_{\alpha_{12}}\partial_{\alpha_{31}} \,,
\nonumber\\
\label{15092018-man03-73-q} \hspace{-0.7cm} G_2 &  =  &   - \frac{D}{4\kappa_2^2e_2} \partial_{B_2}^2  +  e_2 (N_{B_2} + \nu_2)(  N_{B_2} + \nu_2 + 1 )  + u_2
\nonumber\\
&& - \frac{4m_3^2}{D}B_3^2  \partial_{\alpha_{23}}^2 + \frac{4B_1}{q_1}\partial_{\alpha_{12}}^2  +   2\alpha_{31} \partial_{\alpha_{12}}\partial_{\alpha_{23}} \,.
\eeq

\noindent {\bf Step 4}. We make the transformations
\be \label{15092018-man03-75-q}
V^{(5)}  =   (-\frac{4}{q_1}B_1)^{-\nu_1/2}  U_{B_2} V^{(6)}\,, \qquad V^{(6)} = U_{\nu_1,W_1} U_{\nu_2,W_{23}} V^{(7)}\,, \qquad V^{(7)} =   U_{B_2}^{-1}V^{(8)}\,,
\ee
where $U_{B_2}$, $U_{\nu,W}$ are defined in \rf{02092018-man03-13app},\rf{02092018-man03-14app} and we use the notation
\beq
\label{15092018-man03-79-q} &&  \hspace{-1.7cm} W_1   =   2 \alpha_{23} \partial_{\alpha_{12}}\partial_{\alpha_{31}}     - \partial_{\alpha_{12}}^2 - \frac{4m_3^2}{D} B_3^2 \partial_{\alpha_{31}}^2   \,,
\nonumber\\
\label{15092018-man03-80-q} && \hspace{-1.7cm} W_2    =    2 \alpha_{31} \partial_{\alpha_{12}}\partial_{\alpha_{23}}  -  \partial_{\alpha_{12}}^2 - \frac{4m_3^2}{D} B_3^2 \partial_{\alpha_{23}}^2      \,, \qquad W_{23}    =    2 \alpha_{31} \partial_{\alpha_{12}}\partial_{\alpha_{23}}   - \frac{4m_3^2}{D} B_3^2 \partial_{\alpha_{23}}^2  \,.
\eeq
Realizations of $G_1$, $G_2$ \rf{15092018-man03-72-q} on $V^{(6)}$, $V^{(7)}$, $V^{(8)}$ \rf{15092018-man03-75-q} take  the forms
\beq
&& \hspace{-1.4cm} G_1  = u_1 +   q_1 \bigl(N_{B_1}  + \nu_1 +1 \bigr)  \partial_{B_1}  - \frac{q_1}{4B_1}(\nu_1^2 + W_1)\,,
\nonumber\\
\label{15092018-man03-76-q} && \hspace{-1.4cm} G_2  = (B_2^2 - \frac{D}{4\kappa_2^2e_2^2})  \partial_{B_2}^2   + \frac{u_2}{e_2} +    (1- \frac{4\kappa_2^2e_2^2}{D} B_2^2)^{-1} \bigl( \nu_2^2 - 1 + W_2 \bigr) \,, \hspace{1cm} \hbox{ for } V^{(6)};
\\
&& \hspace{-1.4cm} G_1  = u_1 +   q_1 \bigl(N_{B_1}  + \nu_1 +1 \bigr)  \partial_{B_1}  - \frac{q_1}{4B_1}\nu_1^2 \,,
\nonumber\\
\label{15092018-man03-77-q} && \hspace{-1.4cm}  G_2  = (B_2^2 - \frac{D}{4\kappa_2^2e_2^2})  \partial_{B_2}^2    +    (1- \frac{4\kappa_2^2e_2^2}{D} B_2^2)^{-1} \bigl( \nu_2^2 - 1\bigr) + \frac{u_2}{e_2} \,, \hspace{2cm} \hbox{ for } V^{(7)};
\\
&& \hspace{-1.4cm} G_1  = u_1 +   q_1 \bigl(N_{B_1}  + \nu_1 +1 \bigr)  \partial_{B_1}  - \frac{q_1}{4B_1}\nu_1^2 \,,
\nonumber\\
\label{15092018-man03-78-q} && \hspace{-1.4cm}  G_2   =    - \frac{D}{4\kappa_2^2e_2} \partial_{B_2}^2  + e_2 (N_{B_2} + \nu_2)(  N_{B_2} + \nu_2 + 1 ) + u_2 \,,  \hspace{2.6cm} \hbox{ for } V^{(8)}.
\eeq
For $\nu_a$ \rf{02092018-man03-06app-bb1} and $W$-operators \rf{15092018-man03-80-q}, we note the relations which admit us to get $G_1$, $G_2$ in \rf{15092018-man03-77-q},
\beq
&& (\nu_1^2+ W_1)U_{\nu_1,W_1}  = U_{\nu_1,W_1} \nu_1^2\,, \hspace{2.5cm} \nu_1^2 U_{\nu_2,W_{23}}  = U_{\nu_2,W_{23}} \nu_1^2\,,
\nonumber\\
&& (\nu_2^2+ W_2)U_{\nu_1,W_1}  = U_{\nu_1,W_1} (\nu_2^2 + W_{23})\,, \hspace{1cm} (\nu_2^2+ W_{23})U_{\nu_2,W_{23}}  = U_{\nu_2,W_{23}} \nu_2^2\,. \qquad
\eeq
Equations $G_1V^{(8)}=0$ and $G_2V^{(8)}=0$ with $G_1$ and $G_2$ as in \rf{15092018-man03-78-q} constitute a system of two decoupled second-order differential equations w.r.t. $B_1$ and $B_2$. Four independent solutions of these equations are given in \rf{08092018-man03-05-q}.
%
%
%
%
%
%
\be  \label{15092018-man03-84}
\hbox{\bf \large Case \quad  $m_1^2< 0$,\quad $m_2^2< 0$,\quad $m_3^2 > 0$.}
\ee
%
%
%
%
%
{\bf Step 1}. This case corresponds to the second vertex $p_\smp3^-$ \rf{09092018-man03-01-add}. For this case, vertex $V^{(3)}$  is given in \rf{15092018-man03-13}, where the realization of operators $G_a$ \rf{15092018-man03-04}-\rf{15092018-man03-06} on the vertex $V^{(3)}$ \rf{15092018-man03-13} is given in \rf{15092018-man03-08}-\rf{15092018-man03-10}. We make the transformation
\beq
\label{15092018-man03-85} && V^{(3)} = U_\zeta V^{(4)}\,,
\nonumber\\
\label{15092018-man03-86} && U_\zeta  = \exp\Bigl( \frac{\zeta_3}{m_3} ( B_1 - \half \kappa_1  ) \partial_{\alpha_{31}} - \frac{\zeta_3}{m_3} ( B_2 + \half \kappa_2) \partial_{\alpha_{23}} - \frac{\zeta_3}{2m_3} \bigl( m_1^2 - m_2^2  \bigr)\partial_{B_3}\Bigr).\qquad
\eeq
Realization of $G_a$ \rf{15092018-man03-08}-\rf{15092018-man03-10} on $V^{(4)}$ \rf{15092018-man03-85} takes the form%
\footnote{ For massive field corresponding to $a=3$, operators $G_1$, $G_2$ \rf{15092018-man03-87} involve, besides $\alpha_{33}$-term \rf{14092018-man03-04}, $\zeta_3^2$-term, i.e., we get $(\alpha_{33}+\zeta_2^2)$-term which can be ignored by virtue of the second constraint in \rf{01092018-man03-13}.
}
 \beq
\label{15092018-man03-87} \hspace{-0.7cm} G_1 &  =  &   - r_1 \partial_{B_1}^2  + q_1 \bigl(N_{B_1}  + \nu_1 +1 \bigr)  \partial_{B_1} +  e_1 (N_{B_1} + \nu_1)(  N_{B_1} + \nu_1 + 1 )  + u_1
\nonumber\\
\hspace{-0.7cm} & + &  \frac{2}{\kappa_1}\bigl(N_{B_1}  + \nu_1 +1 \bigr) \bigl(B_3\partial_{\alpha_{31}}  - (B_2   - \half \kappa_2) \partial_{\alpha_{12}}\bigr)
\nonumber\\
& - &    ( B_2 + \frac{3}{2}\kappa_2) \partial_{B_1} \partial_{\alpha_{12}} - B_3 \partial_{B_1} \partial_{\alpha_{31}}  +   2\alpha_{23} \partial_{\alpha_{12}}\partial_{\alpha_{31}} \,,
\nonumber\\
\label{15092018-man03-88} \hspace{-0.7cm} G_2 &  =  &   - r_2 \partial_{B_2}^2 + q_2 \bigl(N_{B_2}  + \nu_2 +1 \bigr)  \partial_{B_2} +  e_2 (N_{B_2} + \nu_2)(  N_{B_2} + \nu_2 + 1 )  + u_2
\nonumber\\
\hspace{-0.7cm} & + &  \frac{2}{\kappa_2}\bigl(N_{B_2}  + \nu_2 +1 \bigr) \bigl(( B_1 + \half \kappa_1 ) \partial_{\alpha_{12}} -   B_3 \partial_{\alpha_{23}}\bigr)
\nonumber\\
& - &     B_3 \partial_{B_2} \partial_{\alpha_{23}}  - ( B_1  - \frac{3}{2} \kappa_1) \partial_{B_2} \partial_{\alpha_{12}}   +  2 \alpha_{31} \partial_{\alpha_{12}}\partial_{\alpha_{23}} \,,
\\
\label{15092018-man03-89} G_3 & = &  m_3\partial_{\zeta_3}\,,
\eeq
where we use the quantities \rf{02092018-man03-07app}-\rf{02092018-man03-09app-x1} taken for masses given in \rf{15092018-man03-84}.
From \rf{15092018-man03-89} and equation $G_3V^{(4)}=0$, we see that  $V^{(4)}$ is independent of $\zeta_3$,
\be \label{15092018-man03-90}
V^{(4)} = V^{(4)}(B_a,\alpha_{aa+1})\,.
\ee
\noindent {\bf Step 2}. We make the transformation
\beq
\label{15092018-man03-91}  && V^{(4)}  = U_{\partial B}  U_{\partial\alpha}  V^{(5)}\,, \qquad   U_{\partial B} = \exp( \frac{q_1}{2e_1}\partial_{B_1}  + \frac{q_2}{2e_2}\partial_{B_2} \bigr)\,,
\\
\label{15092018-man03-92} && U_{\partial\alpha} = \exp\Bigl( \frac{ B_1}{\kappa_2 e_2} \partial_{\alpha_{12}}  - \frac{ B_2}{\kappa_1 e_1} \partial_{\alpha_{12}} +  \frac{B_3}{\kappa_1 e_1}  \partial_{\alpha_{31}} -  \frac{B_3 }{\kappa_2 e_2} \partial_{\alpha_{23}} - \frac{h_{12}}{2\kappa_1\kappa_2 e_1 e_2} \partial_{\alpha_{12}} 
\nonumber\\
&& \hspace{2cm} +\,\, \frac{2}{D} h_{12} B_1 B_2 \partial_{\alpha_{12}} + \frac{2}{D} h_{23} B_2 B_3 \partial_{\alpha_{23}} + \frac{2}{D} h_{31} B_3 B_1 \partial_{\alpha_{31}}\Bigr).
\eeq
Realization of $G_1$, $G_2$ \rf{15092018-man03-87} on $V^{(5)}$ \rf{15092018-man03-91} takes the form
\beq
\label{15092018-man03-93} \hspace{-0.7cm} G_1 &  =  &   -  \frac{D}{4\kappa_1^2e_1} \partial_{B_1}^2   +  e_1 (N_{B_1} + \nu_1)(  N_{B_1} + \nu_1 + 1 )  + u_1
\nonumber\\
& + &   \bigl(\frac{4\kappa_2^2 e_2}{D} B_2^2 - \frac{1}{e_2} \bigr) \partial_{\alpha_{12}}^2 - \frac{4m_3^2}{D} B_3^2 \partial_{\alpha_{31}}^2
+  2\alpha_{23} \partial_{\alpha_{12}}\partial_{\alpha_{31}} \,,
\nonumber\\
\label{15092018-man03-94} \hspace{-0.7cm} G_2 &  =  &   -  \frac{D}{4\kappa_2^2e_2} \partial_{B_2}^2   +  e_2 (N_{B_2} + \nu_2)(  N_{B_2} + \nu_2 + 1 )  + u_2
\nonumber\\
& + &  \bigl(\frac{4\kappa_1^2 e_1}{D} B_1^2 - \frac{1}{ e_1 } \bigr) \partial_{\alpha_{12}}^2 - \frac{4m_3^2}{D} B_3^2 \partial_{\alpha_{23}}^2 + 2\alpha_{31} \partial_{\alpha_{12}}\partial_{\alpha_{23}} \,.
\eeq

\noindent {\bf Step 3}. We make the transformations
\be \label{15092018-man03-95}
V^{(5)} = U_{B_1} U_{B_2} V^{(6)},\qquad V^{(6)} = U_{\nu_1,W_1} U_{\nu_2,W_{23}} V^{(7)}, \qquad
V^{(7)} = U_{B_1}^{-1} U_{B_2}^{-1} V^{(8)},
\ee
where $U_{B_a}$, $U_{\nu,W}$ are given in \rf{02092018-man03-13app}, \rf{02092018-man03-14app} and we use the notation
\beq
&& \hspace{-1.4cm} W_1    =     2 \alpha_{23} \partial_{\alpha_{12}}\partial_{\alpha_{31}}   - \partial_{\alpha_{12}}^2 - \frac{4m_3^2}{D} B_3^2 \partial_{\alpha_{31}}^2    \,,
\nonumber\\
\label{15092018-man03-97} &&  \hspace{-1.4cm}  W_2  =     2 \alpha_{31} \partial_{\alpha_{12}}\partial_{\alpha_{23}} -  \partial_{\alpha_{12}}^2 - \frac{4m_3^2}{D} B_3^2 \partial_{\alpha_{23}}^2    \,, \qquad W_{23}   =    2 \alpha_{31} \partial_{\alpha_{12}}\partial_{\alpha_{23}} - \frac{4m_3^2}{D} B_3^2 \partial_{\alpha_{23}}^2  \,.\quad
\eeq
Realizations of $G_1$, $G_2$ \rf{15092018-man03-93} on $V^{(6)}$, $V^{(7)}$, $V^{(8)}$ \rf{15092018-man03-95} take the forms
\beq
\label{15092018-man03-99} && \hspace{-2cm} G_a  =  \bigl(B_a^2 - \frac{D}{4\kappa_a^2e_a^2} \bigr)  \partial_{B_a}^2    +    \bigl( 1- \frac{4\kappa_a^2e_a^2}{D} B_a^2 \bigr)^{-1} \bigl( \nu_a^2 - 1 + W_a \bigr) + \frac{u_a}{e_a} \,, \hspace{1cm} \hbox{for } V^{(6)}\,,
\\
\label{15092018-man03-100} && \hspace{-2cm}  G_a   =   (B_a^2 - \frac{D}{4\kappa_a^2e_a^2})  \partial_{B_a}^2   +    (1- \frac{4\kappa_a^2e_a^2}{D} B_a^2)^{-1} \bigl( \nu_a^2 - 1\bigr) + \frac{u_a}{e_a} \,, \hspace{2.1cm} \hbox{for } V^{(7)}\,,
\\
\label{15092018-man03-101} && \hspace{-2cm}  G_a   =    - \frac{D}{4\kappa_a^2e_a} \partial_{B_a}^2  + e_a (N_{B_a} + \nu_a)(  N_{B_a} + \nu_a + 1 ) + u_a \,,  \hspace{2.8cm} \hbox{for } V^{(8)}\,,
\eeq
where, in \rf{15092018-man03-99}-\rf{15092018-man03-101}, $a=1,2$.
For $\nu_a$ \rf{02092018-man03-06app-bb1} and $W$-operators \rf{15092018-man03-97}, we note the relations which admit us to get $G_a$ in \rf{15092018-man03-100},
\beq
&& (\nu_1^2+ W_1)U_{\nu_1,W_1}  = U_{\nu_1,W_1} \nu_1^2\,, \hspace{2.5cm} \nu_1^2 U_{\nu_2,W_{23}}  = U_{\nu_2,W_{23}} \nu_1^2\,,
\nonumber\\
&& (\nu_2^2+ W_2)U_{\nu_1,W_1}  = U_{\nu_1,W_1} (\nu_2^2 + W_{23})\,, \hspace{1cm} (\nu_2^2+ W_{23})U_{\nu_2,W_{23}}  = U_{\nu_2,W_{23}} \nu_2^2\,. \qquad
\eeq
Equations $G_1V^{(8)}=0$ and $G_2V^{(8)}=0$ with $G_1$ and $G_2$ as in \rf{15092018-man03-101} constitute a system of two decoupled second-order differential equations w.r.t. $B_1$ and $B_2$. Four independent solutions of these equations are given in \rf{09092018-man03-05}.

\appendix{ \large Derivation of vertices $p_\smp3^-$ \rf{10092018-man03-02}, \rf{11092018-man03-02}, \rf{12092018-man03-02}, \rf{13092018-man03-02} }

We consider four vertices in  \rf{10092018-man03-01-add}. For these vertices, we outline the derivation of the respective solutions in \rf{10092018-man03-02}, \rf{11092018-man03-02},\rf{12092018-man03-02},\rf{13092018-man03-02}. We split our derivation in several steps.

\noindent {\bf Realization of $G_a$, $G_\beta$ on $p_\smp3^-$ \rf{10092018-man03-01-add} for arbitrary masses}. First, for $p_\smp3^-$  \rf{10092018-man03-01-add} with arbitrary $m_1$, $m_2$, $m_3$, we find $G_\beta$ and $G_{a,\Po^2}$ in \rf{01092018-man03-52}. We use $M_a^i$, $a=1,2,3$ for continuous-spin field \rf{01092018-man03-22}. Plugging such $M_a^i$ into $\Jbf^{-i\dagger}$ \rf{01092018-man03-47}, we cast $\Jbf^{-i\dagger}|p_\smp3^-\rangle$ into the form given in \rf{01092018-man03-52} with the following $G_\beta$ and $G_{a,\Po^2}$:
\beq
\label{16092018-man03-01} &&   \hspace{-1cm} G_{a,\Po^2} =   G_a +  \Pbf^- \frac{2 \beta}{\beta_a^3} \alpha_a^i \frac{ g_{\upsilon_a} \partial_{\upsilon_a} }{2N_a + d-2 } \partial_{B_a}^2\,,\qquad a=1,2,3
\\
\label{16092018-man03-02} G_a &  =  &    \Bigl(B_{a+2} - \frac{\beta_a}{\beta_{a+2}}  g_{\upsilon_{a+2}} \partial_{\upsilon_{a+2}}   \Bigr)\partial_{\alpha_{a+2 a}}
 -   \Bigl( B_{a+1} + \frac{\beta_a}{\beta_{a+1}}  g_{\upsilon_{a+1}} \partial_{\upsilon_{a+1}}  \Bigr) \partial_{\alpha_{aa+1}}
\nonumber\\
& + & \half \Bigl( \frac{\betach_a}{ \beta_a} m_a^2 +  m_{a+1}^2 - m_{a+2}^2  \Bigr) \partial_{B_a}    +   \upsilon_a g_{\upsilon_a}
\nonumber\\
& + & \frac{g_{\upsilon_a} \partial_{\upsilon_a} }{2N_a+d-2} \Bigl( \frac{2\beta_{a+1}}{\beta_a} B_{a+1} \partial_{B_a} \partial_{\alpha_{aa+1}} + \frac{2 \beta_{a+2}}{\beta_a} B_{a+2} \partial_{B_a} \partial_{\alpha_{a+2a}}
\nonumber\\
&  +  &  2 \alpha_{a+1a+2} \partial_{\alpha_{aa+1}}\partial_{\alpha_{a+2a}}
+ \frac{\beta}{\beta_a^2} \sum_{b=1,2,3} \frac{m_b^2}{ \beta_b} \partial_{B_a}^2 \Bigr),\qquad
\\
\label{16092018-man03-03} G _\beta &  =  &    -  \frac{1}{\beta}  \No_\beta - \sum_{a=1,2,3} \frac{1}{\beta_a^2} g_{\upsilon_a} \partial_{\upsilon_a} \partial_{B_a}\,,
\eeq
where $g_{v_a}$ are given in \rf{02092018-man03-06app-bb2}. Using $G_a$, $G_\beta$ \rf{16092018-man03-02},\rf{16092018-man03-03}, we now consider equations \rf{01092018-man03-53},\rf{01092018-man03-54}.

\noindent {\bf Vertex $V^{(3)}$}. We multiply $G_a$ \rf{16092018-man03-02} on the left by $(2N_a+d-2)/\kappa_a$ and make the transformation%
\footnote{ To investigate equations $G_ap_\smp3^-=0$ it is convenient to use equivalence class for the $G_a$. Namely, the $G_a$ and $(2N_a+d-2)G_a$ are considered to be equivalent.
}
\beq
\label{16092018-man03-04} && \hspace{-1cm} p_\smp3^-  =   U_{\upsilon_1}U_{\upsilon_2}U_{\upsilon_3} V^{(1)}\,, \qquad V^{(1)} = U_{\Gamma_1} U_{\Gamma_2}U_{\Gamma_3} V^{(2)}\,,  \qquad V^{(2)}   = U_\beta  V^{(3)}\,,
\\
\label{16092018-man03-05} && U_\beta   = \exp\Bigl( - \frac{\betach_1}{2\beta_1} \kappa_1 \partial_{B_1} - \frac{\betach_2}{2\beta_2} \kappa_2 \partial_{B_2} - \frac{\betach_3}{2\beta_3} \kappa_3 \partial_{B_3} \Bigr)\,,
\eeq
where $U_{\upsilon_a}$, $U_{\Gamma_a}$ are defined in \rf{02092018-man03-10app},\rf{02092018-man03-12app}. Using notation in \rf{02092018-man03-07app},\rf{02092018-man03-08app}, we find that realization of $G_a$, $G_\beta$ \rf{16092018-man03-02},\rf{16092018-man03-03} on $V^{(3)}$ \rf{16092018-man03-04} takes the form
\beq
\label{16092018-man03-06} \hspace{-0.7cm} G_a &  =  & - r_a \partial_{B_a}^2 + q_a \bigl(N_{B_a}  + \nu_a +1 \bigr)\partial_{B_a}   +  e_a (N_{B_a} + \nu_a)(  N_{B_a} + \nu_a + 1 )  + u_a
\nonumber\\
&+ & \frac{2}{\kappa_a} (N_a + \nu_a +1)\Bigl( (B_{a+2} + \half \kappa_{a+2})\partial_{\alpha_{a+2a}}  - (B_{a+1}   - \half \kappa_{a+1}) \partial_{\alpha_{aa+1}} \Bigr)
\nonumber\\
\hspace{-0.7cm} & - & ( B_{a+1} + \frac{3}{2}\kappa_{a+1}) \partial_{B_a} \partial_{\alpha_{aa+1}} - (B_{a+2} - \frac{3}{2}\kappa_{a+2}) \partial_{B_a} \partial_{\alpha_{a+2a}}  +   2\alpha_{a+1a+2} \partial_{\alpha_{aa+1}}\partial_{\alpha_{a+2a}},\qquad
\\
\label{16092018-man03-07} G _\beta &  =  &    -  \frac{1}{\beta}  \No_\beta  \,.
\eeq

\noindent  {\bf Dependence of $V^{(3)}$ on $\beta_a$}. We fix dependence of $V^{(3)}$ on $\beta_a$. To this end we use Eqs.\rf{01092018-man03-54},\rf{01092018-man03-58} and relation \rf{16092018-man03-07} to note the following equations for the vertex $V^{(3)}$:
\be \label{16092018-man03-08}
\No_\beta V^{(3)} = 0 \,, \qquad \sum_{a=1,2,3}\beta_a\partial_{\beta_a} V^{(3)} = 0\,.
\ee
Equations \rf{16092018-man03-08} imply that $V^{(3)}$ is independent of $\beta_1$, $\beta_2$,$\beta_3$. Thus, the vertex $V^{(3)}$ is given by
\be \label{16092018-man03-09}
V^{(3)} = V^{(3)}(B_a,\alpha_{aa+1})\,.
\ee
We now use results above-obtained for the study of four vertices in \rf{10092018-man03-01-add} in turn.
%
%
%
%
%
\be  \label{16092018-man03-10}
\hbox{\bf \large Case \quad  $m_1=0$,\quad $m_2=0$,\quad $m_3^2 < 0$.}
\ee
%
%
%
%
%
{\bf Step 1}. This case corresponds to the first vertex $p_\smp3^-$ \rf{10092018-man03-01-add}.  In \rf{16092018-man03-04}, we use operators $U_{\Gamma_1}$, $U_{\Gamma_2}$ given in \rf{02092018-man03-11app}. Then, in all remaining relations in \rf{16092018-man03-01}-\rf{16092018-man03-09}, we set $m_1=0$, $m_2=0$ and  use vertex $V^{(3)}$ given in \rf{16092018-man03-09},
where the realization of operators $G_a$ \rf{16092018-man03-06} on the vertex $V^{(3)}$ \rf{16092018-man03-09} takes the form
\beq
\label{16092018-man03-11} \hspace{-0.7cm} G_a &  =  & - r_a \partial_{B_a}^2 + q_a \bigl(N_{B_a}  + \nu_a +1 \bigr)\partial_{B_a}     + u_a
\nonumber\\
&+ & \frac{2}{\kappa_a} (N_{B_a} + \nu_a +1)\Bigl( (B_{a+2} + \half \kappa_{a+2})\partial_{\alpha_{a+2a}}  - (B_{a+1}   - \half \kappa_{a+1}) \partial_{\alpha_{aa+1}} \Bigr)
\nonumber\\
\hspace{-0.7cm} & - & ( B_{a+1} + \frac{3}{2}\kappa_{a+1}) \partial_{B_a} \partial_{\alpha_{aa+1}} - (B_{a+2} - \frac{3}{2}\kappa_{a+2}) \partial_{B_a} \partial_{\alpha_{a+2a}}  +   2\alpha_{a+1a+2} \partial_{\alpha_{aa+1}}\partial_{\alpha_{a+2a}},\qquad
\\
\label{16092018-man03-12} \hspace{-0.7cm} G_3 &  =  & - r_3 \partial_{B_3}^2  +  e_3 (N_{B_3} + \nu_3)(  N_{B_3} + \nu_3 + 1 )  + u_3
\nonumber\\
&+ & \frac{2}{\kappa_3} (N_{B_3} + \nu_3 +1)\Bigl( (B_2 + \half \kappa_2)\partial_{\alpha_{23}}  - (B_1   - \half \kappa_1) \partial_{\alpha_{31}} \Bigr)
\nonumber\\
\hspace{-0.7cm} & - & ( B_1 + \frac{3}{2}\kappa_1) \partial_{B_3} \partial_{\alpha_{31}} - (B_2 - \frac{3}{2}\kappa_2) \partial_{B_3} \partial_{\alpha_{23}}  +   2\alpha_{12} \partial_{\alpha_{31}}\partial_{\alpha_{23}}\,,\qquad
\eeq
where $a=1,2$, and we use quantities \rf{02092018-man03-07app},\rf{02092018-man03-08app} taken for masses given in \rf{16092018-man03-10}.

\noindent {\bf Step 2}. We make the transformation
\beq
\label{16092018-man03-13} && V^{(3)}
= U_{\partial B}U_{\partial \alpha} V^{(4)}\,, \qquad U_{\partial B} = \exp\bigl(-\frac{r_1}{q_1}\partial_{B_1} - \frac{r_2}{q_2}\partial_{B_2} \bigr)\,,
\\
\label{16092018-man03-14}  && U_{\partial\alpha}  = \exp\Bigl( \bigl( 2 B_1B_3 + \kappa_3 B_1 - \kappa_1\kappa_3)\frac{\partial_{\alpha_{31}}}{m_3^2} + \bigl(2B_2B_3  - \kappa_3B_2 - \kappa_2\kappa_3 ) \frac{ \partial_{\alpha_{23}  } }{m_3^2} %
\nonumber\\
&& - \frac{2B_1B_2}{m_3^2}  \partial_{\alpha_{12}}
+ \frac{2B_3}{m_3^2} (\kappa_1\partial_{\alpha_{31}} - \kappa_2 \partial_{\alpha_{23}}) + \frac{2\kappa_1\kappa_2}{m_3^2} \partial_{\alpha_{12}} \Bigr).
\eeq
Realization of $G_a$, $G_3$ \rf{16092018-man03-11},\rf{16092018-man03-12}
on $V^{(4)}$ \rf{16092018-man03-13} takes the form
\beq
\label{16092018-man03-15} \hspace{-0.7cm} G_1 &  =  & q_1 \bigl(N_{B_1}  + \nu_1 +1 \bigr)\partial_{B_1}     + u_1
\nonumber\\
\hspace{-0.7cm} & + & \bigl( \frac{4\kappa_3^2 e_3}{D} B_3^2 - \frac{1}{e_3} \bigr)  \partial_{\alpha_{31}}^2  + \frac{4 }{q_2}B_2 \partial_{\alpha_{12}}^2  +      2\alpha_{23} \partial_{\alpha_{12}}\partial_{\alpha_{31}}\,,
\nonumber\\
\label{16092018-man03-16} \hspace{-0.7cm} G_2 &  =  & q_2 \bigl(N_{B_2}  + \nu_2 +1 \bigr)\partial_{B_2}     + u_2
\nonumber\\
\hspace{-0.7cm} & + &   \bigl( \frac{4\kappa_3^2 e_3}{D} B_3^2 - \frac{1}{e_3} \bigr)   \partial_{ \alpha_{23} }^2  + \frac{4}{q_1} B_1 \partial_{\alpha_{12}}^2 +   2\alpha_{31} \partial_{ \alpha_{23} }\partial_{ \alpha_{12} },\qquad
\nonumber\\
\label{16092018-man03-17}  \hspace{-0.7cm} G_3 &  =  & - \frac{D}{4\kappa_3^2e_3} \partial_{B_3}^2  +  e_3 (N_{B_3} + \nu_3)(  N_{B_3} + \nu_3 + 1 )  + u_3
\nonumber\\
& +  & \frac{4}{q_1} B_1 \partial_{\alpha_{31} }^2
+ \frac{4}{q_2} B_2 \partial_{\alpha_{23} }^2
+   2\alpha_{12} \partial_{\alpha_{31}}\partial_{\alpha_{23}}  \,.
\eeq

\noindent {\bf Step 3}. We make the transformation
\beq
\label{16092018-man03-18}  && \hspace{-1cm} V^{(4)} = \bigl(-\frac{4}{q_1}B_1\bigr)_\vph^{-\nu_1/2} \bigl(-\frac{4}{q_2}B_2\bigr)_\vph^{-\nu_2/2}U_{B_3} V^{(5)}\,, \qquad
V^{(5)} = U_{\nu_1,W_1} U_{\nu_2,W_{23}} U_{\nu_3,W_{312}} V^{(6)}\,,
\nonumber\\
&& \hspace{-1cm}  V^{(6)} = U_{B_3}^{-1} V^{(7)}\,,
\eeq
where $U_{B_3}$, $U_{\nu,W}$ are given in \rf{02092018-man03-13app}, \rf{02092018-man03-14app} and we use the notation
\beq
&& \hspace{-1.6cm} W_a =     2  \alpha_{a+1a+2} \partial_{\alpha_{aa+1}}\partial_{\alpha_{a+2a}}   -   \partial_{\alpha_{aa+1}}^2 -   \partial_{\alpha_{a+2a}}^2 \,,\quad a=1,2,3\,,\qquad
\nonumber\\
\label{16092018-man03-20} && \hspace{-1.6cm} W_{23} =  2 \alpha_{31} \partial_{\alpha_{12}} \partial_{\alpha_{23}}  -    \partial_{\alpha_{23}}^2\,, \quad W_{32} =  2 \alpha_{12} \partial_{\alpha_{23}} \partial_{\alpha_{31}}  -    \partial_{\alpha_{23}}^2\,, \quad W_{312} =  2 \alpha_{12} \partial_{\alpha_{23}} \partial_{\alpha_{31}}\,.
\eeq
Realizations of $G_1$, $G_2$, $G_3$ \rf{16092018-man03-15} on $V^{(5)}$, $V^{(6)}$, $V^{(7)}$ \rf{16092018-man03-18} take the following forms
\beq
&& \hspace{-1.4cm} G_a  =  u_a +  q_a\bigl(N_{B_a}  + 1 \bigr)\partial_{B_a}    - \frac{ q_a }{4B_a}(\nu_a^2 + W_a) \,, \qquad a=1,2\,,
\nonumber\\
\label{16092018-man03-22}  && \hspace{-1.4cm} G_3  =   (B_3^2 -\frac{D}{4 \kappa_3^2 e_3^2 })  \partial_{B_3}^2   +   \bigl(1- \frac{4\kappa_3^2 e_3^2}{D} B_3^2 \bigr)_\vph^{-1} \bigl( \nu_3^2 - 1 + W_3 \bigr) + \frac{u_3}{e_3} \,, \hspace{1cm} \hbox{ for } V^{(5)};
\\
\label{16092018-man03-23} && \hspace{-1.4cm} G_a  =  u_a +  q_a\bigl(N_{B_a}  + 1 \bigr)\partial_{B_a}    - \frac{ q_a }{4B_a}\nu_a^2 \,, \qquad a=1,2\,,
\nonumber\\
\label{16092018-man03-24}  && \hspace{-1.4cm} G_3  =   (B_3^2 -\frac{D}{4 \kappa_3^2 e_3^2 })  \partial_{B_3}^2   +   \bigl(1- \frac{4\kappa_3^2 e_3^2}{D} B_3^2 \bigr)_\vph^{-1} \bigl( \nu_3^2 - 1\bigr) + \frac{u_3}{e_3} \,,  \hspace{2cm} \hbox{ for }V^{(6)};
\\
&& \hspace{-1.4cm}  G_a  =  u_a +  q_a\bigl(N_{B_a}  + 1 \bigr)\partial_{B_a}    - \frac{ q_a }{4B_a}\nu_a^2 \,, \qquad a=1,2\,, \qquad
\nonumber\\
\label{16092018-man03-25}  && \hspace{-1.4cm} G_3  =   - \frac{D}{4\kappa_3^2 e_3} \partial_{B_3}^2  +  e_3 (N_{B_3} + \nu_3)(  N_{B_3} + \nu_3 + 1 )  + u_3\,,  \hspace{3cm} \hbox{ for }V^{(7)}.\qquad
\eeq
For $\nu_a$ \rf{02092018-man03-06app-bb1} and $W$-operators \rf{16092018-man03-20}, we note the relations which admit us to get $G_{1,2,3}$ in \rf{16092018-man03-24},
\beq
&&  \hspace{-0.8cm} (\nu_1^2+ W_1)U_{\nu_1,W_1}  = U_{\nu_1,W_1}\nu_1^2\,, \hspace{2.5cm} \nu_1^2 U_{\nu_2,W_{23}}  = U_{\nu_2,W_{23}} \nu_1^2\,,\qquad
\nonumber\\
\label{16092018-man03-25-b1} &&  \hspace{-0.8cm}\nu_1^2 U_{\nu_3,W_{312}} =  U_{\nu_3,W_{312}} \nu_1^2\,,
\\
&&  \hspace{-0.8cm} (\nu_2^2+ W_2)U_{\nu_1,W_1}  = U_{\nu_1,W_1} (\nu_2^2 + W_{23})\,, \hspace{1cm} (\nu_2^2+ W_{23})U_{\nu_2,W_{23}}  = U_{\nu_2,W_{23}} \nu_2^2\,,
\nonumber\\
\label{16092018-man03-25-b2} &&  \hspace{-0.8cm} \nu_2^2 U_{\nu_3,W_{312}}  = U_{\nu_3,W_{312}} \nu_2^2\,,
\\
&& \hspace{-0.8cm} (\nu_3^2+ W_3)U_{\nu_1,W_1}  = U_{\nu_1,W_1} (\nu_3^2 + W_{32})\,, \hspace{1cm} (\nu_3^2+ W_{32})U_{\nu_2,W_{23}}  = U_{\nu_2,W_{23}} (\nu_3^2 +W_{312})\,, \hspace{1cm}
\nonumber\\
\label{16092018-man03-25-b3} &&  \hspace{-0.8cm} (\nu_3^2 +W_{312}) U_{\nu_3,W_{312}}  = U_{\nu_3,W_{312}} \nu_3^2\,.
\eeq
Equations $G_aV^{(7)}=0$ with $G_1$, $G_2$, $G_3$ as in \rf{16092018-man03-25} constitute a system of three decoupled second-order differential equations w.r.t. $B_1$, $B_2$, $B_3$. Eight independent solutions of these equations are given in \rf{10092018-man03-05}.
%
%
%
%
%
\be  \label{16092018-man03-26}
\hbox{\bf \large Case \quad  $m_1 = m$,\quad $m_2=m$,\quad $m_3=0$,\quad $m^2 < 0$.}
\ee
%
%
%
%
%
{\bf Step 1}. This case corresponds to the second vertex $p_\smp3^-$ \rf{10092018-man03-01-add}.  Now, in \rf{16092018-man03-04}, we use operator $U_{\Gamma_3}$ given in \rf{02092018-man03-11app}. Then, in all remaining relations in \rf{16092018-man03-01}-\rf{16092018-man03-09} we set $m_1=m$, $m_2=m$, $m_3=0$ and use the vertex $V^{(3)}$ given in \rf{16092018-man03-09},
where the realization of operators $G_a$ \rf{16092018-man03-06} on the vertex $V^{(3)}$ \rf{16092018-man03-09} takes the form
\beq
\label{16092018-man03-27} && \hspace{-1cm}  G_a   =  - r_a \partial_{B_a}^2 + q_a \bigl(N_{B_a}  + \nu_a +1 \bigr)\partial_{B_a}   +  e_a (N_{B_a} + \nu_a)(  N_{B_a} + \nu_a + 1 )  + u_a
\nonumber\\
&& \hspace{-0.4cm} +\,\, \frac{2}{\kappa_a} (N_{B_a} + \nu_a +1)\Bigl( (B_{a+2} + \half \kappa_{a+2})\partial_{\alpha_{a+2a}}  - (B_{a+1}   - \half \kappa_{a+1}) \partial_{\alpha_{aa+1}} \Bigr)
\nonumber\\
&& \hspace{-0.4cm}  -\,\, ( B_{a+1} + \frac{3}{2}\kappa_{a+1}) \partial_{B_a} \partial_{\alpha_{aa+1}} - (B_{a+2} - \frac{3}{2}\kappa_{a+2}) \partial_{B_a} \partial_{\alpha_{a+2a}}  +   2\alpha_{a+1a+2} \partial_{\alpha_{aa+1}}\partial_{\alpha_{a+2a}},\qquad
\\
\label{16092018-man03-28}  && \hspace{-1cm} G_3   =  - r_3 \partial_{B_3}^2   +  u_3
  +  \frac{2}{\kappa_3} (N_{B_3} + \nu_3 +1) \Bigl( (B_2 + \half \kappa_2)\partial_{\alpha_{23}}  - (B_1   - \half \kappa_1) \partial_{\alpha_{31}} \Bigr)
\nonumber\\
&& \hspace{0cm}  - \,\, ( B_1 + \frac{3}{2}\kappa_1) \partial_{B_3} \partial_{\alpha_{31}} - (B_2 - \frac{3}{2}\kappa_2) \partial_{B_3} \partial_{\alpha_{23}}  +   2\alpha_{12} \partial_{\alpha_{31}}\partial_{\alpha_{23}}\,,\qquad
\eeq
where $a=1,2$, and we use quantities \rf{02092018-man03-07app},\rf{02092018-man03-08app} taken for masses given in \rf{16092018-man03-26}.

\noindent {\bf Step 2}. We make the transformation
\beq
\label{16092018-man03-29} && \hspace{-1cm} V^{(3)}
= U_{\partial B}U_{\partial \alpha} V^{(4)}\,, \qquad U_{\partial B} = \exp\bigl(\frac{q_1}{2e_1}\partial_{B_1} + \frac{q_2}{2e_2}\partial_{B_2} \bigr)\,,
\\
\label{16092018-man03-30} &&  \hspace{-1cm} U_{\partial\alpha} = \exp\Bigl( \frac{B_3^2}{2\kappa_3 r_3 }(  B_2 \partial_{\alpha_{23}} - B_1\partial_{\alpha_{31}}
)  -  \frac{B_3}{2r_3 }(B_1\partial_{\alpha_{31}}
+ B_2 \partial_{\alpha_{23}})
\nonumber\\
&&  +  \frac{B_3}{\kappa_1 e_1 }  \partial_{\alpha_{31}} - \frac{B_3}{\kappa_2 e_2 }   \partial_{\alpha_{23}} - \frac{B_2}{\kappa_1 e_1 } \partial_{\alpha_{12}}
+ \frac{ B_1}{\kappa_2 e_2 }\partial_{\alpha_{12}}  - \frac{\kappa_1\kappa_2}{m^2 } \partial_{\alpha_{12}} - \frac{\kappa_2\kappa_3}{m^2 } \partial_{\alpha_{23}}- \frac{\kappa_3\kappa_1}{m^2 } \partial_{\alpha_{31}}
\nonumber\\
&& - \frac{1}{2\kappa_1^2 e_1 } B_1 B_2 \partial_{\alpha_{12}}   + \frac{\kappa_3}{8 r_3} (B_2 \partial_{\alpha_{23}} - B_1 \partial_{\alpha_{31}} )  \Bigr)\,.
\eeq
Realization of $G_a$ \rf{16092018-man03-27},\rf{16092018-man03-28} on $V^{(4)}$ \rf{16092018-man03-29} takes the form
\beq
\label{16092018-man03-31} \hspace{-0.7cm} G_1 &  =  &  e_1 (N_{B_1} + \nu_1)(  N_{B_1} + \nu_1 + 1 )  + u_1
\nonumber\\
\hspace{-0.7cm} & - & 2 B_2\partial_{B_1} \partial_{\alpha_{12}} + 2\kappa_3 \partial_{B_1} \partial_{\alpha_{31} }   - \frac{1}{e_2} \partial_{\alpha_{12} }^2  +    2\alpha_{23} \partial_{\alpha_{12}}\partial_{\alpha_{31} }\,,
\nonumber\\
\label{16092018-man03-32} \hspace{-0.7cm} G_2 &  =  &  e_2 (N_{B_2} + \nu_2)(  N_{B_2} + \nu_2 + 1 )  + u_2
\nonumber\\
\hspace{-0.7cm} & - & 2 B_1\partial_{B_2} \partial_{\alpha_{12}} - 2\kappa_3 \partial_{B_2} \partial_{\alpha_{23} }   - \frac{1}{e_1} \partial_{\alpha_{12} }^2  +    2\alpha_{31} \partial_{\alpha_{12}}\partial_{\alpha_{23} }\,,
\nonumber\\
\label{16092018-man03-33} \hspace{-0.7cm} G_3 &  =  &  - r_3\partial_{B_3}^2   +  u_3 +  \frac{1}{\kappa_3} \{\nu_3 , B_2\partial_{\alpha_{23}}  - B_1 \partial_{\alpha_{31}} \}
- \frac{1}{e_1} \partial_{\alpha_{31}}^2  - \frac{1}{e_2} \partial_{\alpha_{23}}^2 +   2\alpha_{12} \partial_{\alpha_{31}}\partial_{\alpha_{23}}.\qquad
\eeq

\noindent {\bf Step 3}. We make the transformation
\be \label{16092018-man03-34}
V^{(4)} =  e_1^{\omega_1/2} e_2^{\omega_2/2} V^{(5)}\,, \qquad  V^{(5)} = U_{\omega_1,W_1} U_{\omega_2,W_{23}} V^{(6)}\,,
\ee
where $e_a$ and $U_{\nu,W}$ are given in \rf{02092018-man03-08app}, \rf{02092018-man03-14app} and we use the notation
\beq
\label{16092018-man03-35} && \omega_a = N_{B_a} + \nu_a + \half\,, \qquad a=1,2\,,
\\
\label{16092018-man03-36} && W_1   =    - 2 B_2\partial_{B_1} \partial_{\alpha_{12}} +  2\kappa_3  \partial_{B_1} \partial_{\alpha_{31} }   -  \partial_{\alpha_{12} }^2  +   2\alpha_{23} \partial_{\alpha_{12}}\partial_{\alpha_{31} }\,,
\\
\label{16092018-man03-37} && W_2   =    -  2 B_1\partial_{B_2} \partial_{\alpha_{12}} -2 \kappa_3  \partial_{B_2} \partial_{\alpha_{23} }   -   \partial_{\alpha_{12} }^2  +   2 \alpha_{31} \partial_{\alpha_{12}}\partial_{\alpha_{23} }\,,
\\
\label{16092018-man03-38} && W_3   =      \frac{1}{ \kappa_3} \{\nu_3 , B_2\partial_{\alpha_{23}}  - B_1 \partial_{\alpha_{31}} \}  -   \partial_{\alpha_{31}}^2  -   \partial_{\alpha_{23}}^2 +   2 \alpha_{12} \partial_{\alpha_{31}}\partial_{\alpha_{23}}\,,
\\
\label{16092018-man03-39} && W_{23}   =     -  2 B_1\partial_{B_2} \partial_{\alpha_{12}} -2 \kappa_3 \partial_{B_2} \partial_{\alpha_{23} }   +   2 \alpha_{31} \partial_{\alpha_{12}}\partial_{\alpha_{23} }\,,
\\
\label{16092018-man03-40} && W_{321}  =    \frac{1}{ \kappa_3} \{\nu_3 , B_2\partial_{\alpha_{23}}  - B_1 \partial_{\alpha_{31}} \}    +   2 \alpha_{12} \partial_{\alpha_{31}}\partial_{\alpha_{23}}\,.
\eeq
Realizations of $G_1$, $G_2$, $G_3$ \rf{16092018-man03-31} on $V^{(5)}$, $V^{(6)}$ \rf{16092018-man03-34} take the forms
\beq
&& \hspace{-2cm} G_a =  (N_{B_a} + \nu_a)(  N_{B_a} + \nu_a + 1 )  + \frac{u_a}{e_a} +  W_a\,, \qquad a=1,2\,,
\nonumber\\
\label{16092018-man03-41} && \hspace{-2cm}  G_3 = u_3 - r_3 \partial_{B_3}^2 +  W_3 \,, \hspace{7.2cm} \hbox{ for } \ \ V^{(5)};
\\
&& \hspace{-2cm}  G_a =  (N_{B_a} + \nu_a)(  N_{B_a} + \nu_a + 1 )  + \frac{u_a}{e_a} \,, \qquad a=1,2\,,
\nonumber\\
\label{16092018-man03-42} && \hspace{-2cm}  G_3 = u_3 - r_3 \partial_{B_3}^2 +  W_{321} \,,  \hspace{7cm} \hbox{ for } \ \ V^{(6)}.
\eeq
For $\omega_a$ and $W$-operators in \rf{16092018-man03-35}-\rf{16092018-man03-40}, we note the relations which admit us to get $G_{1,2,3}$ in \rf{16092018-man03-42},
\beq
&&  \hspace{-1cm} (\omega_1^2 + W_1) U_{\omega_1,W_1} =  U_{\omega_1,W_1} \omega_1^2\,, \hspace{2.5cm} \omega_1^2 U_{\omega_2,W_{23}} =  U_{\omega_2,W_{23}} \omega_1^2\,,
\nonumber\\
&&  \hspace{-1cm} (\omega_2^2 + W_2) U_{\omega_1,W_1} =  U_{\omega_1,W_1}(\omega_2^2 + W_{23})\,, \hspace{1cm} (\omega_2^2 + W_{23}) U_{\omega_2,W_{23}} =  U_{\omega_2,W_{23}}\omega_2^2\,, \qquad
\nonumber\\
&& \hspace{-1cm} W_3 U_{\omega_1,W_1} =  U_{\omega_1,W_1}W_{32}\,, \hspace{3.6cm}
W_{32} U_{\omega_2,W_{23}} =  U_{\omega_2,W_{23}} W_{321}\,.
\eeq
Equations $G_aV^{(6)}=0$ with $G_1$, $G_2$, $G_3$ as in \rf{16092018-man03-42} are three decoupled second-order differential equations w.r.t. $B_1$, $B_2$, $B_3$. All independent solutions of these equations are given in \rf{11092018-man03-05}.
%
%
%
%
%
\be  \label{16092018-man03-43}
\hbox{\bf \large Case \quad  $m_1^2 < 0$, \quad $m_2^2 < 0$,\quad $m_3=0$, \quad $m_1 \ne m_2$.}
\ee
%
%
%
%
%
{\bf Step 1}. This case corresponds to the third vertex $p_\smp3^-$ \rf{10092018-man03-01-add}. In \rf{16092018-man03-04}, we use operator $U_{\Gamma_3}$ given in \rf{02092018-man03-11app}. Then, in all remaining relations in \rf{16092018-man03-01}-\rf{16092018-man03-09}, we set $m_3=0$ and  use vertex $V^{(3)}$ given in \rf{16092018-man03-09},
where the realization of operators $G_a$ \rf{16092018-man03-06} on the vertex $V^{(3)}$ \rf{16092018-man03-09} takes the form
\beq
\label{16092018-man03-44} \hspace{-0.7cm} G_a &  =  & - r_a \partial_{B_a}^2 + q_a \bigl(N_{B_a}  + \nu_a +1 \bigr)\partial_{B_a}   +  e_a (N_{B_a} + \nu_a)(  N_{B_a} + \nu_a + 1 )  + u_a
\nonumber\\
&+ & \frac{2}{\kappa_a} (N_{B_a} + \nu_a +1)\Bigl( (B_{a+2} + \half \kappa_{a+2})\partial_{\alpha_{a+2a}}  - (B_{a+1}   - \half \kappa_{a+1}) \partial_{\alpha_{aa+1}} \Bigr)
\nonumber\\
\hspace{-0.7cm} & - & ( B_{a+1} + \frac{3}{2}\kappa_{a+1}) \partial_{B_a} \partial_{\alpha_{aa+1}} - (B_{a+2} - \frac{3}{2}\kappa_{a+2}) \partial_{B_a} \partial_{\alpha_{a+2a}}  +   2\alpha_{a+1a+2} \partial_{\alpha_{aa+1}}\partial_{\alpha_{a+2a}},\qquad
\nonumber\\
\label{16092018-man03-45} \hspace{-0.7cm} G_3 &  =  & - r_3 \partial_{B_3}^2 + q_3 \bigl(N_{B_3}  + \nu_3 +1 \bigr)\partial_{B_3}   +  u_3
\nonumber\\
& + & \frac{2}{\kappa_3} (N_{B_3} + \nu_3 +1) \Bigl( (B_2 + \half \kappa_2)\partial_{\alpha_{23}}  - (B_1   - \half \kappa_1) \partial_{\alpha_{31}} \Bigr)
\nonumber\\
\hspace{-0.7cm} & - & ( B_1 + \frac{3}{2}\kappa_1) \partial_{B_3} \partial_{\alpha_{31}} - (B_2 - \frac{3}{2}\kappa_2) \partial_{B_3} \partial_{\alpha_{23}}  +   2\alpha_{12} \partial_{\alpha_{31}}\partial_{\alpha_{23}}\,,\qquad
\eeq
where, in \rf{16092018-man03-44}, $a=1,2$. In \rf{16092018-man03-44} and below, we use quantities \rf{02092018-man03-07app}-\rf{02092018-man03-09app-x1} taken for masses given in \rf{16092018-man03-43}.

\noindent {\bf Step 2}. We make the transformation
\beq
\label{16092018-man03-46} && V^{(3)}
= U_{\partial B}U_{\partial \alpha} V^{(4)}\,,   \qquad U_{\partial B} = \exp\bigl(\frac{q_1}{2e_1}\partial_{B_1} + \frac{q_2}{2e_2}\partial_{B_2} - \frac{r_3}{q_3}\partial_{B_3} \bigr)\,,
\\
\label{16092018-man03-47}  && U_{\partial \alpha} = \exp\Bigl( \frac{2 B_1B_3\partial_{ \alpha_{31} } - 2B_2 B_3 \partial_{ \alpha_{23} } }{\kappa_3 q_3}    + \frac{ B_3 \partial_{ \alpha_{31} } - B_2 \partial_{ \alpha_{12} }}{\kappa_1 e_1} + \frac{ B_1 \partial_{ \alpha_{12} } - B_3 \partial_{ \alpha_{23} }}{\kappa_2 e_2}
\nonumber\\
&& \hspace{2cm} -\,\,\, \frac{h_{12}}{2\kappa_1\kappa_2 e_1 e_2} \partial_{\alpha_{12}}  - \frac{\kappa_1}{q_3} \partial_{\alpha_{31}}  + \frac{\kappa_2}{q_3} \partial_{\alpha_{23}}
\nonumber\\
&& \hspace{2cm} + \,\, \frac{2h_{12}}{\kappa_3^2 q_3^2}B_1 B_2 \partial_{\alpha_{12}} - \frac{2m_1^2}{\kappa_3 q_3^2} B_1\partial_{\alpha_{31}} + \frac{2m_2^2}{\kappa_3 q_3^2} B_2\partial_{\alpha_{23}} \Bigr).
\eeq
Realization of $G_1$, $G_2$, $G_3$ \rf{16092018-man03-45} on $V^{(4)}$ \rf{16092018-man03-46} takes the form
\beq
\label{16092018-man03-48} \hspace{-0.7cm} G_1 &  =  & - \frac{D}{4\kappa_1^2 e_1} \partial_{B_1}^2  +  e_1 (N_{B_1} + \nu_1)(  N_{B_1} + \nu_1 + 1 )  + u_1
\nonumber\\
\hspace{-0.7cm} & + & \Bigl(\frac{ 4 \kappa_2^2 e_2 }{\kappa_3^2b_3^2} B_2^2 - \frac{1}{ e_2 }\Bigr) \partial_{\alpha_{12}}^2  + \frac{4}{q_3} B_3 \partial_{\alpha_{31}}^2 +  2\alpha_{23} \partial_{\alpha_{12}} \partial_{\alpha_{31} }\,,
\nonumber\\
\label{16092018-man03-49} \hspace{-0.7cm} G_2 &  =  & - \frac{D}{4\kappa_2^2 e_2} \partial_{B_2}^2  +  e_2 (N_{B_2} + \nu_2)(  N_{B_2} + \nu_2 + 1 )  + u_2
\nonumber\\
\hspace{-0.7cm} & + & \Bigl(\frac{ 4 \kappa_1^2 e_1 }{\kappa_3^2b_3^2} B_1^2 - \frac{1}{ e_1 }\Bigr) \partial_{\alpha_{12}}^2  + \frac{4}{q_3} B_3 \partial_{\alpha_{23}}^2 +  2\alpha_{31} \partial_{\alpha_{12}} \partial_{\alpha_{23} }\,,
\nonumber\\
\label{16092018-man03-50} \hspace{-0.7cm} G_3 &  =  &  q_3 \bigl(N_{B_3}  + \nu_3 +1 \bigr)\partial_{B_3}   +  u_3
\nonumber\\
& + & \bigl(\frac{4 \kappa_1^2 e_1}{\kappa_3^2 q_3^2} B_1^2 - \frac{1}{ e_1 } \bigr) \partial_{\alpha_{31}}^2 +  \bigl(\frac{4 \kappa_2^2 e_2}{\kappa_3^2 q_3^2} B_2^2 - \frac{1}{ e_2 } \bigr) \partial_{\alpha_{23}}^2
+  2\alpha_{12} \partial_{\alpha_{31}} \partial_{\alpha_{23}}\,.\qquad
\eeq

\noindent {\bf Step 3}. We make the transformations
\beq
\label{16092018-man03-51} && V^{(4)} = \bigl( -\frac{4}{q_3} B_3 \bigr)^{-\nu_3/2} U_{B_1} U_{B_2} V^{(5)}\,, \qquad
V^{(5)} = U_{\nu_1,W_1} U_{\nu_2,W_{23}} U_{\nu_3,W_{312}}V^{(6)}\,,\qquad
\nonumber\\
\label{16092018-man03-52} &&  V^{(6)} = U_{B_1}^{-1} U_{B_2}^{-1} V^{(7)}\,,
\eeq
where $U_{B_a}$, $U_{\nu,W}$ are given in \rf{02092018-man03-13app}, \rf{02092018-man03-14app} and we use the notation
\beq
&& \hspace{-1.9cm} W_a =     2  \alpha_{a+1a+2} \partial_{\alpha_{aa+1}}\partial_{\alpha_{a+2a}}   -   \partial_{\alpha_{aa+1}}^2 -   \partial_{\alpha_{a+2a}}^2 \,,\quad a=1,2,3\,,\qquad
\nonumber\\
\label{16092018-man03-54} && \hspace{-1.9cm} W_{23} =  2 \alpha_{31} \partial_{\alpha_{12}} \partial_{\alpha_{23}}  -    \partial_{\alpha_{23}}^2\,, \quad W_{32} =  2 \alpha_{12} \partial_{\alpha_{23}} \partial_{\alpha_{31}}  -    \partial_{\alpha_{23}}^2\,, \quad W_{312} =  2 \alpha_{12} \partial_{\alpha_{23}} \partial_{\alpha_{31}}\,.
\eeq
Realizations of $G_1$, $G_2$, $G_3$ \rf{16092018-man03-48} on $V^{(5)}$, $V^{(6)}$, $V^{(7)}$ \rf{16092018-man03-52} take the forms
\beq
&& \hspace{-2cm} G_a  =   (B_a^2 -\frac{D}{4 \kappa_a^2 e_a^2 })  \partial_{B_a}^2   +   \bigl(1- \frac{4\kappa_a^2 e_a^2}{D} B_a^2 \bigr)_\vph^{-1} \bigl( \nu_a^2 - 1 + W_a \bigr) + \frac{u_a}{e_a} \,,
\nonumber\\
\label{16092018-man03-56} && \hspace{-2cm} G_3  =  u_3 +  q_3\bigl(N_{B_3}  + 1 \bigr)\partial_{B_3}    - \frac{ q_3 }{4B_3}(\nu_3^2 + W_3) \,,  \hspace{4.6cm} \hbox{ for } V^{(5)}\,;
\\
&&  \hspace{-2cm} G_a  =   (B_a^2 -\frac{D}{4 \kappa_a^2 e_a^2 })  \partial_{B_a}^2   +   \bigl(1- \frac{4\kappa_a^2 e_a^2}{D} B_a^2 \bigr)_\vph^{-1} \bigl( \nu_a^2 - 1\bigr) + \frac{u_a}{e_a} \,,
\nonumber\\
\label{16092018-man03-57} && \hspace{-2cm} G_3  =  u_3 +  q_3\bigl(N_{B_3}  + 1 \bigr)\partial_{B_3}    - \frac{ q_3 }{4B_3}\nu_3^2 \,, \hspace{6cm} \hbox{ for } V^{(6)}\,;
\\
&& \hspace{-2cm} G_a  =   - \frac{D}{4\kappa_a^2 e_a} \partial_{B_a}^2  +  e_a (N_{B_a} + \nu_a)(  N_{B_a} + \nu_a + 1 )  + u_a\,,
\nonumber\\
\label{16092018-man03-58} && \hspace{-2cm} G_3  =  u_3 +  q_3\bigl(N_{B_3}  + 1 \bigr)\partial_{B_3}    - \frac{ q_3 }{4B_a}\nu_3^2 \,, \hspace{6cm} \hbox{ for } V^{(7)}\,;
\eeq
where, in \rf{16092018-man03-56}-\rf{16092018-man03-58}, $a=1,2$.
For $\nu_a$ \rf{02092018-man03-06app-bb1} and $W$-operators \rf{16092018-man03-54}, we note the relations \rf{16092018-man03-25-b1}-\rf{16092018-man03-25-b3} which admit us to get $G_1$, $G_2$, $G_3$ in \rf{16092018-man03-57}.
Equations $G_aV^{(7)}=0$ with $G_1$, $G_2$, $G_3$ as in \rf{16092018-man03-58} constitute a system of three decoupled second-order differential equations w.r.t. $B_1$, $B_2$, $B_3$. Eight independent solutions of these equations are given in \rf{12092018-man03-05}.

%
%
%
%
\be  \label{16092018-man03-59}
\hbox{\bf \large Case \quad  $m_1^2 <0$,\quad $m_2^2 < 0$,\quad $m_3^2 < 0$.}
\ee
%
%
%
%
%
{\bf Step 1}. This case corresponds to the fourth vertex $p_\smp3^-$ in \rf{13092018-man03-01-add}.  Vertex $V^{(3)}$ for this case is given in \rf{16092018-man03-09}, where the realization of operators $G_a$ \rf{16092018-man03-02} on the vertex $V^{(3)}$ is given in \rf{16092018-man03-06}.
We make the transformations
\beq
\label{16092018-man03-60} && \hspace{-1.8cm}  V^{(3)} = U_{\partial B} U_{\partial \alpha} V^{(4)}\,, \qquad U_{\partial B}  = \exp\bigl( \sum_{a=1,2,3} \frac{q_a}{2e_a}\partial_{B_a} \bigr)\,,
\\
\label{16092018-man03-61} && \hspace{-1.8cm} U_{\partial\alpha} = \exp\Bigl( \sum_{a=1,2,3}\frac{1}{ \kappa_a e_a }( B_{a+2} \partial_{ \alpha_{a+2a} } - B_{a+1} \partial_{ \alpha_{aa+1} } ) - \frac{h_{a+1a+2} }{ 2\kappa_{a+1}\kappa_{a+2}e_{a+1}e_{a+2} }\partial_{\alpha_{a+1a+2} }\Bigr) U_D\,,
\eeq
where $U_D$ is defined in \rf{02092018-man03-15app}. Realization of $G_a$ \rf{16092018-man03-06} on $V^{(4)}$ \rf{16092018-man03-60} takes the form
\beq
\label{16092018-man03-62} \hspace{-0.7cm} G_a &  =  & - \frac{D}{4\kappa_a^2 e_a} \partial_{B_a}^2  +  e_a (N_{B_a} + \nu_a)(  N_{B_a} + \nu_a + 1 )  + u_a
\nonumber\\
& + &  \bigl(\frac{4\kappa_{a+1}^2 e_{a+1} }{D} B_{a+1}^2 - \frac{1}{e_{a+1}}\bigr) \partial_{\alpha_{aa+1}}^2 + \bigl(\frac{4\kappa_{a+2}^2 e_{a+2} }{D} B_{a+2}^2  - \frac{1}{e_{a+2} }\bigr) \partial_{\alpha_{a+2a}}^2
\nonumber\\
& + &  2\alpha_{a+1a+2} \partial_{\alpha_{aa+1}}\partial_{\alpha_{a+2a}}\,, \qquad a=1,2,3\,,
\eeq
where, here and below, we use quantities defined in \rf{02092018-man03-07app}-\rf{02092018-man03-09app-x1}.

\noindent {\bf Step 2}. We make the transformations
\be \label{16092018-man03-63}
V^{(4)} =   U_{B_1} U_{B_2}U_{B_3} V^{(5)}\,, \qquad
V^{(5)} = U_{\nu_1,W_1} U_{\nu_2,W_{23}} U_{\nu_3,W_{312}}V^{(6)},
\qquad V^{(6)} = U_{B_1}^{-1} U_{B_2}^{-1}U_{B_3}^{-1} V^{(7)}\,,
\ee
where $U_{B_a}$, $U_{\nu,W}$ are given in \rf{02092018-man03-13app}, \rf{02092018-man03-14app} and we use the notation
\beq
&& \hspace{-1.6cm} W_a =     2  \alpha_{a+1a+2} \partial_{\alpha_{aa+1}}\partial_{\alpha_{a+2a}}   -   \partial_{\alpha_{aa+1}}^2 -   \partial_{\alpha_{a+2a}}^2 \,,\quad a=1,2,3\,,\qquad
\nonumber\\
\label{16092018-man03-64} && \hspace{-1.6cm} W_{23} =  2 \alpha_{31} \partial_{\alpha_{12}} \partial_{\alpha_{23}}  -    \partial_{\alpha_{23}}^2\,, \quad W_{32} =  2 \alpha_{12} \partial_{\alpha_{23}} \partial_{\alpha_{31}}  -    \partial_{\alpha_{23}}^2\,, \quad W_{312} =  2 \alpha_{12} \partial_{\alpha_{23}} \partial_{\alpha_{31}}\,.
\eeq
Realizations of $G_a$ \rf{16092018-man03-62} on $V^{(5)}$, $V^{(6)}$, $V^{(7)}$ \rf{16092018-man03-63} take the forms
\beq
\label{16092018-man03-67}  && \hspace{-2cm} G_a  =   (B_a^2 -\frac{D}{4 \kappa_a^2 e_a^2 })  \partial_{B_a}^2   +   \bigl(1- \frac{4\kappa_a^2 e_a^2}{D} B_a^2 \bigr)_\vph^{-1} \bigl( \nu_a^2 - 1 + W_a \bigr) + \frac{u_a}{e_a} \,,\hspace{1cm} \hbox{ for } V^{(5)}\,;
\\
\label{16092018-man03-68} &&  \hspace{-2cm} G_a  =   (B_a^2 -\frac{D}{4 \kappa_a^2 e_a^2 })  \partial_{B_a}^2   +   \bigl(1- \frac{4\kappa_a^2 e_a^2}{D} B_a^2 \bigr)_\vph^{-1} \bigl( \nu_a^2 - 1\bigr) + \frac{u_a}{e_a} \,, \hspace{2cm} \hbox{ for } V^{(6)}\,;
\\
\label{16092018-man03-69} &&  \hspace{-2cm}  G_a   =    - \frac{D}{4\kappa_a^2e_a} \partial_{B_a}^2  + e_a (N_{B_a} + \nu_a)(  N_{B_a} + \nu_a + 1 ) + u_a \,, \hspace{2.8cm} \hbox{ for } V^{(7)}\,;
\eeq
where, in \rf{16092018-man03-67}-\rf{16092018-man03-69}, $a=1,2,3$. For $\nu_a$ \rf{02092018-man03-06app-bb1} and $W$-operators \rf{16092018-man03-64}, we note the relations \rf{16092018-man03-25-b1}-\rf{16092018-man03-25-b3} which admit us to get $G_a$ in \rf{16092018-man03-68}.
Equations $G_aV^{(7)}=0$ with $G_{1,2,3}$ as in \rf{16092018-man03-69} constitute a system of three decoupled second-order differential equations w.r.t. $B_1$, $B_2$, $B_3$. Eight independent solutions of these equations are given in \rf{13092018-man03-05}.

\newsection{ \large One continuous-spin massless field, one arbitrary spin massless field, and one arbitrary spin massive field}

We use the shortcut $(0,\kappa)_\smCSF$ for a continuous-spin massless  field, while the shortcuts $(0,s)$ and $(m,s)$ are used for the respective  spin-$s$ massless and spin-$s$ massive field. We now consider cubic vertices for the following three fields:
\beq
\label{04092018-man03-01-q} && (0,s_1)\hbox{-}(m_2,s_2)\hbox{-}(0,\kappa_3)_\smCSF^\vph \hspace{0.8cm} \hspace{1cm} m_2^2 > 0\,,
\nonumber\\
&&\hbox{\small one massless field, one massive field, and one continuous-spin massless field.}\quad
\eeq
Relation \rf{04092018-man03-01-q} tells us that  spin-$s_1$ massless and spin-$s_2$  massive fields carry the respective external line indices $a=1$ and $a=2$, while the continuous-spin massless field corresponds to $a=3$.

For fields \rf{04092018-man03-01-q}, we find the following general solution to cubic vertex $p_\smp3^-$
\beq
\label{04092018-man03-02-q} p_\smp3^-  & = & U_\upsilon U_\Gamma U_\beta  U_\zeta U_{\partial B} U_{\partial \alpha} U_{JB} U_W   V_\pm^{(6)} \,,
\\
\label{04092018-man03-03-q} && p_\smp3^- = p_\smp3^-  (\beta_a, B_a, \alpha_{aa+1}\,, \zeta_2\,, \upsilon_3)\,,
\\
\label{04092018-man03-04-q} && V_\pm^{(6)} = V_\pm^{(6)}(B_2,B_3, \alpha_{aa+1})\,.
\eeq
In \rf{04092018-man03-02-q}, we introduce two vertices $V_\pm^{(6)}$ labelled by the superscripts $\pm$. In \rf{04092018-man03-03-q} and  \rf{04092018-man03-04-q}, the arguments of the generic vertex $p_\smp3^-$ and the vertices $V_\pm^{(6)}$ are shown explicitly. The definition of the arguments $B_a$ and $\alpha_{ab}$ may be found in \rf{02092018-man03-04app}. Various quantities $U$ appearing in \rf{04092018-man03-02-q} are differential operators w.r.t. the $B_a$ and $\alpha_{aa+1}$. These quantities will be presented below. For two vertices $V_\pm^{(6)}$ \rf{04092018-man03-04-q}, we find the following solution:
\beq
\label{04092018-man03-05-q} && \hspace{-1 cm} V_+^{(6)}  = I_{\nu_3}(\sqrt{z_3})V_+\,,\qquad  V_-^{(6)}  = K_{\nu_3}(\sqrt{z_3})V_-\,, \qquad V_\pm = V_\pm (B_2,\alpha_{12},\alpha_{23},\alpha_{31})\,,\quad
\\
\label{04092018-man03-06-q} &&   \hspace{-1 cm}  z_3 = \frac{4\kappa_3B_3}{m_2^2}\,,
\eeq
where, in \rf{04092018-man03-05-q}, the $I_\nu$ and $K_\nu$ are the modified Bessel functions. In \rf{04092018-man03-05-q}, in place of the variable $B_3$, we use new variable $z_3$ \rf{04092018-man03-06-q}. Operator $\nu_3$ is defined below.

Generic vertex $p_\smp3^-$ \rf{04092018-man03-03-q} depends on the eleven variables, while,  the vertices $V_\pm$ \rf{04092018-man03-05-q} depend only on the four variables. By definition, the vertices $V_\pm$ \rf{04092018-man03-05-q} are expandable in the variables $B_2$, $\alpha_{12}$, $\alpha_{23}$, $\alpha_{31}$. The general solution \rf{04092018-man03-02-q} for the vertex $p_\smp3^-$ is expressed  in terms of the operators $U$, $\nu_3$ acting on the vertices $V_\pm$  \rf{04092018-man03-05-q}. To complete the description of the vertex $p_\smp3^-$ we now provide expressions for the operators $U$, $\nu_3$ given by
\beq
\label{04092018-man03-08-q} && \hspace{-1.5cm} U_\upsilon = \upsilon_3^{N_3}\,, \qquad  \qquad N_3 = N_{B_3} + N_{ \alpha_{31} }  + N_{ \alpha_{23} }\,,
\\
\label{04092018-man03-09-q} && \hspace{-1.5cm} U_\Gamma   =  \Bigl( \frac{2^{N_3} \Gamma( N_3 + \frac{d-2}{2}) }{   \Gamma(N_3 + 1) } \Bigr)^{1/2}\,,
\\
\label{04092018-man03-10-q} && \hspace{-1.5cm} U_\beta   = \exp\bigl(- \frac{\betach_2}{2\beta_2} \zeta_2 m_2 \partial_{B_2} - \frac{\betach_3}{2\beta_3} \kappa_3 \partial_{B_3} \bigr)\,,
\\
&& \hspace{-1.5cm} U_\zeta  = \exp\Bigl(-\frac{\zeta_2}{m_2} B_1\partial_{\alpha_{12}} + \frac{\zeta_2}{m_2}( B_3 - \half \kappa_3 ) \partial_{\alpha_{23}} \Bigr)\,,
\\
&& \hspace{-1.5cm} U_{\partial B} = \exp\bigl( \frac{1}{2 }\kappa_3  \partial_{ B_3}  \bigr)
\\
\label{04092018-man03-11-q} && \hspace{-1.5cm} U_{\partial\alpha} = \exp\Bigl(  \frac{2\kappa_3 }{m_2^2} B_2  \partial_{ \alpha_{23} }  + \frac{2 B_1 B_2}{m_2^2}  \partial_{\alpha_{12}} + \frac{B_2 B_3}{m_2^2}\partial_{\alpha_{23}} - \frac{2 B_3 B_1}{m_2^2}  \partial_{\alpha_{31}} \Bigr)\,,
\\
\label{04092018-man03-12-q} && \hspace{-1.5cm} U_{JB} = \bigl(   \frac{4\kappa_3B_3}{m_2^2 }\bigr)^{ -\nu_3/2 }
\,,
\\
\label{04092018-man03-13-q} && \hspace{-1.5cm} U_W  = \sum_{n=0}^\infty \frac{\Gamma(\nu + n)}{4^nn!\Gamma(\nu +2n) } W^n\,,
\\
\label{04092018-man03-14-q} && \hspace{-1.5cm} W_3 =    - \frac{4 B_2^2}{ m_2^2 } \partial_{\alpha_{23}}^2  + 2 \alpha_{12} \partial_{\alpha_{31}} \partial_{\alpha_{23}}\,,  \qquad \nu_3  =   N_{\alpha_{23}} +   N_{\alpha_{31}} + \frac{d-4}{2}\,,
\eeq
where quantities $\betach_a$, $N_{B_a}$, $N_{\alpha_{ab}}$, $N_a$ appearing in  \rf{04092018-man03-08-q}-\rf{04092018-man03-14-q} are defined in \rf{02092018-man03-04app}-\rf{02092018-man03-06app}.

Expressions \rf{04092018-man03-02-q}-\rf{04092018-man03-14-q} provide the complete generating form description of cubic vertices for coupling of one continuous-spin massless field to  two chains of massless and massive fields \rf{01092018-man03-15}. Now we describe cubic vertices for coupling of one continuous-spin massless field to massless and massive fields having the respective arbitrary but fixed spin-$s_1$ and spin-$s_2$ values. Using the first algebraic constraints in \rf{01092018-man03-13},\rf{01092018-man03-14}, it is easy to see that vertices we are interested in must satisfy the algebraic constraints
\be \label{04092018-man03-16-q}
( N_{\alpha_1} - s_1 )|p_\smp3^-\rangle  = 0\,,\qquad ( N_{\alpha_2} + N_{\zeta_2} - s_2 )|p_\smp3^-\rangle  = 0\,.
\ee
In terms of the vertices $V_\pm$ \rf{04092018-man03-05-q}, constraints \rf{04092018-man03-16-q} take the form
\be \label{04092018-man03-17-q}
( N_{ \alpha_{12} } + N_{ \alpha_{31} } - s_1) V =0 \,,
\qquad ( N_{B_2} + N_{ \alpha_{12} } + N_{ \alpha_{23} } - s_2) V =0 \,, \qquad V =V_\pm\,,
\ee
where to simplify our presentation we drop the superscripts $\pm$ and use a vertex $V$ in place of the vertices $V_\pm$, i.e., we use $V=V_\pm$. Doing so, we note that the general solution to constraints \rf{04092018-man03-17-q} can be presented as
\beq
\label{04092018-man03-18-q} && V  =   V(s_1,s_2;\,n,k) \,, \qquad  V(s_1,s_2;\, n,k)  =  B_2^{s_2-s_1 + k -n}\alpha_{12}^{s_1-k} \alpha_{23}^n  \alpha_{31}^k \,,\qquad
\\
\label{04092018-man03-19-q} && 0 \leq k \leq s_1\,, \hspace{2.5cm} 0 \leq n \leq s_2 - s_1+k\,.
\eeq
The integers $n$, $k$ \rf{04092018-man03-18-q} are the freedom of our solution for the vertex $V$. In order for vertices \rf{04092018-man03-18-q} to be sensible, the integers $n$, $k$ should satisfy restrictions \rf{04092018-man03-19-q} which amount to the requirement that the powers of all variables $B_2$, $\alpha_{12}$,  $\alpha_{23}$,  $\alpha_{31}$ in \rf{04092018-man03-18-q} be non--negative.
Expressions for cubic interaction vertices given in  \rf{04092018-man03-02-q}-\rf{04092018-man03-14-q}, \rf{04092018-man03-18-q} and values of $n$, $k$ given in \rf{04092018-man03-19-q} provide the complete description and classification of parity invariant cubic interaction vertices that can be constructed for one spin-$s_1$ massless field, one spin-$s_2$ massive field and one continuous-spin massless field.


\end{document}